\documentclass[twocolumn]{aa} 
\usepackage{graphicx} 
\usepackage{natbib} 
\bibpunct{(}{)}{;}{a}{}{,} 
\usepackage{txfonts} 
\begin{document}
   \title{The impact of Galactic synchrotron emission 
   on CMB anisotropy measurements} 
    \subtitle{I. Angular power spectrum analysis of total intensity 
    all-sky surveys.} 


   \author{L.~La~Porta\inst{1},  
          C.~Burigana\inst{2,3}, 
      W.~Reich\inst{1}, 
      \and
      P.~Reich\inst{1} 
          }

   \offprints{L.~La~Porta}

   \institute{Max-Planck-Institut f\"ur Radioastronomie,
              Auf dem H\"ugel, 69, D-53121 Bonn, Germany\\
             \and
              INAF-IASF Bologna,
              via P.~Gobetti, 101, I-40129 Bologna, Italy\\
             \and
              Dipartimento di Fisica, 
              Universita' degli Studi di Ferrara,
              Via Saragat, 1, I-44100 Ferrara, Italy\\
             }

   \date{Received 7 August 2007 / Accepted 8 November 2007}


  \abstract
   { Galactic foreground emission fluctuations are a limiting factor for precise
   cosmic microwave background (CMB) anisotropy measurements.}
   { We intend to improve current estimates of the influence of Galactic 
   synchrotron emission on the analysis of CMB anisotropies.  } 
   { We perform an angular power spectrum analysis (APS) of all-sky total 
   intensity maps at 408~MHz and 1420~MHz, which are dominated 
  by synchrotron emission out of the Galactic plane. 
  We subtract the brighter sources from the maps, which 
  turns out to be essential for the results obtained. 
  We study the APS as a function of Galactic latitude by considering 
  various cuts and as a function of sky position 
   by dividing the sky into patches of $\sim 15^{\circ}\times 15^{\circ}$ in size. } 
   { The APS of the Galactic radio diffuse synchrotron emission 
   is best fitted by a power law, $C_{\ell} \sim k \ell^{\alpha}$, 
   with $\alpha \in [-3.0,-2.6]$, where the lower values 
   of $\alpha$ typically correspond to the higher 
   latitudes. Nevertheless, the analysis of the patches 
   reveals that strong local variations exist. 
   A good correlation is found between the APS normalized amplitude, 
   $k_{100}=k \times 100^\alpha$, at 408~MHz and 1420~MHz. 
   The mean APS for $\ell \in [20,40]$ is used to determine 
   the mean spectral index between 408 MHz and 1420 MHz,  
    $\beta_{(0.408-1.4){\rm GHz}} \in [-3.2,-2.9]$ 
   ($C_{\ell}(\nu) \propto \nu^{-2\beta}$), which is then adopted to 
   extrapolate the synchrotron APS results to the microwave range. 
   } 
   { We use the 408 MHz and 1420 MHz APS results to predict  
   the Galactic synchrotron emission fluctuations at frequencies 
   above 20~GHz. 
   A simple extrapolation to 23~GHz of the synchrotron emission APS 
   found at these radio frequencies does not explain
   all the power in the WMAP synchrotron component even 
   at middle/high Galactic latitudes. 
   This suggests a significant microwave contribution 
   (of about $50\%$ of the signal)
   by other components such as free-free or spinning dust emission.
   The comparison between the extrapolated synchrotron APS and 
   the CMB APS shows that a mask excluding the 
   regions with 
   $|b_{gal}| \lesssim 5^{\circ}$ 
   would reduce the foreground fluctuations 
   to about half of the cosmological ones 
   at 70~GHz 
   even at the lowest multipoles. 
   The main implications of our analysis for the cosmological 
   exploitation of microwave temperature anisotropy maps are discussed.   
 }

\authorrunning{La Porta et al.}

\titlerunning{APS analysis of total intensity all-sky surveys}

   \maketitle
%

\section{Introduction}

  The possibility to study the primordial phases of our 
  Universe and its properties and evolution through CMB anisotropies 
  relies on our capability to precisely extract the cosmological signal 
  from observations \citep{teg00_fore}. 
  Maps of the microwave sky include many 
  Galactic and extragalactic astrophysical contributions. 
  A correct recovery of the CMB anisotropy field requires 
  an accurate removal of those foreground signals from the 
  observed maps. Current knowledge of the foreground components 
  permits one to retrieve the bulk of the cosmological 
  information encoded in the CMB anisotropy, and, in particular, 
  its angular power spectrum (APS). 
  Nevertheless, a deeper understanding of the foregrounds is 
  crucial to settle important cosmological 
  issues, which arose from the 
  {\sc WMAP}~\footnote{http://lambda.gsfc.nasa.gov/} results 
  \citep{naselski06_wmap1_et_fore,naselski06_wmap_ell4,chiang07_lowLanom} 
  and could potentially be addressed by the forthcoming 
  {\sc Planck}~\footnote{http://www.rssd.esa.int/Planck} 
  mission \citep{tauber04_cospar}. 
  Moreover, it would enable a precise reconstruction of the 
  individual foreground components and therefore 
  a complete astrophysical exploitation of the satellite data.

Galactic synchrotron emission is the major source of 
contamination at frequencies below 50~GHz 
for intermediate and large angular scales,  
as recently confirmed by the impressive {\sc WMAP} results 
 \citep{ben03_wmap_1yr_fore,hins06_wmap_3yr_temp}.  
Synchrotron radiation \citep{rybicki79} arises 
from cosmic ray electrons gyrating in the magnetic field 
of our Galaxy. 
The energy spectrum and the density of the cosmic ray electrons
as well as the magnetic field strength vary across the Galaxy, 
therefore the observed synchrotron emission will depend
on the frequency and on the region of the sky. 
Radio observations at $\nu \lesssim 5$~GHz provide 
the clearest picture of the Galactic synchrotron morphology, 
since at these frequencies the diffuse non-thermal radiation 
clearly dominates over all other emission components 
outside the Galactic plane. 
In the past, the 408~MHz all-sky map \citep{haslam82_408mhz} 
has extensively been used as a template for the 
Galactic synchrotron emission in foreground separation attempts 
(e.g. Bouchet \& Gispert~1999; Bennett et al.~2003). 
In addition, that map as well as other less suited 
surveys have been exploited to find an appropriate 
parametrization of the synchrotron emission APS to be  
$i)$ extrapolated to the microwave range for estimating the 
contamination of CMB anisotropies at different angular scales 
\citep{giard01_hardtrao,bacci01_synch} 
and  
$ii)$ used to set priors in foreground separation applications   
 \citep{teg96_lf,bouchet99_wf,bouchet99_foresim}. 
The outcome of these analyses is that the APS of the 
synchrotron emission computed over large portions of the sky 
can be modelled 
by a power law, i.e. $C_{\ell} \sim k\; \ell^{\alpha}$ 
($\ell \sim 180^{\circ}/{\theta}$), with a spectral 
 index $\alpha \sim [-3,-2.5]$ for $\ell \lesssim 200$, corresponding 
 to angular scales $\theta \gtrsim 1^{\circ}$.  
Clearly, such a general result is just a first step as it 
does not describe the complexity of the synchrotron emission APS,  
whose parameters are expected to change with frequency 
and with sky direction.

We have carried out a detailed analysis of all-sky radio 
maps and improved on previous attempts by providing an 
accurate characterization of the synchrotron 
emission APS. 
The results obtained for the new 1.4~GHz polarization 
all-sky survey (Reich et al., in prep.) will be 
reported in a forthcoming companion paper. 
The analysis presented in this paper focuses on 
the synchrotron emission APS in total intensity. 
A substantial improvement was possible by using 
 a new all-sky map at 1.42~GHz (Reich et al., in prep.),   
 which has a higher angular resolution and better sensitivity,  
 in addition to the all-sky map at 408~MHz. 
 These maps are currently the best suited data 
for studying the Galactic synchrotron emission 
at largest angular scales. 
A more detailed description of some technical aspects related to this work
can be found in \citet{phdthesis}.   
We extensively used the 
{\tt HEALPix}\footnote{http://healpix.jpl.nasa.gov/} 
software package \citep{gorski05_healpix}. 
{\tt HEALPix} (Hierarchical Equal Area
isoLatitude Pixelization) is a curvilinear partition of
the sphere optimized for fast spherical harmonics 
transforms and angular power spectrum estimation. 
The latter task is performed by the facility {\tt Anafast}. 
We also made use of the data reduction package 
based on the NOD2-software \citep{haslam74_nod}. \\

The paper is organized as follows.
Sect.~2 describes the characteristics of the 408~MHz 
and 1.42~GHz total intensity surveys, their projection 
onto {\tt HEALPix} maps and noise considerations. 
In Sect.~3 the Galactic radio emission APS 
over large areas is examined, which reveals 
the necessity of a discrete source subtraction 
 for a correct evaluation of the diffuse synchrotron APS. 
The two all-sky maps are decomposed into a map of the 
diffuse component and a map of discrete sources. 
Their angular power spectra are derived and discussed. 
The results obtained by fitting the angular power 
spectra of the diffuse component maps are presented. 
In Sect. 4 the radio survey angular power spectra 
are extrapolated to the microwave range for a 
comparison with the WMAP 3-yr results. 
Sect. 5 is dedicated to a local analysis of the radio map APS. 
We summarize our results and conclusions in Sect.~6. \\ 

\section{The data}

The present analysis focuses on the APS of the 
Galactic synchrotron emission at radio frequencies. 
However, the 23 GHz synchrotron component obtained 
by \citet{hins06_wmap_3yr_temp} 
 using the WMAP 3-yr data has also been 
considered to some extent and will be further 
discussed in Sect.~\ref{res_extrapolation}.\\ 

\subsection{The 408~MHz and 1420~MHz surveys}

The 408~MHz map \citep{haslam82_408mhz} was produced by merging 
different datasets obtained with large parabolic reflector  
telescopes (Jodrell Bank 76~m, 
 Effelsberg 100~m and Parkes 64~m telescopes - 
see Fig.~1 of Haslam et al.~1982), using a similar  
observing strategy and the same calibration procedure.  
The final map is characterized by an angular resolution of 
 $\theta_{HPBW} \sim 0\fdg85$ and a $20\arcmin$ pixel rms-noise 
of about 670~mK. 
The version used in the present analysis was corrected 
for a zero level problem concerning the portion of the sky observed
from Jodrell Bank \citep{reich88_betasynchr}.\\ 
The total intesity map at 1420~MHz has been obtained by combining a northern and
a southern sky survey (Reich et al., in preparation). 
 Both surveys are on an absolute temperature scale and zero level 
by using low resolution sky horn measurements \citep{testori01_ssky}. 
This includes a correction for far-side lobe contamination 
for single-dish telescopes.  
Receiving systems were used, which provide total intensities unaffected by 
linear polarization. 
The northern sky survey 
was observed with the Stockert 25-m telescope and 
 extends in declination from $-19^{\circ}$ to $90^{\circ}$
\citep{reich82_stock,reich86_stock}. 
The southern sky survey was carried out with the Villa Elisa 30-m telescope
in Argentina and covers the latitude range 
$\delta \in [-90^{\circ}$ to $-10^{\circ}]$ 
\citep{reich01_ssky}. 
Both have an angular resolution $\theta_{HPBW} \sim 36'$
and overlap for latitudes in the range [$-19^{\circ}$,$-10^{\circ}$].
The resulting map has a $15\arcmin$ pixel rms-noise of $\sim 17 {\rm mK}$.
The original maps are provided 
as equidistant cylindrical ({\tt ECP}) maps. 
For the present analysis these maps have been projected 
into the {\tt HEALPix} \citep{gorski05_healpix} 
pixelization scheme adopted by the {\sc WMAP} and 
{\sc Planck } Consortia.  
For this purpose a simple regridding algorithm has been implemented, 
which is described in detail by \citet{techrep}. 
The reliability of the projection provided by this simple
approach has been verified by successfully performing forward
and backward transformations between the two tessellation 
schemes. 
The produced {\tt HEALPix} maps have a pixel size 
of $\sim 7'$ (the number of pixels for an all-sky map 
is $N_{pix}=12\,n_{side}^2$; here we used $n_{side}=512$). 

\subsection{Noise estimate}

The authors of the radio maps have estimated the rms-noise 
directly on the {\tt ECP} maps 
by calculating the standard deviation of 
low emission regions. 
Going from a Cartesian representation of the sphere to the 
{\sc HEALPix} scheme, the rms-noise per pixel should theoretically decrease 
toward the polar caps, according to the formula: 
$$\sigma_{pixel,{\rm {\tt HEALPix}}} \sim \frac{1}{\sqrt N} \times \sigma_{pixel,{\rm {\tt ECP}} }$$
where $N$ is the number of the {\tt ECP} pixels corresponding to each
{\tt HEALPix} pixel at a fixed latitude. Such a relation 
 holds under the hypothesis that the noise is Gaussian 
and uncorrelated among the pixels. Both these assumptions 
are not necessarily satisfied in the examined surveys, 
 in primis because the pixel size is about half the angular resolution.
Furthermore, the rms-noise quoted for the 408~MHz and 1420~MHz surveys
quantifies the temperature fluctuations per pixel, which is 
due not only to the 
receiver white noise, but also to the 
contribution of unresolved sources 
and of residual systematics effects (as ``scanning strategy effects''),  
that could not be fully eliminated in the data 
reduction procedure. 
The overall rms-noise in the {\tt HEALPix} maps 
is therefore probably higher than the guess determined 
in this way, thus making this formula suitable for 
deriving a lower limit. 
With this formula we constructed 
 a map of rms-noise at both frequencies. 
The rms-noise per pixel decreases for increasing latitude  
and varies between $\sim 10$~mK and $\sim 18.7$~mK at 1420 MHz and 
$\sim 0.5$~K and $\sim 0.7$~K at 408 MHz. 
Due to varying integration times, 
the rms-noise is not constant over the {\tt ECP} map,
but it is expected to diminish toward the celestial poles.
Also taking this effect into account, while applying
the above formula, the derived estimate 
decreases at most by a factor $\sim 2$.
This way we obtain an optimistic lower limit for the rms-noise
in the {\tt HEALPix} maps. 
The corresponding APS is computed as: 
 $$C^{noise}_{\ell} \sim c^{noise} \sim 4\pi \sigma^2/N_{pix}$$ 
 where $\sigma$ is the mean value of the rms-noise 
 in the considered area and $N_{pix}$ is the total 
number of pixels in the {\tt HEALPix} map.
A generous upper limit for the noise APS is provided instead by 
the high multipole tail of the APS.
This way we have bracketed the noise APS for each 
considered sky region. 

\subsection{Statistically significant multipole range}

The {\tt HEALPix} maps produced at 408~MHz and 1420~MHz contain precise
statistical information only for angular scales that are:
\begin{itemize}
\item[-] larger than the beam,
   i.e. for $\theta \gtrsim \theta_{HPBW}$, therefore the maximum multipole
   relevant in the APS analysis of these surveys is
   $\ell_{max} \sim 180^{\circ}/\theta_{HPBW}[^{\circ}]$;
\item[-] smaller than the maximum angular extent of the considered
    area, $\theta_{cov}$, so that the coverage sets the minimum
    multipole. A safe choice is
    $\ell_{min} \sim 5 \times 180^{\circ}/\theta_{cov}[^{\circ}]$.
\end{itemize}

\section{The APS of large areas: analysis for various cuts}

\begin{figure}[!t]
  \vskip -0.17cm
  \hskip +0.8cm
  \includegraphics[width=5cm,height=8cm,angle=90]{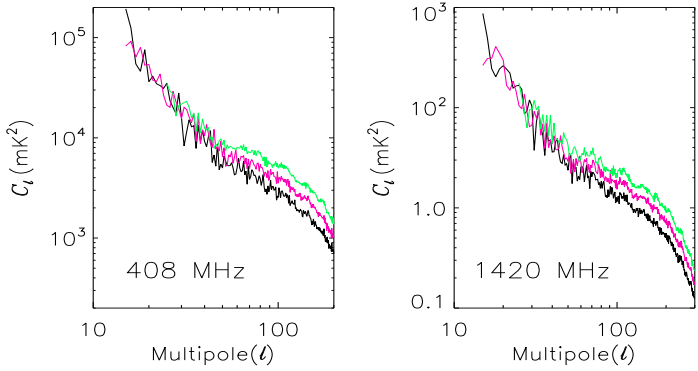}
  \vskip +0.3cm
  \caption{ Angular power spectra of the Galactic plane cut-offs 
   ($|b_{gal}| \ge b_{cut}$) at radio frequencies.  
   Each line refers to a certain $b_{cut}$ (from the top: black $\to 40^{\circ}$, 
   fuxia $\to 50^{\circ}$, green $\to 60^{\circ}$). 
   }
   \label{simmaps}
 \end{figure}
To investigate the dependence of the synchrotron emission APS parameters 
on latitude, the Galactic plane has been ``cut off'' from the maps at 
different latitudes by setting to zero pixels with $|b_{gal}| \le b_{cut}$,   
where $b_{cut}=5^{\circ},10^{\circ},20^{\circ}, 
30^{\circ},40^{\circ},50^{\circ},60^{\circ}$. At the same time, 
this approach preserves the largest possible coverage, 
important to keep the widest range of statistically significant multipoles. 
We also considered ``asymmetric cuts'', i.e. regions 
with $b_{gal} \ge b_{cut}$ (northern cuts) 
and $b_{gal} \le -b_{cut}$ (southern cuts), 
thus taking into account the difference 
between the two Galactic hemispheres. 
In fact, the northern hemisphere contains a large and bright feature of  
the radio sky, i.e. the North Polar Spur (NPS), which is   
 widely believed to be 
 an old supernova remnant in the Solar System 
 vicinity \citep{salter83,egger95}.   
 
\begin{figure}[!t]
\vskip -0.3cm
\hskip +0.3cm
\includegraphics[width=6cm,height=9.5cm,angle=90]{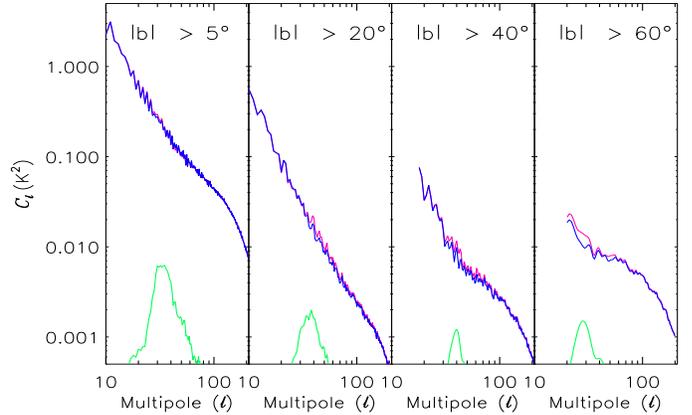}
\caption{Comparison between the cut-off angular power spectra of the original  
(fuxia lines at the top) and destriped (blue) version of the 408~MHz map.
 The cut-off angular power spectra of the difference map are also shown 
 (green lines at the bottom).} 
\label{haslamvsplatania}
\end{figure}

We computed the corresponding APS by using the facility 
{\tt Anafast} of the {\tt HEALPix} package and renormalized 
it to account for the incomplete sky coverage. 
The angular power spectra derived for 
$|b_{gal}| \ge b_{cut}$ with $b_{cut} \ge 40^{\circ}$ 
are shown in Fig.~\ref{simmaps}, as representative examples. 
All the recovered angular power spectra 
flatten towards higher multipoles.  
Such a behaviour of the APS might be due to noise, 
systematic effects (``stripes''), discrete sources or might be an 
intrinsic characteristic of the synchrotron emission 
fluctuation field. 
Instrumental white noise can be discarded because its APS should be
constant, whereas after the flattening the angular power spectra 
decrease with $\ell$ as 
in the presence of beam smoothing. 
``Stripes'' are systematic baseline 
distortions in the telescope's scanning direction
and are mainly due to the limited stability of the
receiving system and to the influence of weather conditions. 
They could also be cancelled from the list of possible 
causes from the comparison with a destriped version 
of the 408~MHz map  
\citep{plat03_destr}.
The cut-off APS of the two versions of the 408~MHz map 
present only marginal differences at intermediate multipoles 
(see Fig.~\ref{haslamvsplatania}).\\

\subsection{Discrete source subtraction}
\label{DS}

 \begin{figure}[!t]
   \begin{center}
   \begin{tabular}{c}
   \includegraphics[width=5cm,angle=90,clip=]{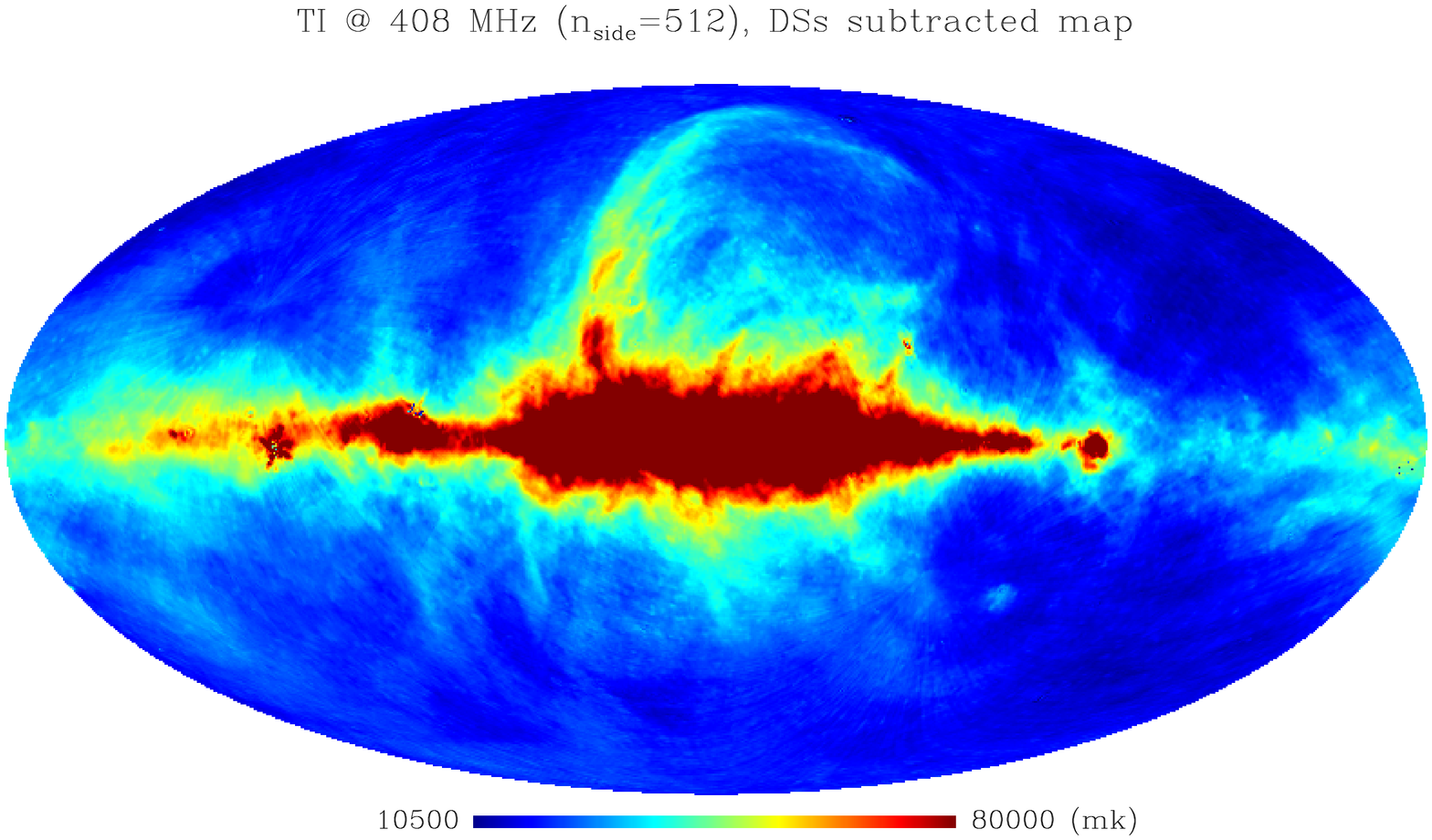}\\
   \includegraphics[width=5cm,angle=90,clip=]{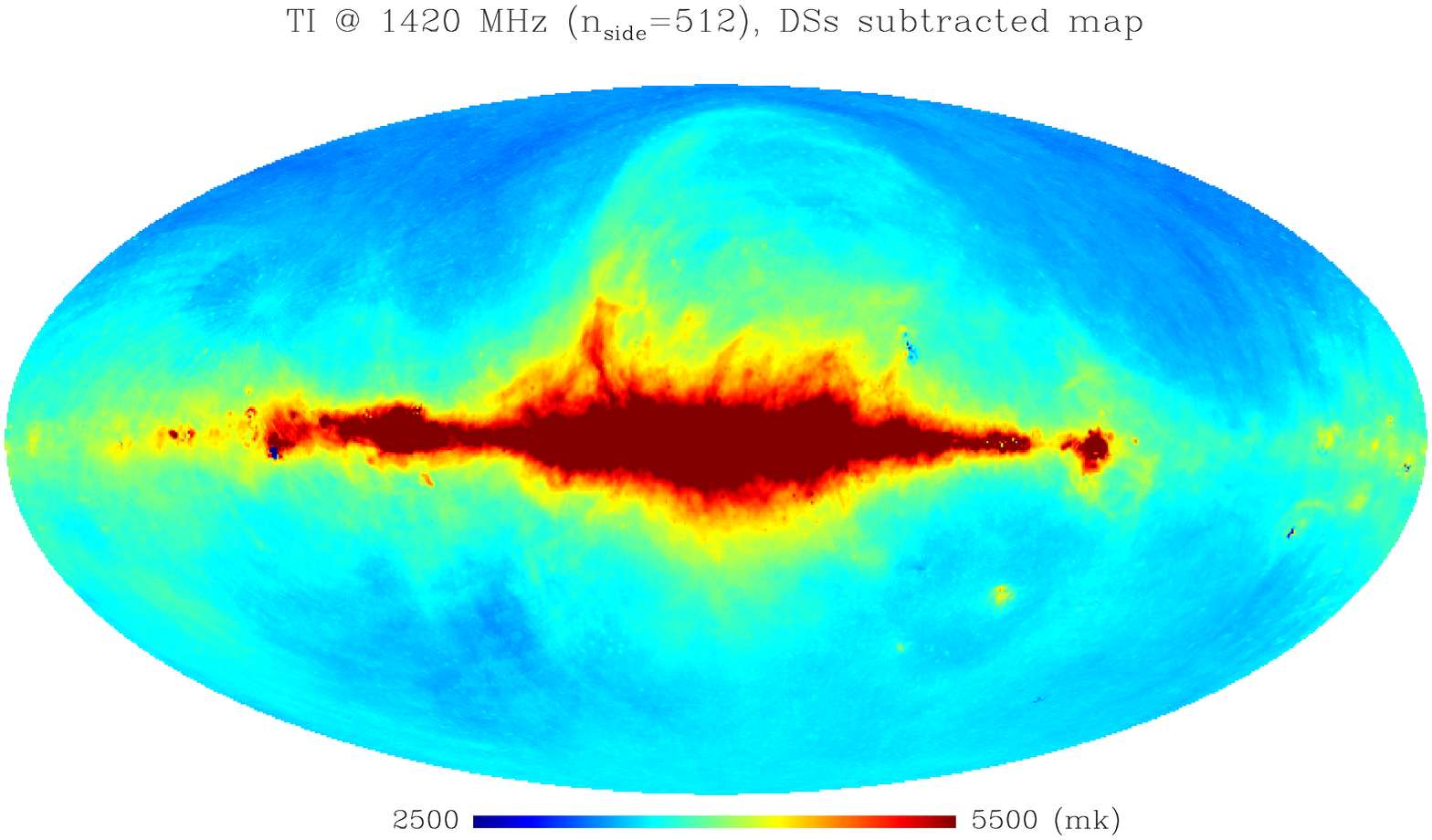} \\ 
   \end{tabular}
   \end{center}
   \caption{ Mollweide projection of the {\tt HEALPix} maps produced 
   at 408 MHz (top) and 1420 MHz (bottom) 
   by subtracting the discrete sources. 
   The maps are in Galactic coordinates. 
   The center of the maps is $l = 0 , b = 0$. 
   The Galactic longitude increases 
   toward the left up to $180^{\circ}$. }
   \label{nosrc_et_src_maps_radio}
\end{figure}
Beside diffuse emission, a large number of discrete sources (DSs)  
are visible in the radio maps. A DS subtraction has been done 
by performing a 2-dimensional Gaussian fitting that 
also provides an estimate of the diffuse 
background, which is approximated by a tilted plane. 
Such an estimate has been used to fill the pixels originally 
corresponding to the subtracted DSs. 
Where the background emission shows strong gradients 
the source fitting is more difficult. 
Consequently, the flux limit above which all discrete 
sources most likely have been subtracted is different close to the
plane and far out of it. 
Namely, for $|b| \gtrsim 45^{\circ}$ all the sources with
peak flux above $\sim 0.8$~Jy (respec. $\sim 6.4$~Jy) 
have been subtracted from the 1420~MHz (respec. 408~MHz), 
whereas for $|b| \lesssim 45^{\circ}$ such a 
source detection threshold 
is $\sim 4.6$~Jy (respec. $\sim 63.8$~Jy). 
All discrete sources that could be reasonably fitted by 
a Gaussian profile have been eliminated and two new maps have been  
generated at 408~MHz and at 1420~MHz (Fig.~\ref{nosrc_et_src_maps_radio}). 
  \begin{figure}[!t]
    \vskip +0.2cm
   \centering
   \begin{tabular}{cc}
   \includegraphics[width=2.5cm,angle=90]{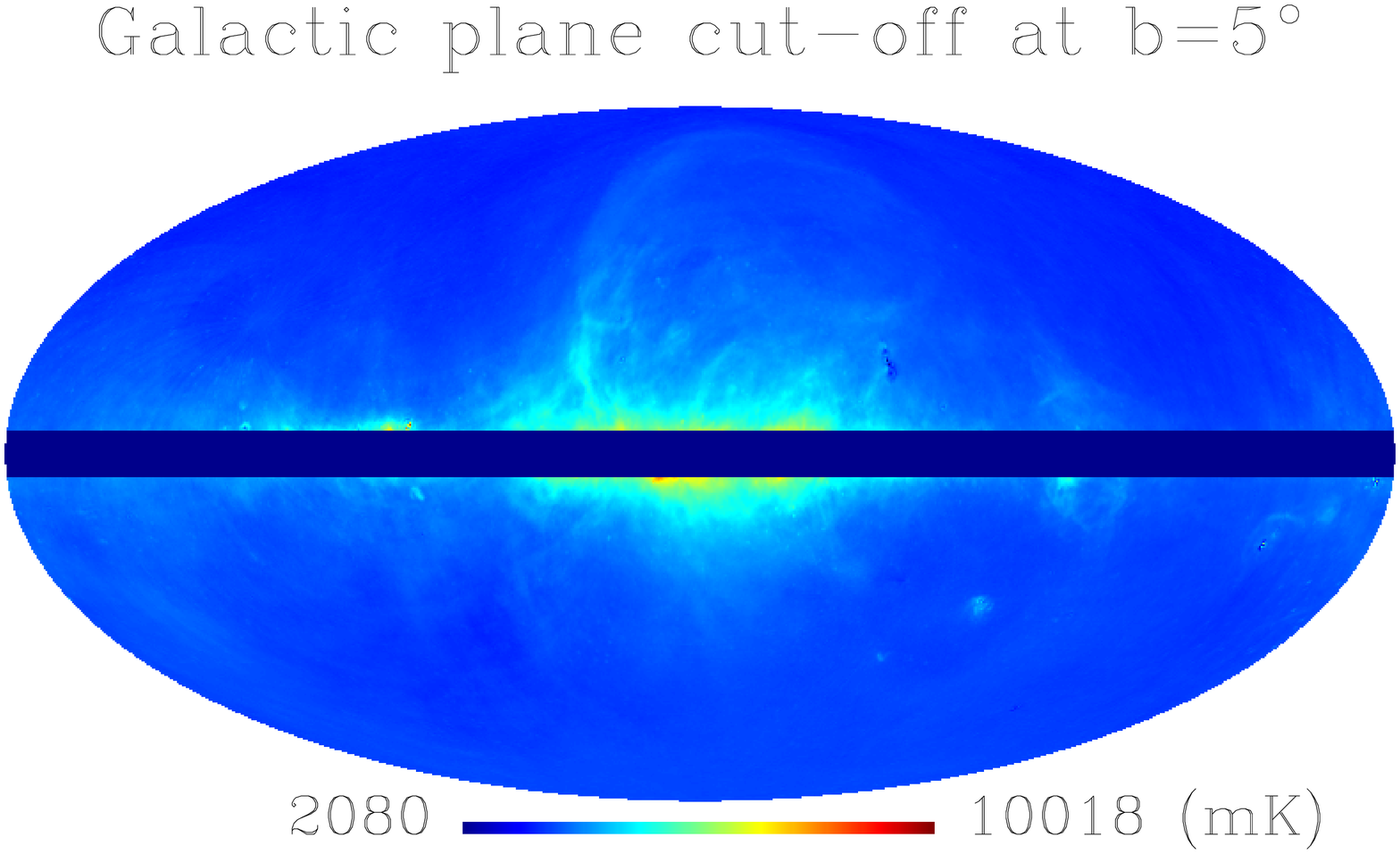}&
   \includegraphics[width=2.5cm,angle=90]{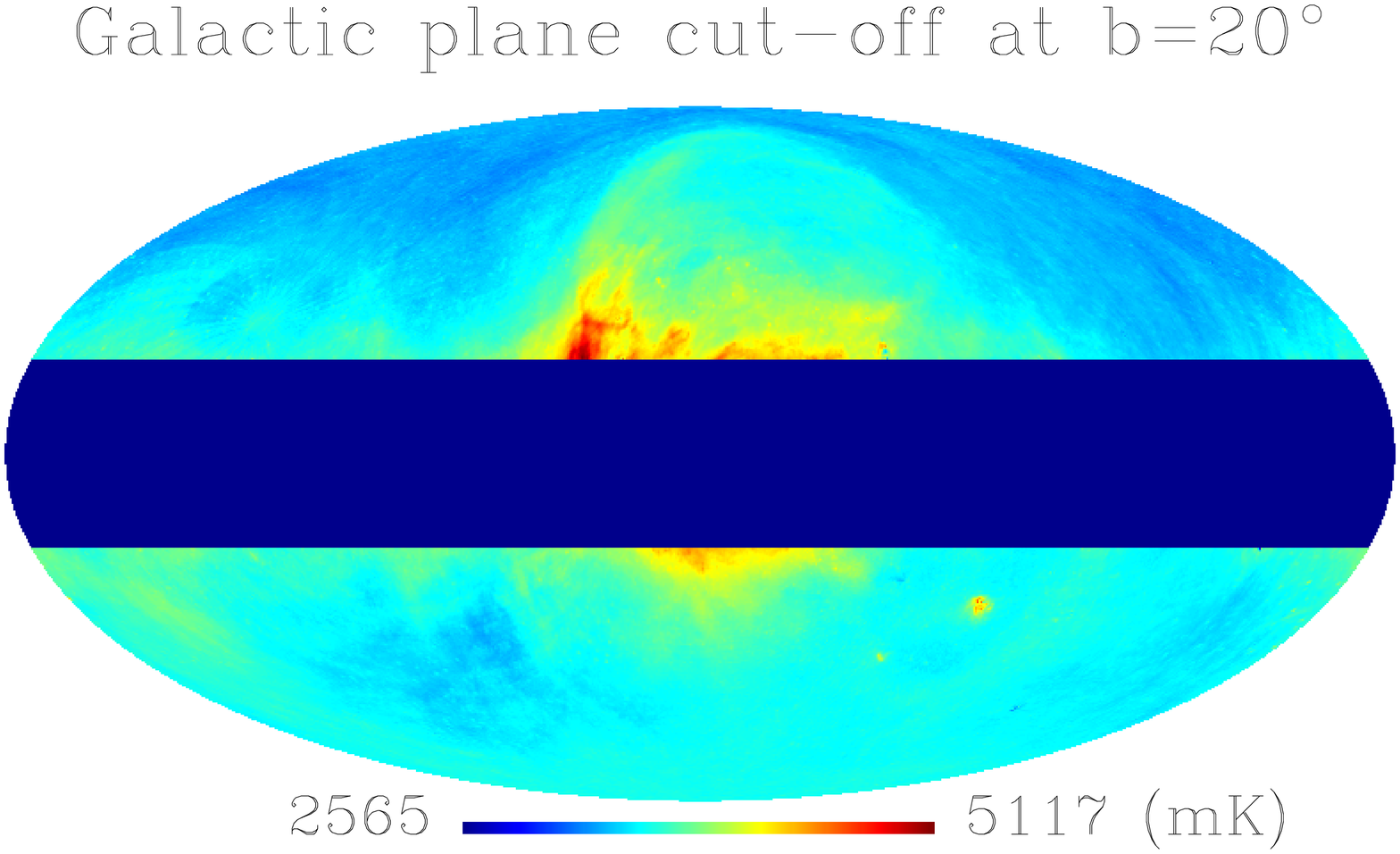}\\
   \includegraphics[width=3cm]{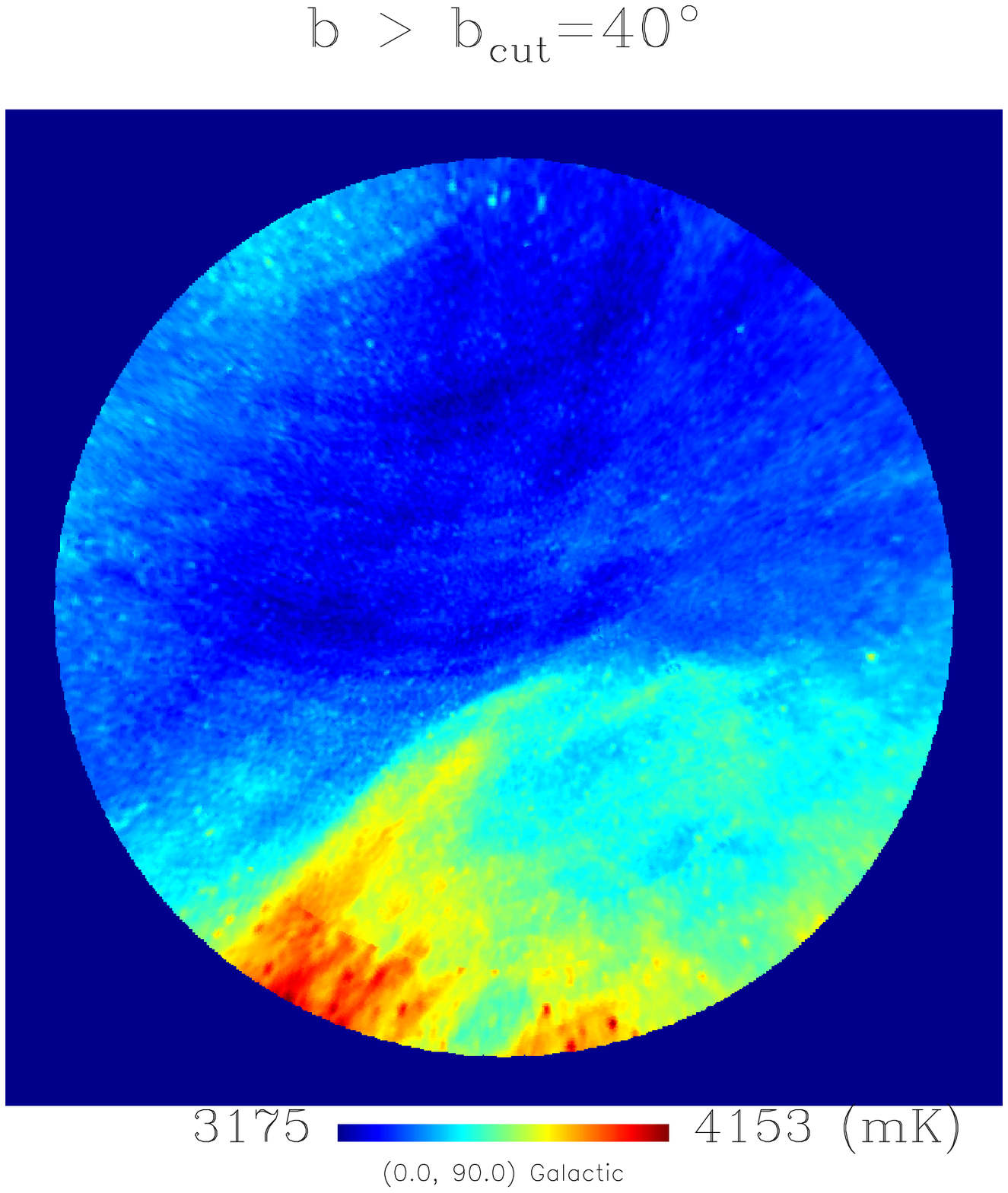} &
   \includegraphics[width=3cm]{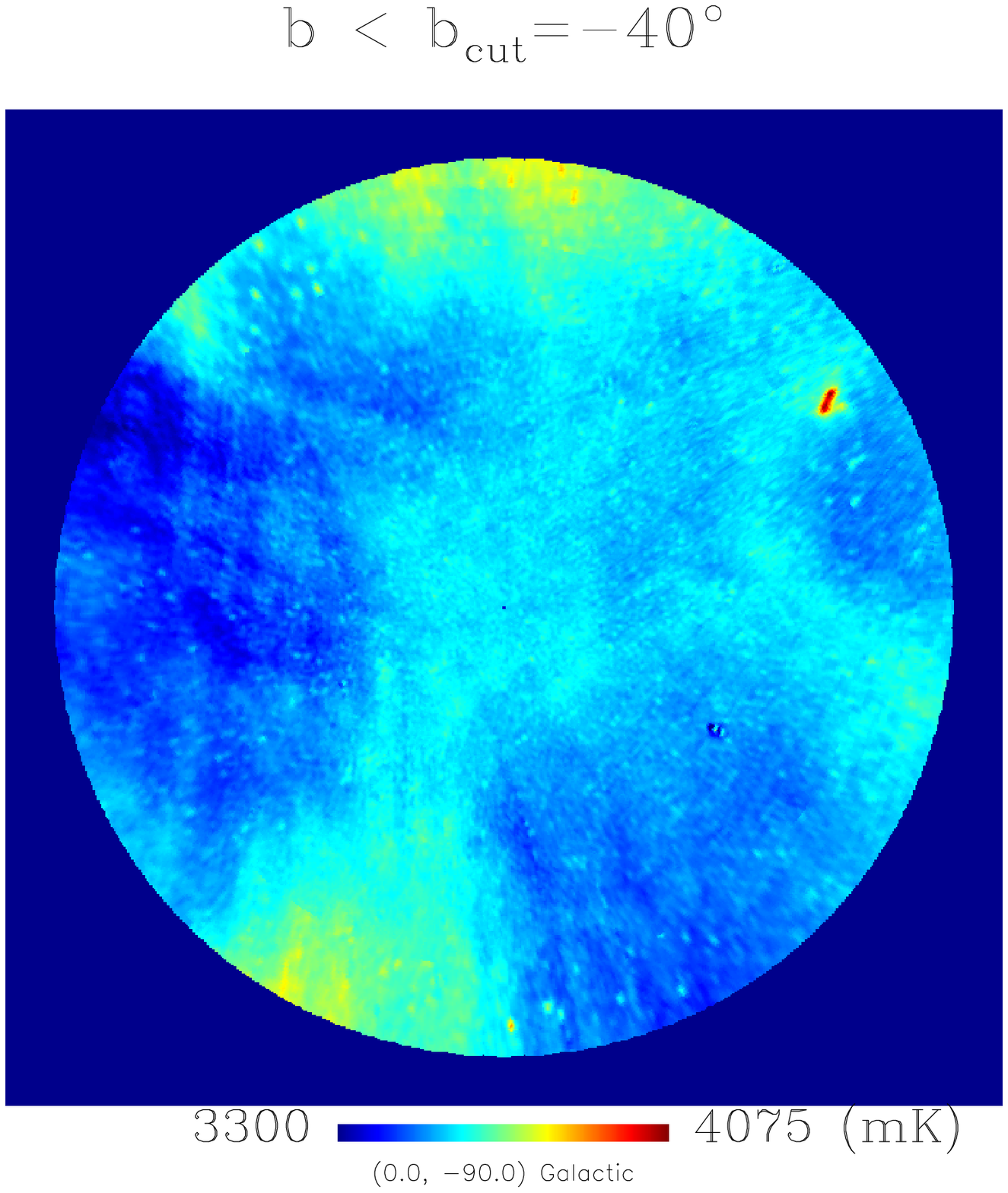} \\
   \includegraphics[width=3cm]{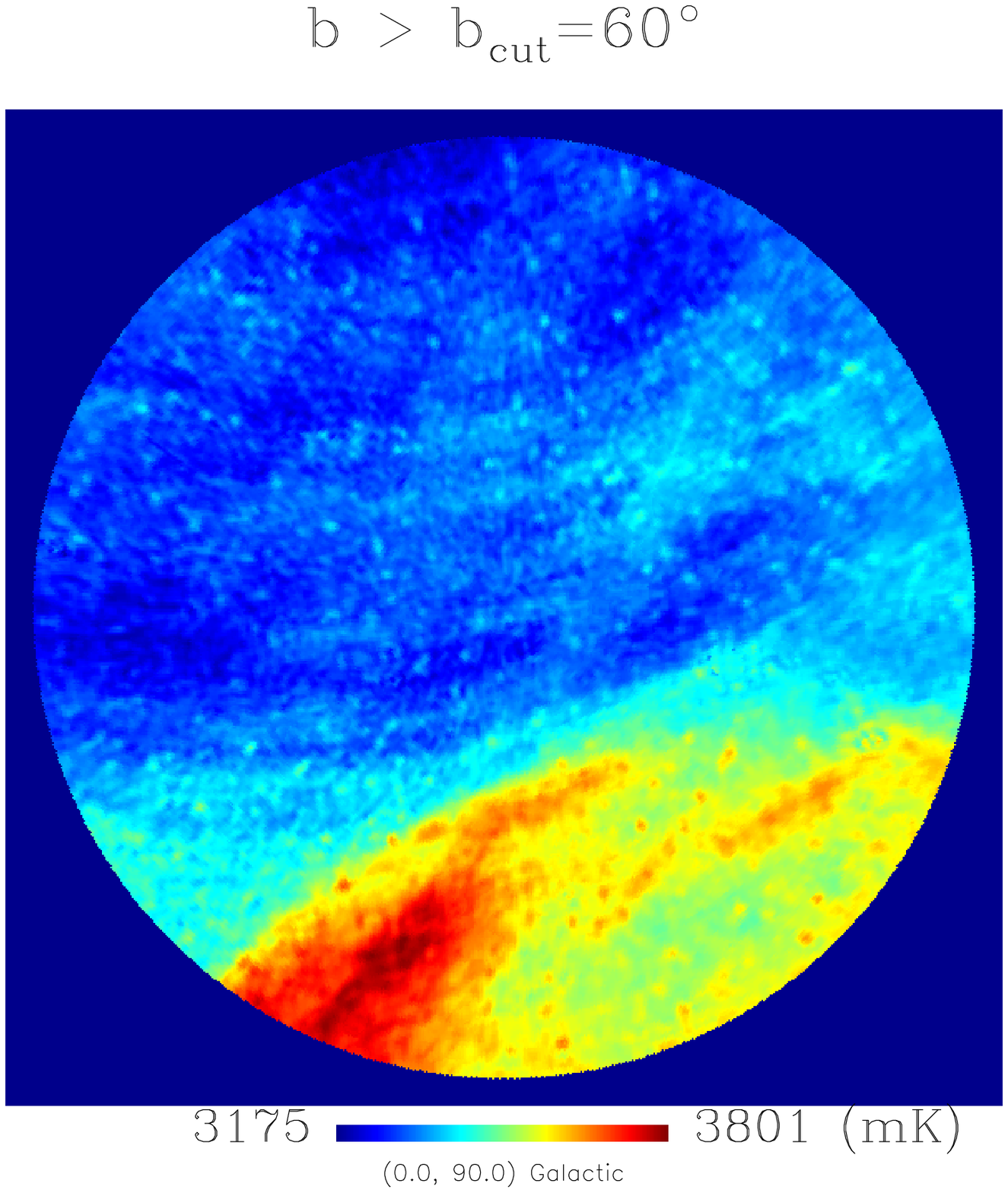} &
   \includegraphics[width=3cm]{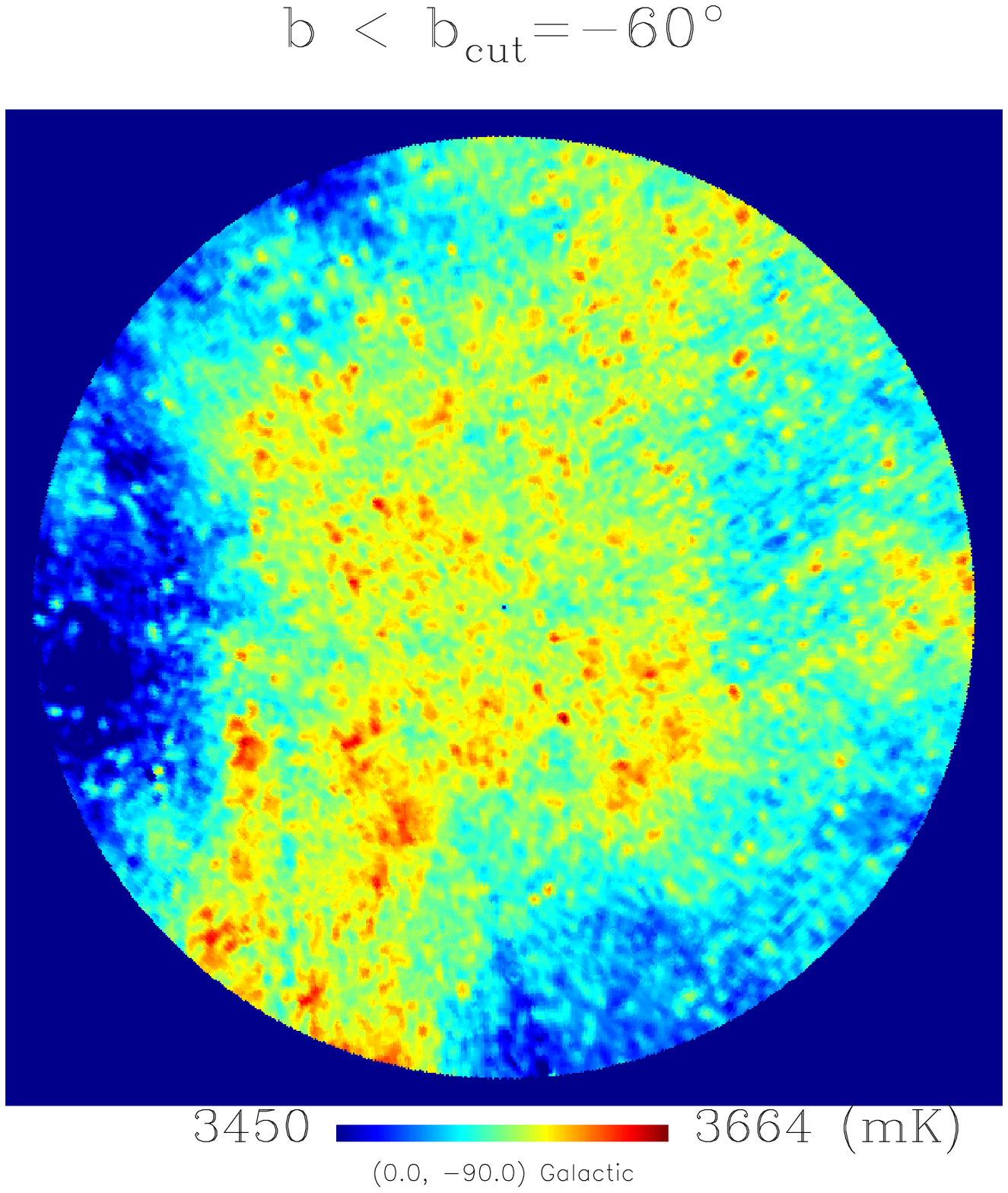} \\
   \end{tabular}
   \caption{ {\tt HEALPix} maps of the Galactic plane cut-offs 
   extracted from the 1.4~GHz all-sky map after DS subtraction.
   A mollweide projection is displayed for the cuts with 
   $|b_{cut}|\le 20^{\circ}$ and a gnomonic view (centered on a 
   Galactic pole) 
    for those with $|b_{cut}|\ge 40^{\circ}$. 
   }
   \label{mapcut1420nosrc}
\end{figure}
  \begin{figure}[!h]
    \vskip +0.8cm
   \centering
   \begin{tabular}{cc}
   \includegraphics[width=2.5cm,angle=90]{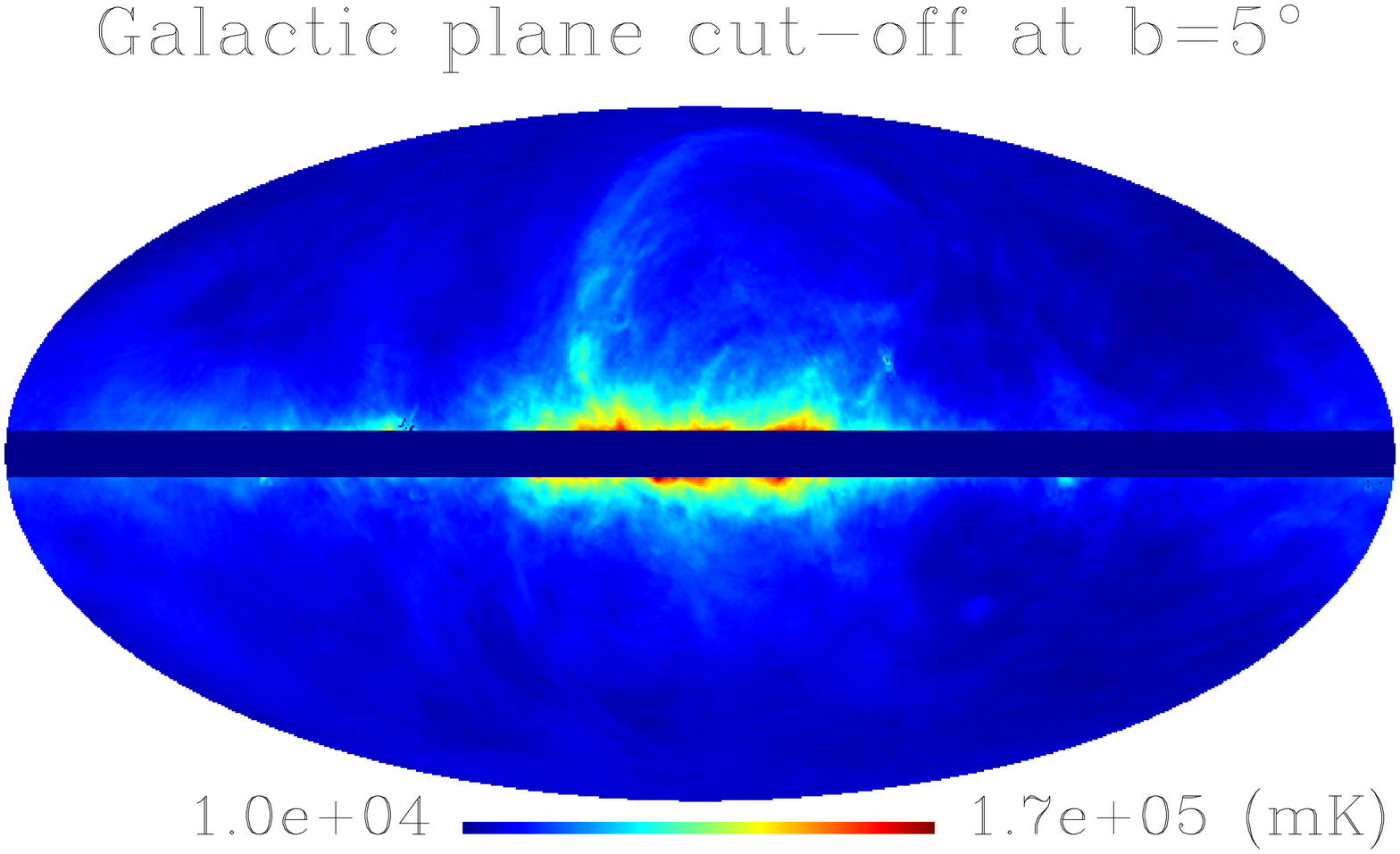}&
   \includegraphics[width=2.5cm,angle=90]{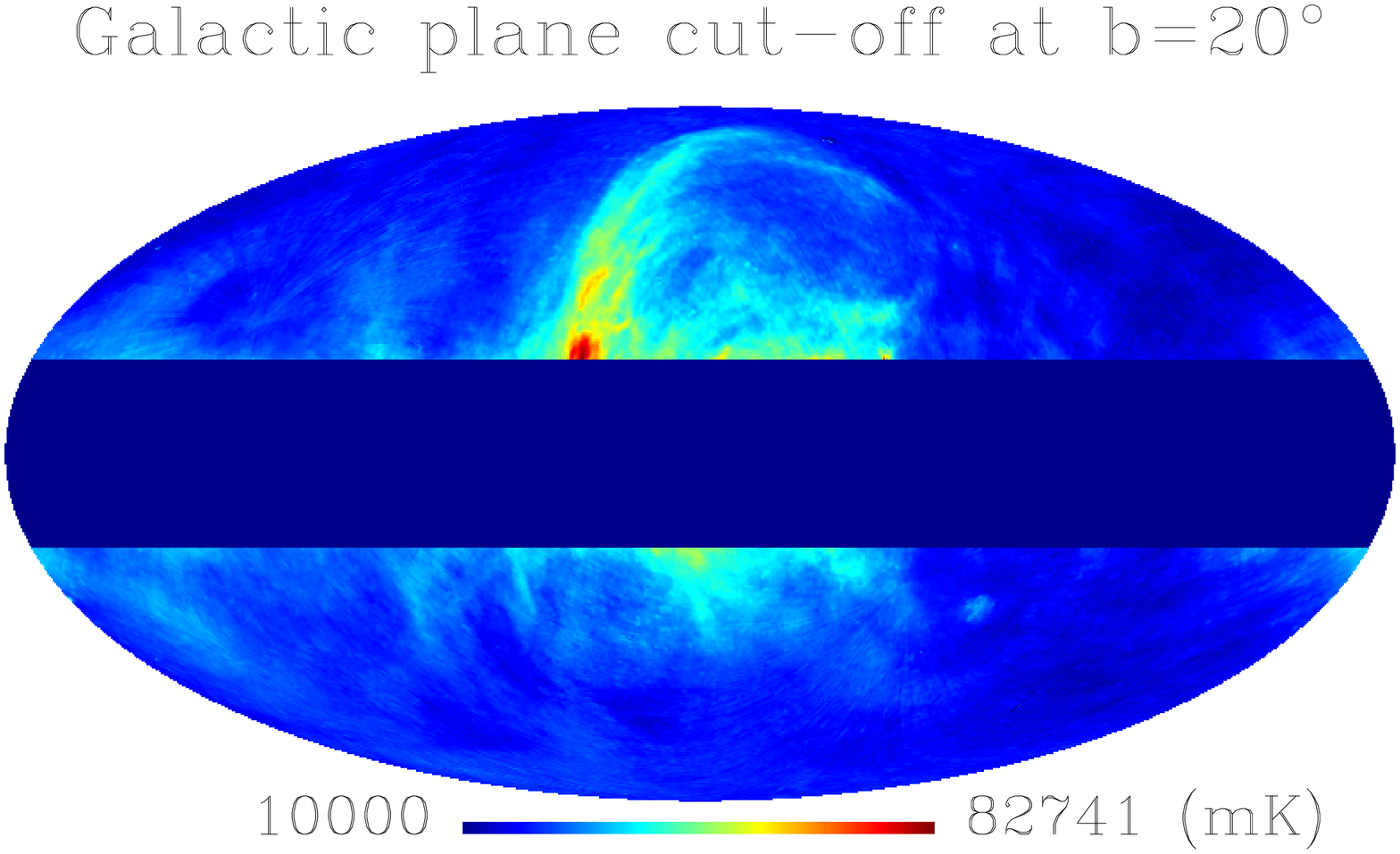}\\
   \includegraphics[width=3cm]{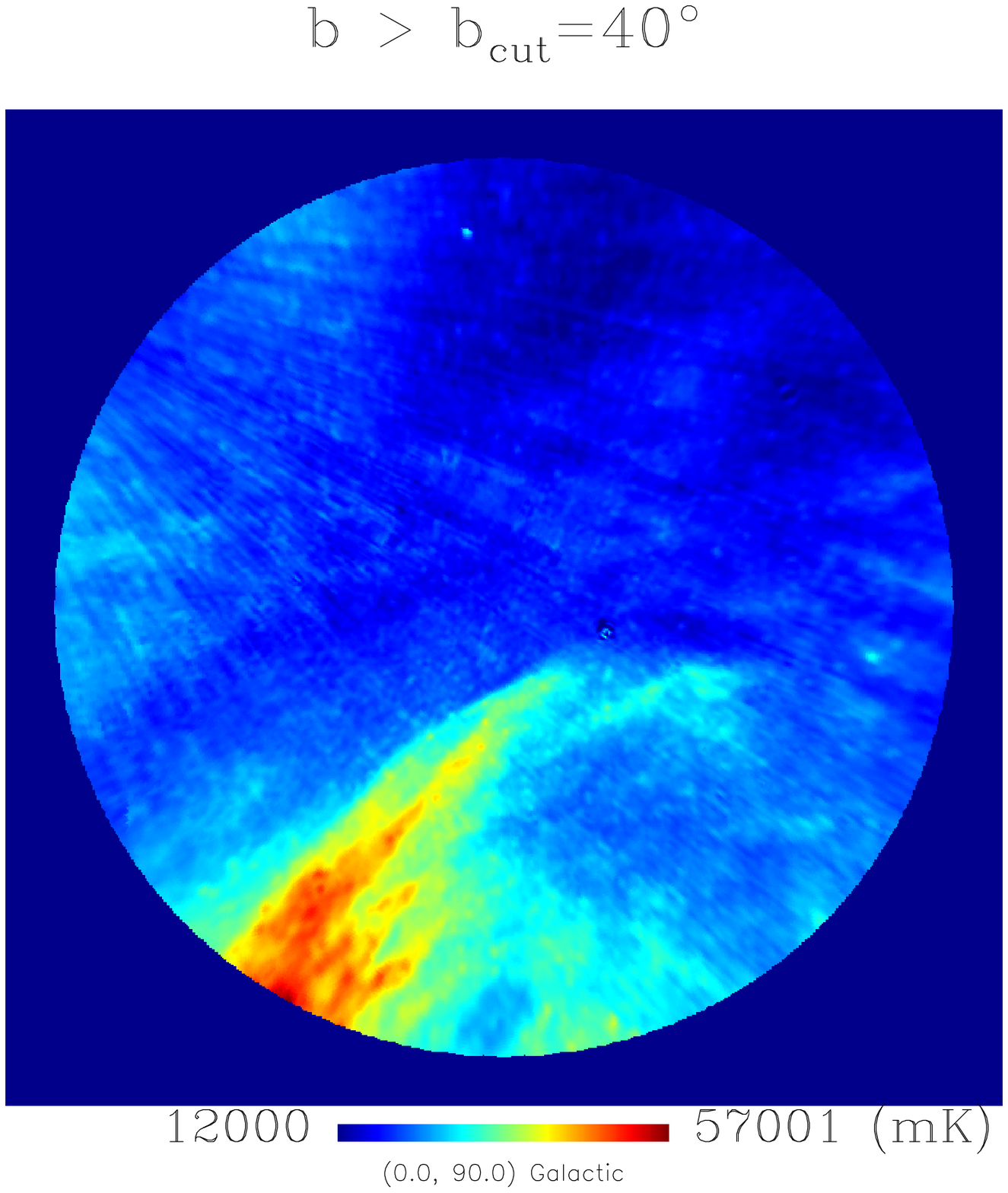} &
   \includegraphics[width=3cm]{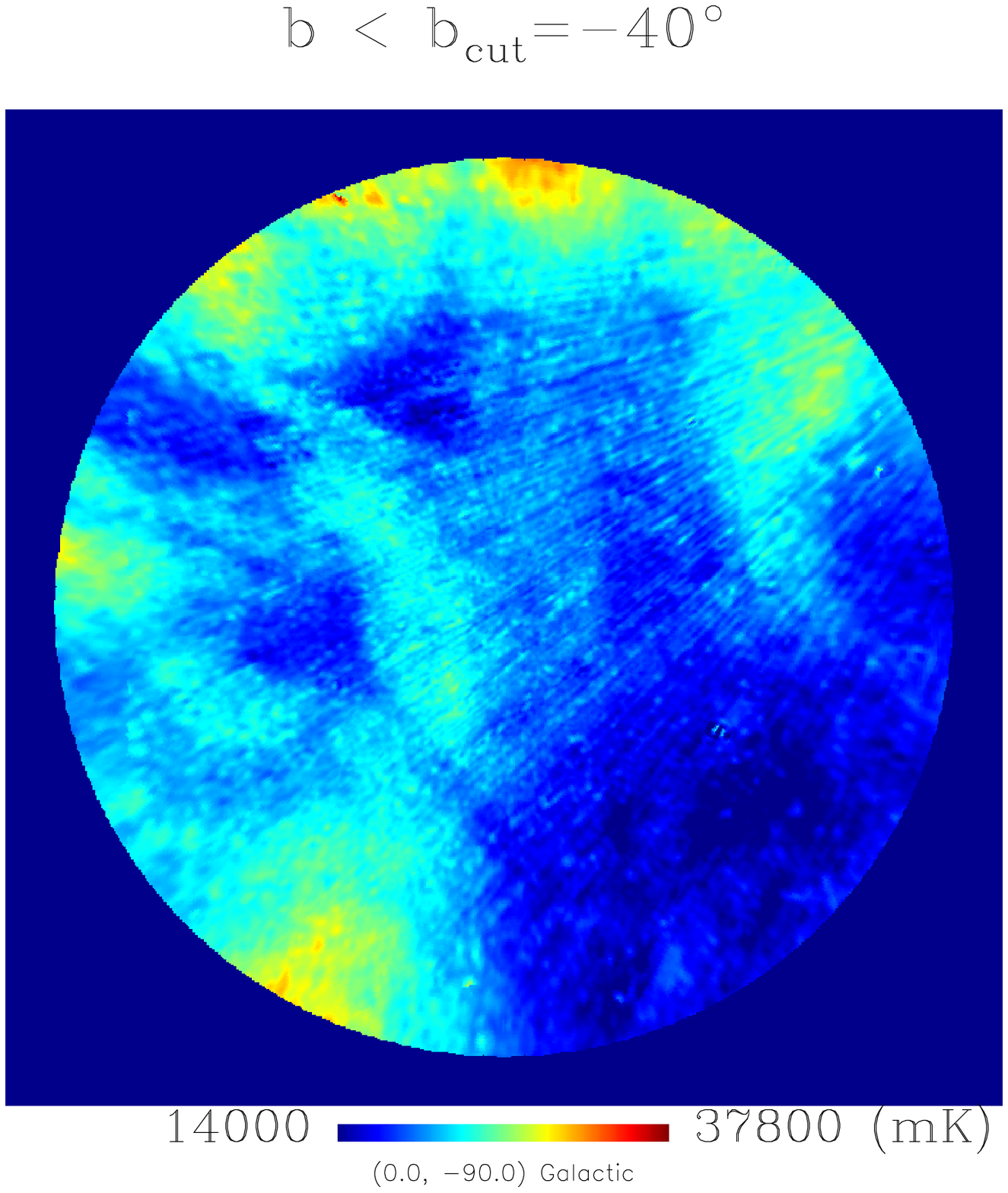} \\
   \includegraphics[width=3cm]{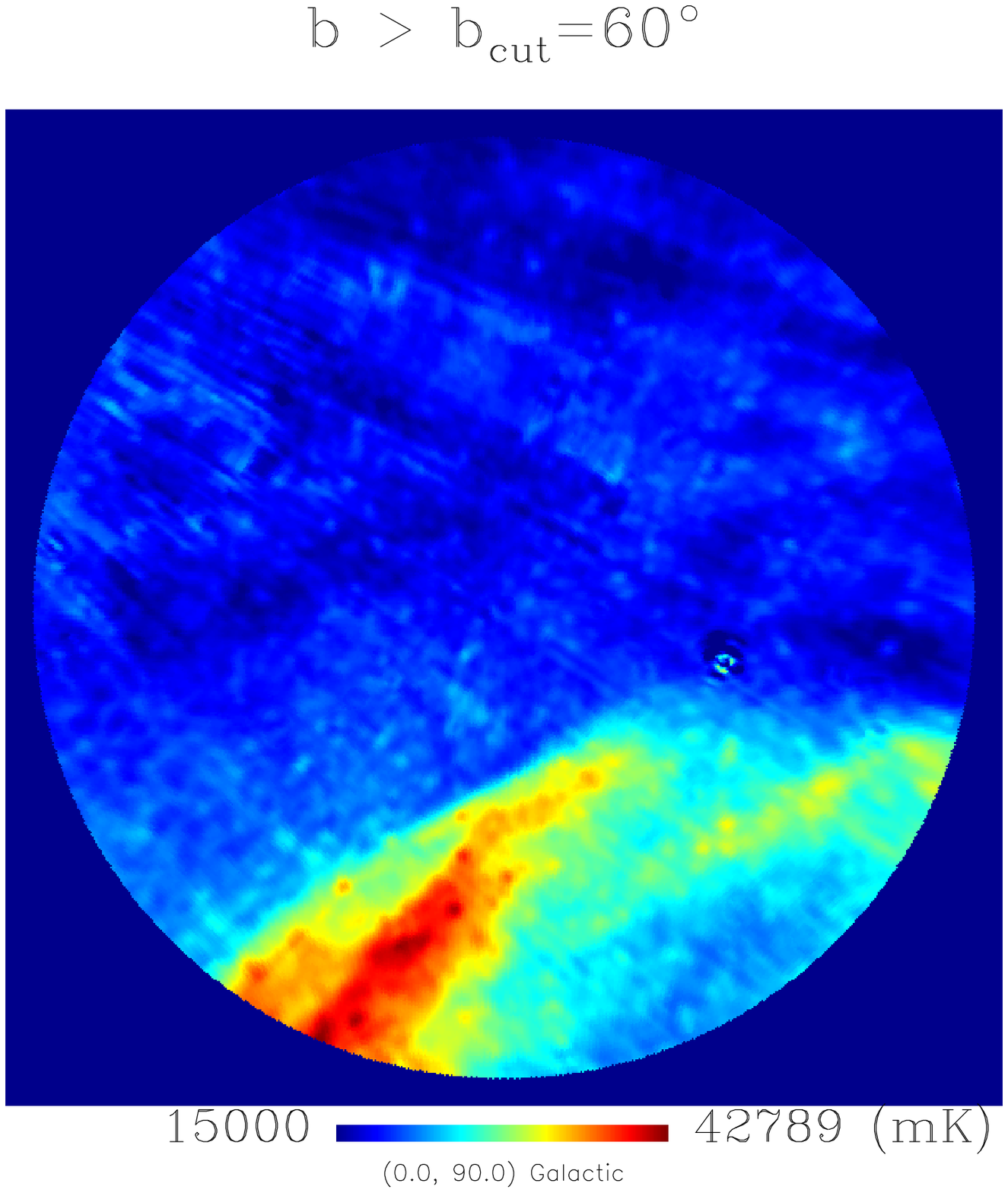} &
   \includegraphics[width=3cm]{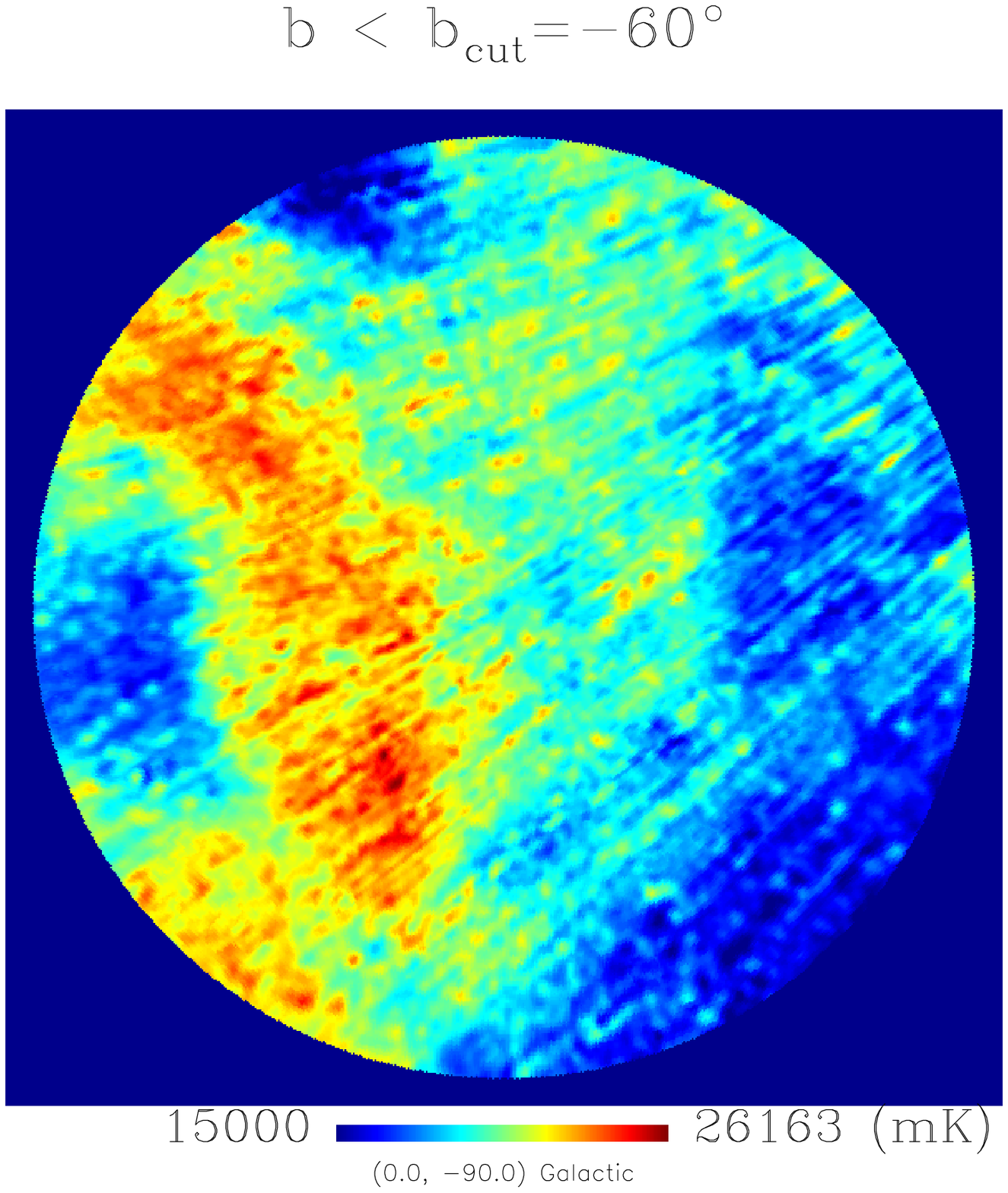} \\
   \end{tabular}
   \vskip -0.3cm 
   \caption{ As in Fig.~\ref{mapcut1420nosrc}, but at 408~MHz.}
   \label{mapcut408nosrc}
\end{figure}

The subtracted DSs (see Burigana et al.~2006 for 
a map) are mostly point sources, 
except for some rather extended objects, as 
for instance the radiogalaxy Centaurus A, 
appearing in the original radio maps right of the 
Galactic center at $b_{gal} \sim 20^{\circ}$. 
Such extended objects are among the brighter subtracted DSs and concentrate 
along or in the proximity of the Galactic plane. 
They are mainly Galactic sources, i.e. {\tt HII}-regions 
or supernova remnants.
On the contrary, the DSs subtracted at medium and high latitudes
are nearly all extragalactic sources. 

\begin{figure}[!h]
   \hskip +1.0cm
   \includegraphics[width=5cm,height=8cm,angle=90]{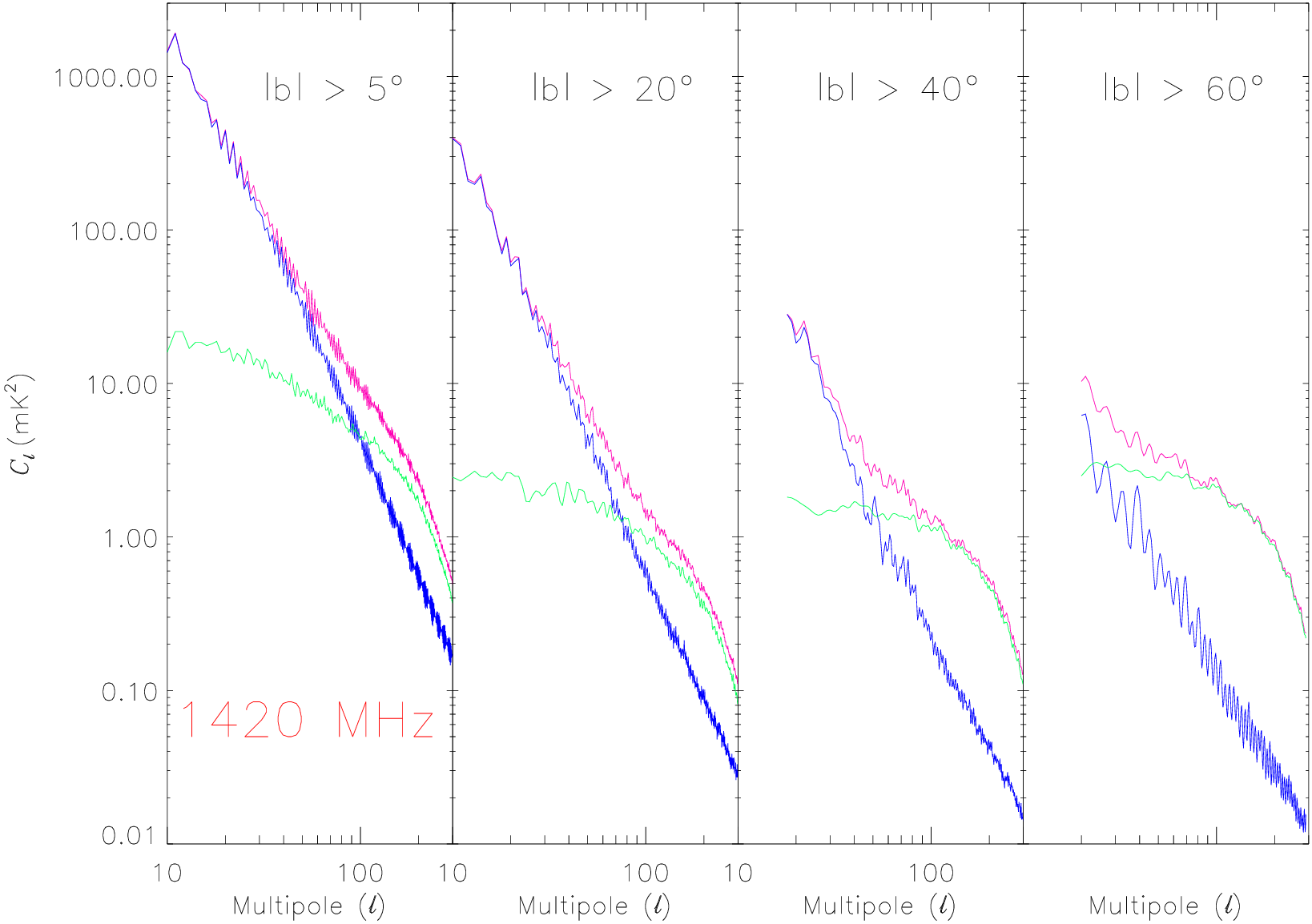} 
   \vskip +0.7cm
   \caption{ APS of some Galactic plane cut-off for the 1420 MHz maps: 
    (from the top in each panel) 
    original ($\to$ fuxia), after discrete source subtraction ($\to$ blue)
   and DSs only ($\to$ green). }
   \label{allapsaftersub1420}
\end{figure}

\begin{figure}[!b]
   \vskip +0.2cm 
   \hskip +1cm
   \includegraphics[width=5cm,height=8cm,angle=90]{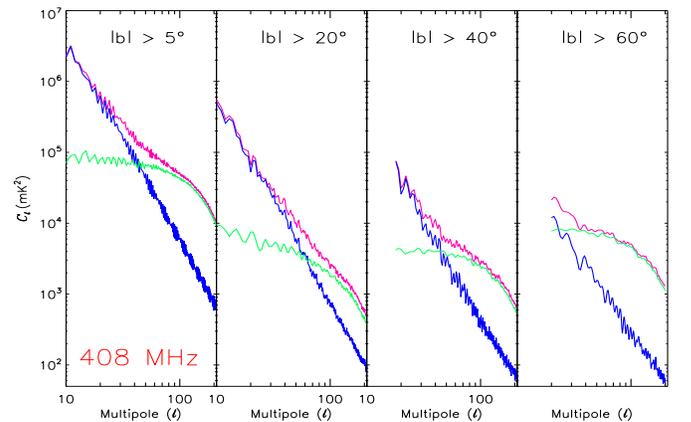} 
   \vskip +0.5cm
   \caption{ As in Fig.~\ref{allapsaftersub1420}, but at 408 MHz.} 
   \label{allapsaftersub408}
\end{figure}

The maps of some Galactic plane cut-offs at 1420~MHz (respec. 408~MHz) 
after DS subtraction are shown in 
Fig.~\ref{mapcut1420nosrc} (respec. Fig.~\ref{mapcut408nosrc}). 
The maps are displayed adopting different scales, in order 
to emphasize the relative importance of the various components.  
Note that ``scanning strategy effects'' are clearly visible 
in the southern sky at high latitude in the map at 408 MHz. 
As discussed above,
the angular power spectra of the destriped and original  
versions of the Haslam map do not exhibit significant differences,
thus implying that ``stripes'' are not 
an issue for the APS analysis at 1420 MHz. 
Figure~\ref{allapsaftersub1420} (respec. Fig.~\ref{allapsaftersub408})
shows the APS of the Galactic plane cut-offs at 1420~MHz (respec. 408~MHz) 
for the original map, for the DS-subtracted map and 
for the map of subtracted DSs. 
The APS of the DS maps almost perfectly matches the flat 
part of the original map APS at large $\ell$. This result  
 identifies DSs as the reason for the flattening of 
the original APS and also confirms that the major contribution 
from source contamination has been eliminated in the 
DS-subtracted maps. 
At high latitude the APS of the Galactic fluctuation 
field is dominated by the DS contribution
for $\ell \gtrsim 100$, which is due to the enhanced 
relative contribution of the DSs respect to the weak diffuse 
background emission. 

\begin{figure}
   \hskip +0.7cm

   \includegraphics[width=4cm,height=9cm,angle=90]{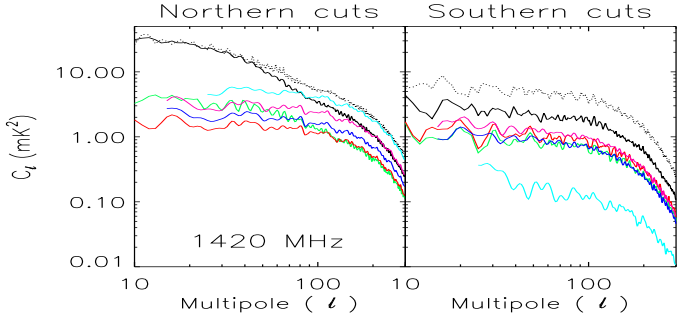}\\

   \caption{Angular power spectra of the northern ($b_{gal} \ge b_{cut}\;\to$ 
   left panel) and southern ($b_{gal} \le -b_{cut}\;\to$ right) 
   cuts for the map of discrete sources at 1420~MHz. 
   Color legend (see online version): 
   black (dotted) $\to |b_{cut}|=5^{\circ}$,
   black $\to |b_{cut}|=10^{\circ}$,
   green $\to |b_{cut}|=20^{\circ}$, red $\to |b_{cut}|=30^{\circ}$,
   dark blue $\to |b_{cut}|=40^{\circ}$, fuxia $\to |b_{cut}|=50^{\circ}$,
   light blue $\to |b_{cut}|=60^{\circ}$.
   }
 \label{allNScut_aps_src_408_1420}
\end{figure}

Figure~\ref{allNScut_aps_src_408_1420} shows 
the APS of the various cuts for the DS map at 1420 MHz. 
Note that for all the southern cuts and for the northern cuts 
with $b_{cut} \ge 30^{\circ}$ the DS angular power spectra 
are rather flat up to $\ell \sim 100$ and then 
decrease as for a beam smoothing. 
On the contrary, for the northern cut at $5^{\circ} - 20^{\circ}$ 
the DS angular power spectra  present a power law behaviour 
at lower multipoles, that implies the existence of significant 
fluctuations also at the larger angular scales, as expected 
in presence of relatively extended discrete structures.  
The DS APS of the cut at $20^{\circ}$ is superimposed 
on that of the cut at $30^{\circ}$ for $\ell \gtrsim 100$,  
whereas it exhibits a power law behaviour at lower multipoles. 
The difference between the angular power spectra of the northern cuts 
at $20^{\circ}$ and $30^{\circ}$ is then due to the DSs located in 
the portion of the sky characterized by 
 $20^{\circ} \le b_{gal} \le 30^{\circ}$,  
 which includes Centaurus A,  
thus explaining the power law behaviour of the APS at the 
lower multipoles.  
The same situation was found at 408 MHz. 

\subsection{The APS after source subtraction}

\begin{figure}[!t]
 \hskip +0.8cm
   \includegraphics[height=9cm,angle=90]{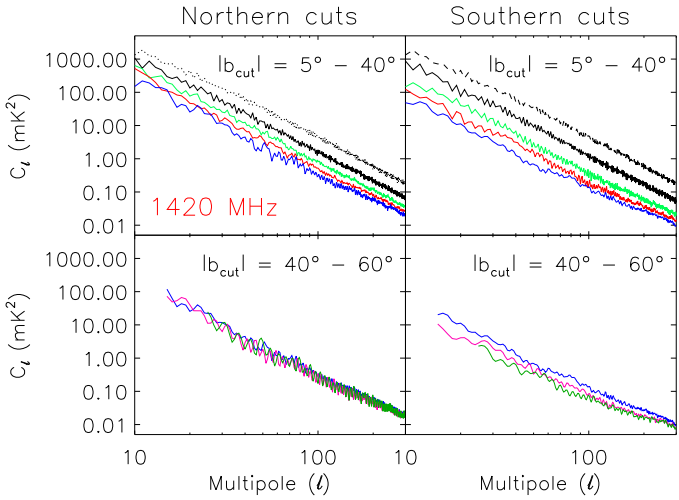}\\
   \caption{
   Comparison between the APS of the cuts for the DS-subtracted map 
  at 1420~MHz.  
  The left (respec. right) panels display the angular power spectra 
  of the northern (respec. southern) cuts. 
   First (respec. second) row panels: 
  $b_{cut}=5^{\circ}-40^{\circ}$ (respec. $b_{cut}=40^{\circ}-60^{\circ}$)~.
  Color legend (see online version): 
  black dotted $\to |b_{cut}|=5^{\circ}$, black $\to |b_{cut}|=10^{\circ}$,
  green $\to |b_{cut}|=20^{\circ}$, red $\to |b_{cut}|=30^{\circ}$,
  dark blue $\to |b_{cut}|=40^{\circ}$, fuxia $\to |b_{cut}|=50^{\circ}$,
  dark green $\to |b_{cut}|=60^{\circ}$.
  }
   \label{allNScut_aps_nosrc_408_1420}
\end{figure}

\begin{figure}
   \centering
   \includegraphics[width=4cm,height=9cm,angle=90]{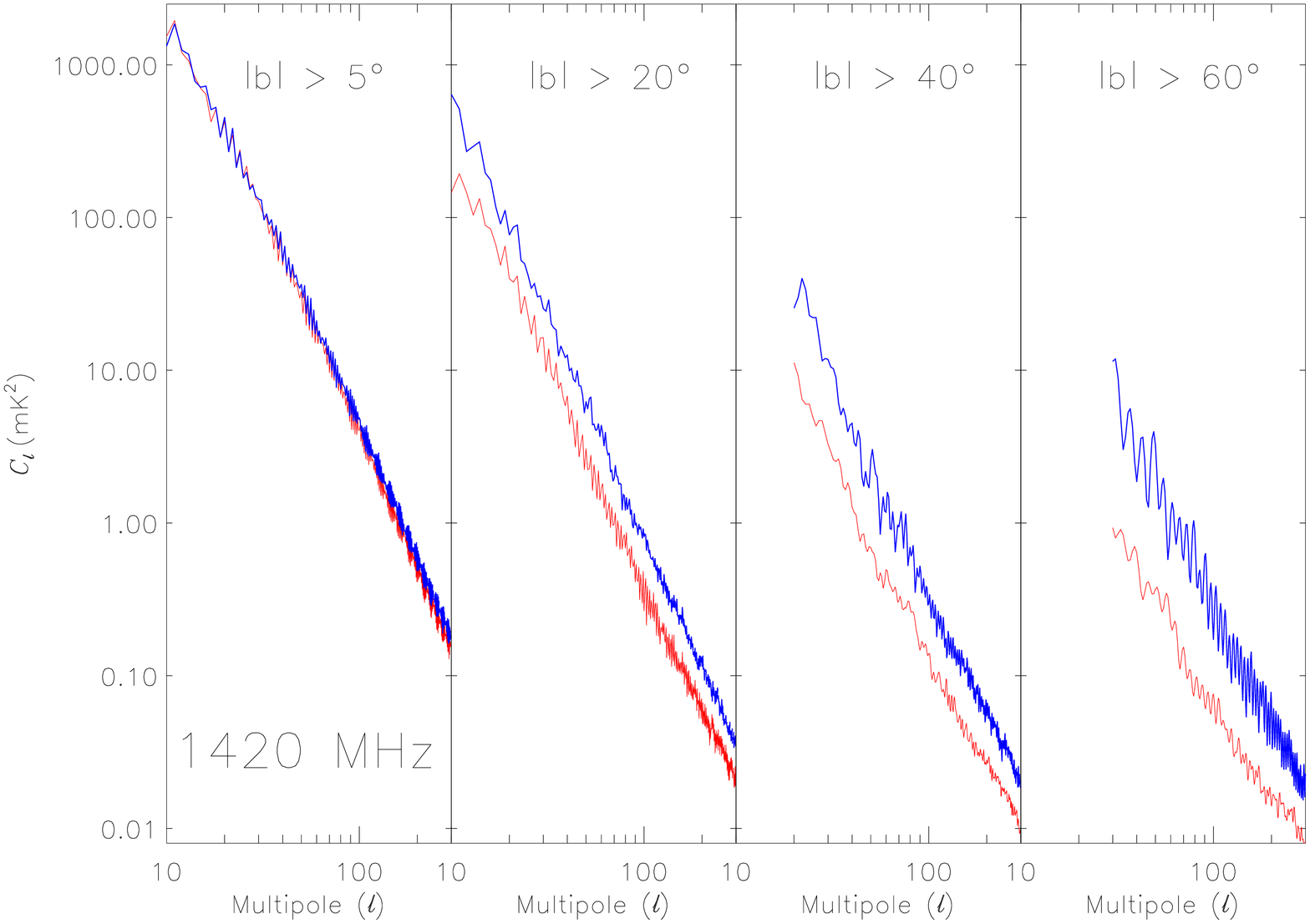}\\
   \caption{Comparison between the angular power spectra of the northern 
   ($b_{gal} > b_{cut}$, $\to$ blue - upper lines) and 
   southern ($b_{gal} < -b_{cut}$ $\to$ red - lower)
   cuts for the DS-subtracted map at 1420 MHz. 
   }
   \label{aps_NvsS_nosrc_1420}
\end{figure}

The angular power spectra of the maps after source subtraction 
approximately follow a power law, as expected for the diffuse 
Galactic synchrotron emission. 
For $|b_{cut}| \le 40^{\circ}$ the angular power spectra of 
all symmetric ($|b| \ge b_{cut}$) and 
asymmetric ($b \le  -b_{cut}$, $b \ge b_{cut}$) Galactic cuts 
are very similar to each other and appear progressively shifted 
downward (see top panels of Fig.~\ref{allNScut_aps_nosrc_408_1420}). 
This result reflects the fact that the Galactic 
diffuse emission becomes weaker for increasing latitude.   
The angular power spectra of the symmetric cuts 
with $|b_{cut}| \ge 40^{\circ}$ are superimposed. 
The same result holds for the angular power 
spectra of the northern cuts, whereas in the southern hemisphere 
the APS amplitude decreases for increasing $|b_{cut}|$. 
This discrepancy leads to the conclusion that the angular 
power spectra of the symmetric cuts with $|b_{cut}| \ge 40^{\circ}$ 
are mainly influenced by the northern hemisphere. 
Indeed, the angular power spectra of the northern 
cuts at $b_{cut} \ge 20^{\circ}$ have amplitudes 
larger than those of the southern cuts at both frequencies. 
As an example, Fig.~\ref{aps_NvsS_nosrc_1420} shows the 
comparison between the angular power spectra of the 
northern and the southern cuts at 1420 MHz.  
The angular power spectra of the two Galactic hemispheres 
 can reasonably be expected to be similar for the Galactic diffuse 
synchrotron emission, while they turn out to be different in 
amplitude and to some extent (mostly at smaller scales) 
in shape. That difference results from the combination of two effects.  
In the southern sky, the angular power spectra of the DS-subtracted maps 
tend to flatten at $\ell \sim 150-200$ due to the presence 
of unsubtracted sources, whose relative contribution to the fluctuation 
field increases because of the low background signal. 
In the northern hemisphere, the Galactic diffuse synchrotron 
emission is strongly influenced by the radiation of the NPS. 
\begin{figure}[!b]
   \hskip +1.0cm
   \includegraphics[width=5.5cm,height=8.5cm,angle=90]{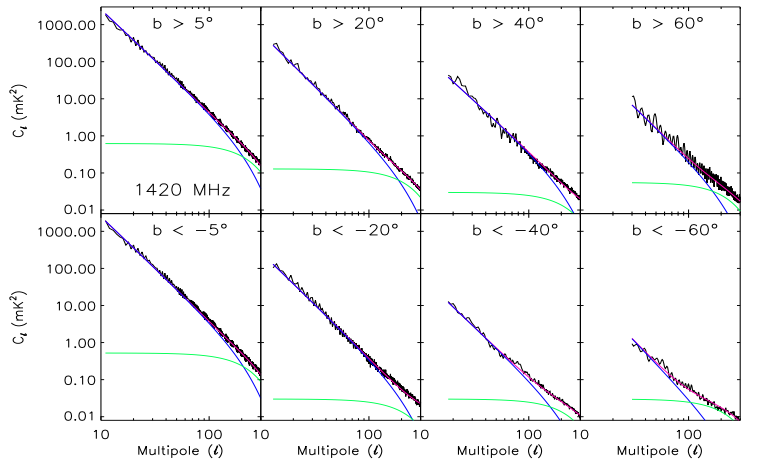}\\
   \caption{Angular power spectra of the northern (top panels)  
   and southern cuts (bottom) for the DS-subtracted map at 1420 MHz   
   together with the best fit curves obtained. The individual contributions  
    of synchrotron emission (blue lines) and of sources (green lines)   
    are also plotted (smoothed by the beam).  
   }
   \label{bfit1420}
\end{figure}

\subsection{Fit of the APS after source subtraction}
\label{sect_fit_aps_nos}

The radio maps after source subtraction include 
two astrophysical components: 
the Galactic diffuse emission and the 
(mainly) extragalactic source contribution, which  
 are convolved with the telescope beam and 
contaminated by the instrument noise, that 
can be approximately treated as white noise. 
We therefore express the corresponding APS as 
\begin{equation}
C^{map}_{\ell} \sim ( C_{\ell}^{synch}+ C_{\ell}^{src} ) W_{\ell} + c^{noise} 
\label{aps_best_model}
\end{equation}
where 
$W_{\ell}={\rm e}^{-\ell (\ell+1)\sigma_{b}^2}$ 
is the window function of the symmetric and Gaussian 
beam\footnote{The pixel window function has been 
also taken into account, but its effect is not 
important here, because the pixel size is 
significantly smaller than $\sigma_{b}$.}, 
 with 
$\sigma_{b}=\theta_{HPBW}[rad]/\sqrt{8{\rm ln}2)}$. 
The synchrotron emission APS is empirically modelled 
as $C_{\ell}^{synch}=k \ell^{\alpha}$. 
We note that such an empirical choice qualitatively can be explained by 
 magnetohydrodynamic turbulence arguments 
\citep{chep98_synch_aps,cho02_synch_aps,cho03_synch_aps}. 
The contribution of the unsubtracted DSs is approximated 
by a constant term, according to the formalism of Poisson 
fluctuations from extragalactic point sources \citep{franc89}. 
The contribution of fluctuations due to source clustering 
is expected to be negligible with respect to the Poisson term 
at the source detection threshold achieved in our 
maps \citep{toffo98_src}. 
\begin{table}[!b]
\newcommand\T{\rule{0pt}{2.6ex}}
\newcommand\B{\rule[-1.2ex]{0pt}{0pt}}
\begin{center}
\begin{tabular}{|c|c|c|c|c|}
\hline
Coverage \T \B & \multicolumn{4}{|c|}{Best Fit parameters @ 1420 MHz}\\
\hline
   \T \B &  $k_{100}$ (mK$^2$) & $\alpha$ & $c^{src}$ (mK$^2$) & $c^{noise}$ (mK$^2$)\\
   \hline
$b_{gal} \ge 5^{\circ}$ \T \B & $ 4.57 ^{+ 0.95}_{-0.13 }$  &
                           $-2.75 ^{+ 0.16}_{-0.03 }$  &
                           $ 0.621 ^{-0.621}_{ +0.079 }$  &
                           $0.0164 ^{+0.1222 }_{-0.0164}$ \\
\hline
$b_{gal} \le -5^{\circ}$ \T \B & $ 4.00 ^{+ 0.78}_{-0.09 }$  &
                            $-2.79 ^{+ 0.15}_{-0.02 }$  &
                            $ 0.522 ^{-0.522 }_{ +0.069 }$  &
                            $0.0164 ^{+0.1003 }_{-0.0164}$ \\
\hline
$b_{gal} \ge 10^{\circ}$ \T \B & $ 1.61 ^{+ 0.38}_{-0.10 }$  &
                           $-2.88 ^{+ 0.21}_{-0.12 }$  &
                           $ 0.227 ^{-0.227}_{ +0.061 }$  &
                           $0.0151 ^{+0.0425 }_{-0.0151}$ \\
\hline
$b_{gal} \le -10^{\circ}$ \T \B & $ 1.41 ^{+ 0.13}_{-0.27 }$  &
                            $-2.74 ^{+ 0.06}_{-0.28 }$  &
                            $ 0.128 ^{-0.128}_{ +0.107 }$  &
                            $0.0158 ^{+0.0279 }_{-0.0158}$ \\
\hline
$b_{gal} \ge 20^{\circ}$ \T \B & $ 0.78 ^{+ 0.22}_{-0.07 }$  &
                           $-2.88 ^{+ 0.21}_{-0.16 }$  &
                           $ 0.128 ^{-0.128}_{ +0.042 }$  &
                           $0.0077 ^{+0.0242 }_{-0.0077}$ \\
\hline
$b_{gal} \le -20^{\circ}$ \T \B & $ 0.41 ^{+ 0.10}_{-0.09 }$  &
                            $-2.83 ^{+ 0.13}_{-0.20 }$  &
                            $ 0.030 ^{-0.030}_{ +0.067 }$  &
                            $0.0146 ^{+0.0047 }_{-0.0146}$ \\
\hline
$b_{gal} \ge 30^{\circ}$ \T \B & $ 0.43 ^{+ 0.17}_{-0.05 }$  &
                           $-3.02 ^{+ 0.39}_{-0.02 }$  &
                           $ 0.128 ^{-0.128}_{ +0.004 }$  &
                           $0.0008 ^{+0.0251 }_{-0.0008}$ \\
\hline
$b_{gal} \le -30^{\circ}$ \T \B & $ 0.22 ^{+ 0.02}_{-0.06 }$  &
                            $-2.77 ^{+ 0.12}_{-0.28 }$  &
                            $ 0.030 ^{-0.030}_{ +0.034 }$  &
                            $0.0065 ^{+0.0051 }_{-0.0065}$ \\
\hline
$b_{gal} \ge 40^{\circ}$ \T \B & $ 0.40 ^{+ 0.01}_{-0.13 }$  &
                           $-2.66 ^{+ 0.25}_{-0.34 }$  &
                           $ 0.030 ^{-0.030}_{ +0.065 }$  &
                           $0.0113 ^{+0.0071 }_{-0.0113}$ \\
\hline
$b_{gal} \le -40^{\circ}$ \T \B & $ 0.11 ^{+ 0.05}_{-0.02 }$  &
                            $-2.77 ^{+ 0.25}_{-0.34 }$  &
                            $ 0.030 ^{-0.030}_{ +0.022 }$  &
                            $0.0049 ^{+0.0050 }_{-0.0049}$ \\
\hline
$b_{gal} \ge 50^{\circ}$ \T \B & $ 0.21 ^{+ 0.09}_{-0.02 }$  &
                           $-2.86 ^{+ 0.46}_{-0.26 }$  &
                           $ 0.056 ^{-0.056}_{ +0.027 }$  &
                           $0.0063 ^{+0.0086 }_{-0.0063}$ \\
\hline
$b_{gal} \le -50^{\circ}$ \T \B & $ 0.05 ^{+ 0.04}_{-0.00 }$  &
                            $-3.00 ^{+ 0.63}_{-0.07 }$  &
                            $ 0.030 ^{-0.030}_{ +0.007 }$  &
                            $ 0.0026 ^{+0.0066 }_{-0.0026}$ \\
\hline
$b_{gal} \ge 60^{\circ}$ \T \B & $ 0.23 ^{+ 0.06}_{-0.05 }$  &
                           $-2.81 ^{+ 0.40}_{-0.33 }$  &
                           $ 0.056 ^{-0.056}_{ +0.030 }$  &
                           $0.0062 ^{+0.0103 }_{-0.0062}$ \\
\hline
$b_{gal} \le -60^{\circ}$ \T \B & $ 0.03 ^{+ 0.03}_{-0.00 }$  &
                            $-3.02 ^{+ 0.81}_{-0.15 }$  &
                            $ 0.030 ^{-0.030}_{ +0.006 }$  &
                            $0.0021 ^{+0.0067 }_{-0.0021}$ \\
\hline
\hline
\end{tabular}
\end{center}
\caption{ Best fit parameters obtained by modelling 
        the angular power spectra of the northern and southern 
         cuts at 1420 MHz according to Eq.~\ref{aps_best_model}. 
          }
\label{BFpar_tab1420}
\end{table}

\begin{figure}[!t]
   \hskip +0.5cm
   \includegraphics[width=5.5cm,height=8.5cm,angle=90]{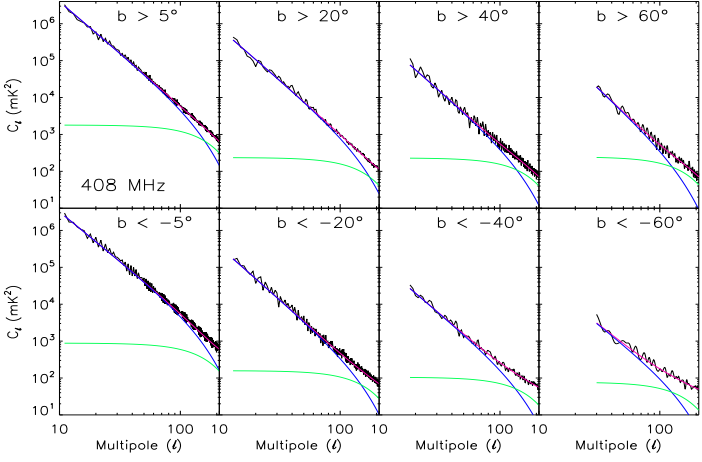}\\
   \vskip +0.2cm
   \caption{ As in Fig.~\ref{bfit1420}, but at 408 MHz. }
   \label{bfit408}
\end{figure}

\begin{figure}[!hb]
   \begin{tabular}{cc}
   \includegraphics[width=4.5cm,height=4cm]{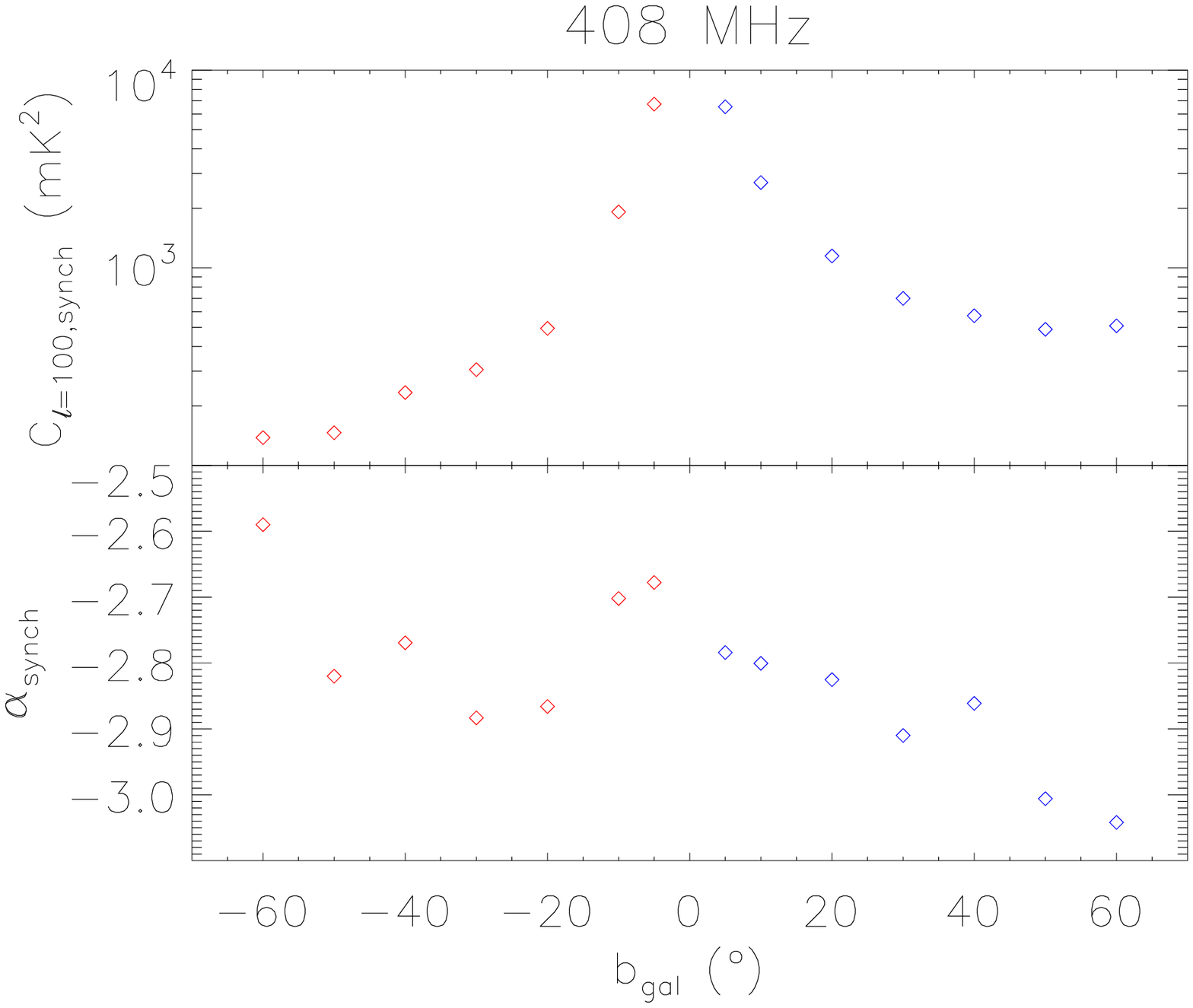}&
   \includegraphics[width=4.5cm,height=4cm]{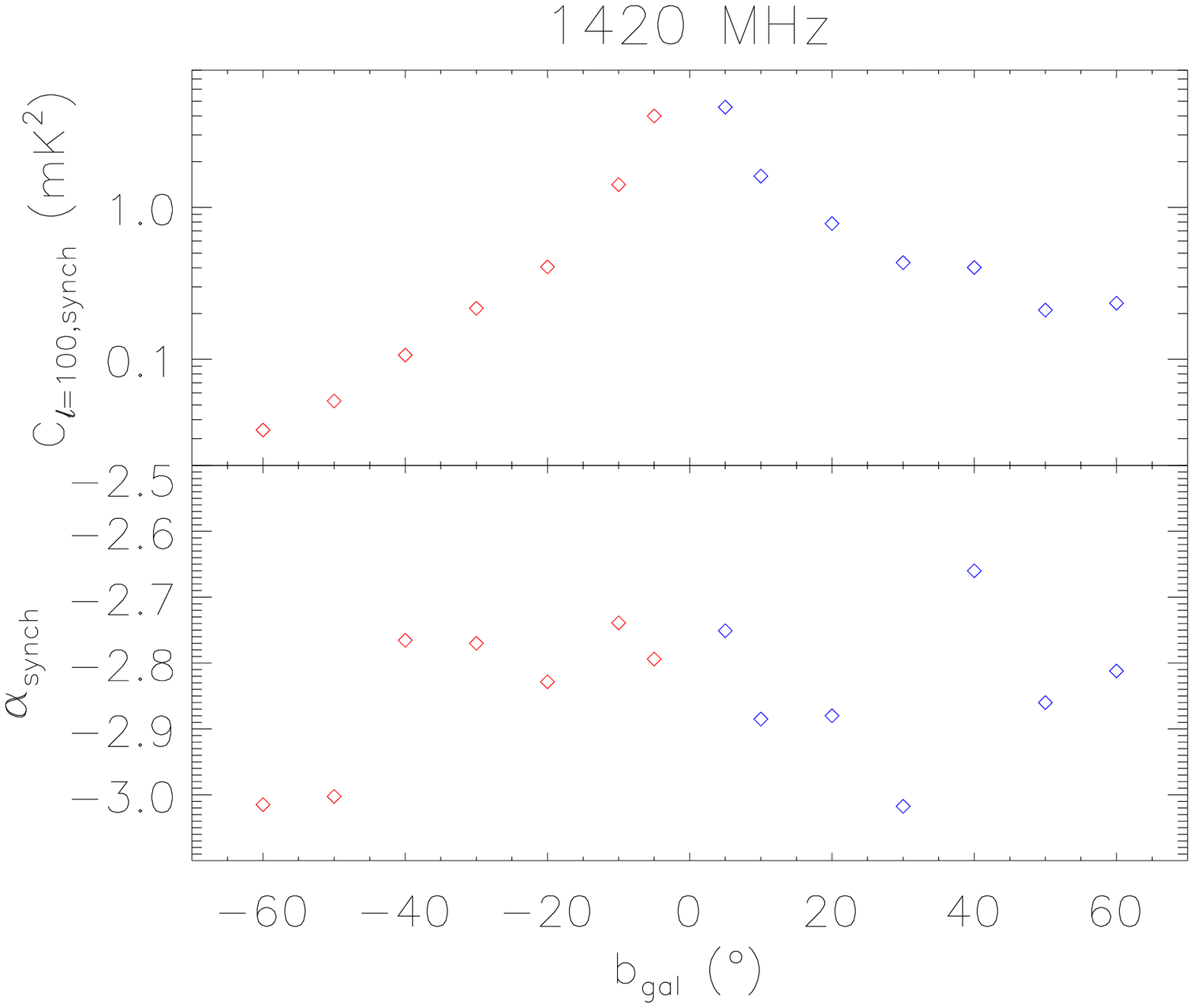}\\
   \end{tabular}
 \vskip +0.4cm   
   \caption{ Best-fit parameters obtained for the Galactic radio synchrotron emission APS  
   against Galactic latitude. }
   \label{bf_vs_lat_ti408_ti1420}
\end{figure}

In order to derive the range of variability of the synchrotron
emission amplitude and slope, two extreme  
cases have been considered (see Appendix C of La~Porta~2007 
for 
details).  
The flattest synchrotron APS compatible with the data 
is found by neglecting  
the source term in Eq.~\ref{aps_best_model} 
and the steepest one is recovered by assuming a 
null noise contribution 
and maximizing the source term. 
We performed a least-square fit to the APS by exploring the 
parameter space on adaptive grids\footnote{For this 
purpose we implemented a specific algorithm and 
tested its reliability with the MINUITS package of the CERN 
libraries \citep{james75_minuits}.}. 
The uncertainties on the best fit parameters are 
derived as the difference with those obtained in
 the two extreme cases. 
Figures~\ref{bfit1420} and \ref{bfit408} 
show the angular power spectra and the best fit curves 
corresponding to the best model at the two frequencies, 
while Tables~\ref{BFpar_tab1420} and \ref{BFpar_tab408}
list the obtained parameters and their uncertainties. 
For the synchrotron term, we quote the value of the normalized 
amplitude $k_{100}=k \times 100^\alpha$, which 
 corresponds to a physical quantity.  
In fact, $k_{100}=C_{\ell=100}$, thus implying that  
the normalized amplitude gives the mean temperature 
fluctuations at angular scales of $\sim 2^{\circ}$. 
Figure~\ref{bf_vs_lat_ti408_ti1420} shows the best 
fit parameters of the synchrotron APS as a function of the 
Galactic latitude. The normalized amplitude, $k_{100}$, 
is maximum when the considered cut includes the lower 
latitudes, where the Galactic radio emission peaks. 
In particular, at 408 MHz (respec. at 1420 MHz) 
$k_{100} \in [488,6527]\;{\rm mK}^2$ 
(respec. $[0.21,4.57]\;{\rm mK}^2$) 
for the northern cuts and 
$k_{100} \in [138,6734]\;{\rm mK}^2$ 
(respec. $[0.03,4.00]\;{\rm mK}^2$) 
for the southern cuts. 
The mean error on $k_{100}$ is of 
$\sim 18\%$ for the cuts at the lower 
latitude ($|b_{cut}| \in [5^{\circ},30^{\circ}]$) 
and of $\sim 30-40\%$ for the others. The 
uncertainty is larger for the cuts at higher 
latitude due to the reduced multipole range suitable 
for the fitting procedure. 
The slope of the synchrotron APS for the northern cuts 
varies in the interval $\sim [-3.0,-2.8]$ at 408 MHz and 
$\sim [-3.0,-2.7]$ at 1420 MHz, while in   
 the southern cuts $\alpha \sim [-2.9,-2.6]$  
and $\alpha \sim [-3.0,-2.7]$, respectively. The errors 
on $\alpha$ are on average $\sim (5-7)\%$ 
for $|b_{cut}| < 30^{\circ}-40^{\circ}$ and 
typically increase to $\sim 18\%$ for cuts at higher 
latitude. \\ 
At both frequencies there is no evidence of a 
systematic dependence of the synchrotron  
emission APS slope on latitude 
(see Fig.~\ref{bf_vs_lat_ti408_ti1420}). \\

The source term $c^{src}$ increases   
for decreasing latitude, as expected given that the 
source subtraction is less complete below $\sim 45^{\circ}$. 
From extragalactic source counts at 1.4~GHz in total intensity 
\citep{prand01_src_counts}, the expected source contribution is 
 $c^{src} \simeq 0.06 {\rm mK}^2$ for flux densities below $\sim 1$~Jy
($c^{src}\simeq [0.03 - 0.3] {\rm~mK}^2$ including 
the quoted $1\sigma$ errors and also considering  
the effect of the finite sampling) and 
$c^{src} \simeq 0.30$~mK$^2$ ($c^{src}\simeq [0.15 - 1.50]$~mK$^2$ ) 
for flux densities below $\sim 5$~Jy. 
At 408~MHz the available source counts \citep{jamr04_src_counts}  
lead to  $c^{src} \simeq 200$~mK$^2$ for 
flux densities below $\sim 6.4$~Jy 
($c^{src}\simeq [150 - 360$~mK$^2$) and 
$c^{src} \simeq 660$~mK$^2$ for flux densities below $\sim 64$~Jy
($c^{src}\simeq [260 - 1150]\,{\rm mK}^2$). 
The values resulting from the fits are consistent with the
above estimates. 

\begin{table}[!h]
\newcommand\T{\rule{0pt}{2.6ex}}
\newcommand\B{\rule[-1.2ex]{0pt}{0pt}} 
\begin{center}
\begin{tabular}{|c|c|c|c|c|}
\hline
Coverage \T \B & \multicolumn{4}{|c|}{Best Fit parameters @ 408 MHz}\\
\hline
  \T \B        &  $k_{100}$ (mK$^2$) & $\alpha$ & $c^{src}$ (mK$^2$) & $c^{noise}$ (mK$^2$)\\
\hline
$b_{gal} \ge 5^{\circ}$ \T \B & $ 6527. ^{ +1306.}_{-1349. }$  &
                           $-2.78 ^{ +0.14}_{-0.22 }$  &
                           $1797. ^{-1797.}_{+1223. }$  &
                           $187.09 ^{+377.84  }_{-187.09}$ \\
\hline
$b_{gal} \le -5^{\circ}$ \T \B & $ 6734. ^{+  743.}_{ -789. }$  &
                            $-2.68 ^{ +0.06}_{-0.11 }$  &
                            $ 882. ^{-882.}_{+1244. }$  &
                            $291.87 ^{+143.62 }_{-291.87}$ \\
\hline
$b_{gal} \ge 10^{\circ}$ \T \B & $ 2700. ^{ + 467.}_{ -444. }$  &
                            $-2.80 ^{ +0.12}_{-0.21 }$  &
                            $ 530. ^{-530.}_{ +520. }$  &
                            $ 95.16 ^{+106.22  }_{-95.16}$ \\
\hline
$b_{gal} \le -10^{\circ}$ \T \B & $ 1919. ^{+  111.}_{ -227. }$  &
                            $-2.70 ^{+ 0.04}_{-0.10 }$  &
                            $ 211. ^{-211.}_{ +438. }$  &
                            $103.99 ^{+ 53.09 }_{-103.99}$ \\
\hline
$b_{gal} \ge 20^{\circ}$ \T \B & $ 1147. ^{ + 215.}_{ -200. }$  &
                           $-2.83 ^{ +0.16}_{-0.19 }$  &
                           $ 236. ^{-236.}_{ +231. }$  &
                           $ 48.98 ^{ +39.85  }_{-48.98}$ \\
\hline
$b_{gal} \le -20^{\circ}$ \T \B & $  493. ^{ + 103.}_{  -86. }$  &
                            $-2.87 ^{ +0.13}_{-0.19 }$  &
                            $ 158. ^{-158.}_{ +129. }$  &
                            $ 31.71 ^{ +37.20 }_{-31.71}$ \\
\hline
$b_{gal} \ge 30^{\circ}$ \T \B &  $ 700. ^{ + 135.}_{ -104. }$  &
                           $-2.91 ^{+ 0.17}_{-0.16 }$  &
                           $ 181. ^{-181.}_{ +148. }$  &
                           $ 32.95 ^{ +38.54  }_{-32.95}$ \\
\hline
$b_{gal} \le -30^{\circ}$ \T \B & $  305. ^{+  112.}_{  -52. }$  &
                            $-2.88 ^{ +0.26}_{-0.14 }$  &
                            $ 155. ^{-155.}_{ +87. }$  &
                            $ 21.81 ^{ +32.50 }_{-21.81}$ \\
\hline
$b_{gal} \ge 40^{\circ}$ \T \B & $  572. ^{  +205.}_{ -111. }$  &
                            $-2.86 ^{ +0.34}_{-0.18 }$  &
                            $ 230. ^{-230.}_{ +106. }$  &
                            $ 19.56 ^{ +41.28  }_{-19.56}$ \\
\hline
$b_{gal} \le -40^{\circ}$ \T \B & $  234. ^{ +  97.}_{  -66. }$  &
                            $-2.77 ^{ +0.35}_{-0.29 }$  &
                            $ 105. ^{-105.}_{ +123. }$  &
                            $ 32.47 ^{ +15.08 }_{-32.47}$ \\
\hline
$b_{gal} \ge 50^{\circ}$ \T \B & $  488. ^{ + 246.}_{  -73. }$  &
                            $-3.01 ^{ +0.46}_{-0.11 }$  &
                            $ 208. ^{-208.}_{ +117. }$  &
                            $ 27.07 ^{ +23.67  }_{-27.07}$ \\
\hline
$b_{gal} \le -50^{\circ}$ \T \B & $  146. ^{ + 122.}_{  -34. }$  &
                            $-2.82 ^{ +0.66}_{-0.22 }$  &
                            $ 137. ^{-137.}_{  +77. }$  &
                            $ 19.95 ^{ +22.47 }_{-19.95}$ \\
\hline
$b_{gal} \ge 60^{\circ}$ \T \B & $  509. ^{  +226.}_{  -80. }$  &
                           $-3.04 ^{ +0.39}_{-0.26 }$  &
                           $ 246. ^{-246.}_{  +91. }$  &
                           $ 19.11 ^{ +45.89  }_{-19.11}$ \\
\hline
$b_{gal} \le -60^{\circ}$ \T \B & $  138. ^{   +71.}_{  -76. }$  &
                            $-2.59 ^{ +0.53}_{-0.80 }$  &
                            $  77. ^{-77.}_{ +131. }$  &
                            $ 32.67 ^{ +11.66 }_{-32.67}$ \\
\hline
\hline
\end{tabular}
\end{center}
\caption{As in Table~\ref{BFpar_tab1420} but at 408 MHz.
         }
\label{BFpar_tab408}
\end{table}

\section{Extrapolation to the microwave range}
\label{res_extrapolation}

In this section we extrapolate the results obtained 
from the analysis of the 408 MHz and 1420 MHz surveys 
over large areas to the microwave range, in order to 
make a direct comparison with the WMAP 3-yr results. 
The main objective of the WMAP mission was the 
realization of a CMB anisotropy total intensity map
and the estimation of the corresponding APS. 
By-products of the mission are maps of the foregrounds 
contaminating the cosmological signal at 
the five frequencies observed by 
the satellite ($\nu \sim 23, 33, 41, 61, 94$~GHz), i.e. 
the microwave emission from the Milky Way, characterized 
by diffuse (dust, free-free, synchrotron) 
and discrete (e.g, {\sc HII} regions, SNRs) components, 
and from extragalactic sources. 
The maps were worked out \citep{hins06_wmap_3yr_temp} by using  
templates of the various astrophysical components,    
constructed by exploiting ancillary data.   
Then a pixel-by-pixel (MEM based) fit of all the maps 
(see Bennett et al.,~2003 for a description of the method), 
i.e. templates and {\sc WMAP} frequency maps after subtraction 
  of the CMB anisotropy field, 
  was performed posing some priors on the 
 spectral behaviour of the foregrounds.\\ 

\subsection{Comparison with the WMAP K-band synchrotron component}
\label{extrap_kmem_comp}

The 23 GHz (K-band) synchrotron map by \citet{hins06_wmap_3yr_temp} 
is considered in this section for a comparison 
with the 408 MHz and 1420 MHz data.
It is evident that such a map provides a picture of the global
non-thermal emission observed by the satellite at 23 GHz, rather
than the Galactic synchrotron component only. Several
extragalactic sources are clearly recognizable. 
Furthermore, as pointed out by \citet{hins06_wmap_3yr_temp}, 
the diffuse non-thermal emission is concentrated at low 
latitudes and appears remarkably well correlated with the 
dust component~\footnote{Such a tight synchrotron-dust 
correlation holds at all WMAP frequencies.}.
This might suggest the presence at 23 GHz 
of anomalous dust emission, as due for example to spinning 
dust grains \citep{draine98_spinndust}. In fact, most dusty active star-forming 
regions are localized along the Galactic plane. 
This hypothesis seems further 
 supported by the joint analysis of the {\sc WMAP} maps 
 (1-yr release) and the Green Bank Galactic Plane 
 Survey by Finkbeiner~(2004). 
De Oliveira-Costa et al.~(2004) estimated the fluctuations
expected at 10 GHz and 15 GHz for the foreground component traced
by the {\tt K-MEM} synchrotron map by Bennett et al.~(2003) (1-yr
  results), by cross-correlating the latter with the 
 Tenerife 10 GHz and 15 GHz CMB maps and all the {\sc WMAP} CMB maps.
They found values one order of magnitude below what is expected 
 for the synchrotron emission and 
concluded that the {\tt K-MEM} synchrotron component 
by Bennett et al.~(2003) is dominated by anomalous dust 
emission even at $|b_{gal}|\gtrsim 20^{\circ}$. 
\citet{hilde07_cosmosomas} also found evidence for 
anomalous microwave emission at high Galactic latitudes  
by cross-correlating the COSMOSOMAS 11 GHz observations 
with the {\sc WMAP} K- and Ka-band map.   
The same conclusion also was reached by \citet{dav06_foresepa}, 
who cross-correlated the {\sc WMAP} 1-yr map with foreground templates
in a dozen small patches located at medium and high latitude.
However, the origin of the spatial correlation  
found at {\sc WMAP} frequencies between synchrotron 
and dust emission is still a matter of debate. 
Bennett et al.~(2003) claim that 
the observed correlation is 
the result of a spatially varying synchrotron 
spectral index, which significantly alters the 
morphology of the synchrotron emission with frequency. 
\citet{hins06_wmap_3yr_temp} affirm that the issue is 
 left open also by the {\sc WMAP} 3-yr results and that
high quality and large coverage surveys at
$\nu \sim 5 - 15\;{\rm GHz}$ are needed
for a decisive test of both the above discussed
explanations of the synchrotron-dust correlation. \\
The separation of the free-free and synchrotron
emission in low latitude regions is also very uncertain. 
On one hand, the 408 MHz map used as a template of the non-thermal
emission contains a non-negligible contribution ($\lesssim 10\%$) of 
 free-free at lower latitudes \citep{dickinson03_fftemplate,paladini05_ff}.  
 On the other hand, the H-$\alpha$ map used as 
 template for the free-free emission \citep{fink03_halpha_map} 
 cannot be properly corrected for dust extinction
 for $|b_{gal}| \lesssim 5^{\circ}$, thus potentially leading 
 to an underestimation of the expected 
 thermal emission at 23 GHz. 
The situation in the vicinity of the Galactic plane
is extremely complicated and remains unclear.\\ 

The extrapolation of the angular power  
spectra derived at 408 MHz and 1420 MHz
 to the microwave range is a delicate issue. 
The astrophysical components contributing to
the fluctuation field APS of the radio 
surveys scale with frequency in a different way. 
For the Galactic radio emission between 408 MHz and 1420 MHz 
\citet{reich04_betagalemiss} compiled a map of the 
spectral index $\beta$ ($T_{b} \propto \nu^{\beta}$) that 
reveals a complex structure, due to superposition 
of the spectral behaviour of the map components  
(synchrotron emission, sources, free-free). 
The situation is further complicated by 
 a possible but not well known 
 steepening of the diffuse synchrotron emission 
power spectrum above 10~GHz, due to the steepening 
of the cosmic ray electron energy spectrum 
\citep{band90_crspec,band91_crspec,strong07_CRrev}. 
Last but not least, the fact that the astrophysical  
components of the map scale with frequency in a different way 
 may imply a change in the overall 
 APS shape, since the relative weight of the 
 foreground contribution to the APS could  
 vary significantly. \\

Given the complexity of the open issues discussed above, 
the following analysis merely aims to verify the consistency 
between the information about the non-thermal radiation 
APS coming from the 408 MHz and 1420 MHz data and 
from the 23 GHz {\sc WMAP} data. 
We focused on what happens at medium and high 
Galactic latitudes, since the problems in interpreting the Galactic 
emission are more complicated close to the plane. \\ 

\begin{figure}[!t]
   \vskip +0.5cm
   \centering
   \hskip +0.8cm
   \includegraphics[width=6cm,height=8cm,angle=90]{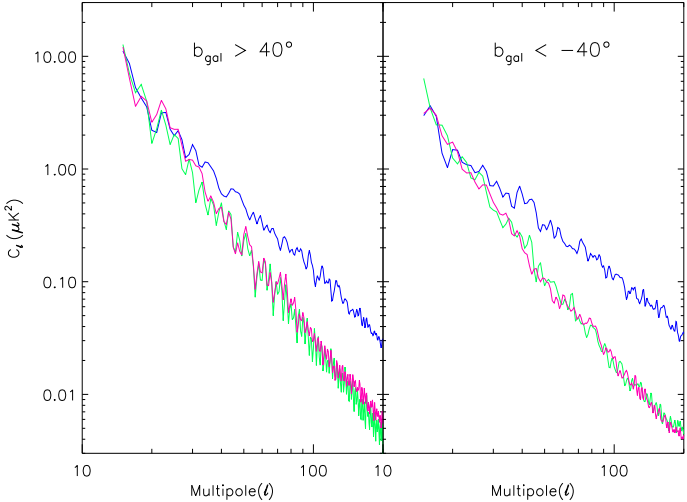}
   \vskip +0.6cm
   \caption{ Angular power spectra of the DS-subtracted radio maps 
    (green $\to$ 408 MHz, fuxia $\to$ 1420 MHz) extrapolated 
    to 23 GHz for a direct comparison with those of the WMAP 3-yr 
    DS-subtracted synchrotron component (blue). 
    The radio angular power spectra have been smoothed to $1^{\circ}$ 
    to match the angular resolution of the 23 GHz map. 
    The frequency spectral indices 
    adopted in the extrapolation are $\beta_{(0.408-23){\rm GHz}}=-2.95$ 
    and $\beta_{(1.4-23){\rm GHz}}=-2.90$ for the northern cut and 
    $\beta_{(0.408-23){\rm GHz}}=-2.90$ and $\beta_{(1.4-23){\rm GHz}}=-2.83$ 
    for the southern one.}
   \label{radioaps_vs_wmap}
\end{figure}

\begin{figure}[!t] 
  \hskip +1.5cm
  \vskip +0.5cm 
  \includegraphics[width=6cm,height=10cm,angle=90]{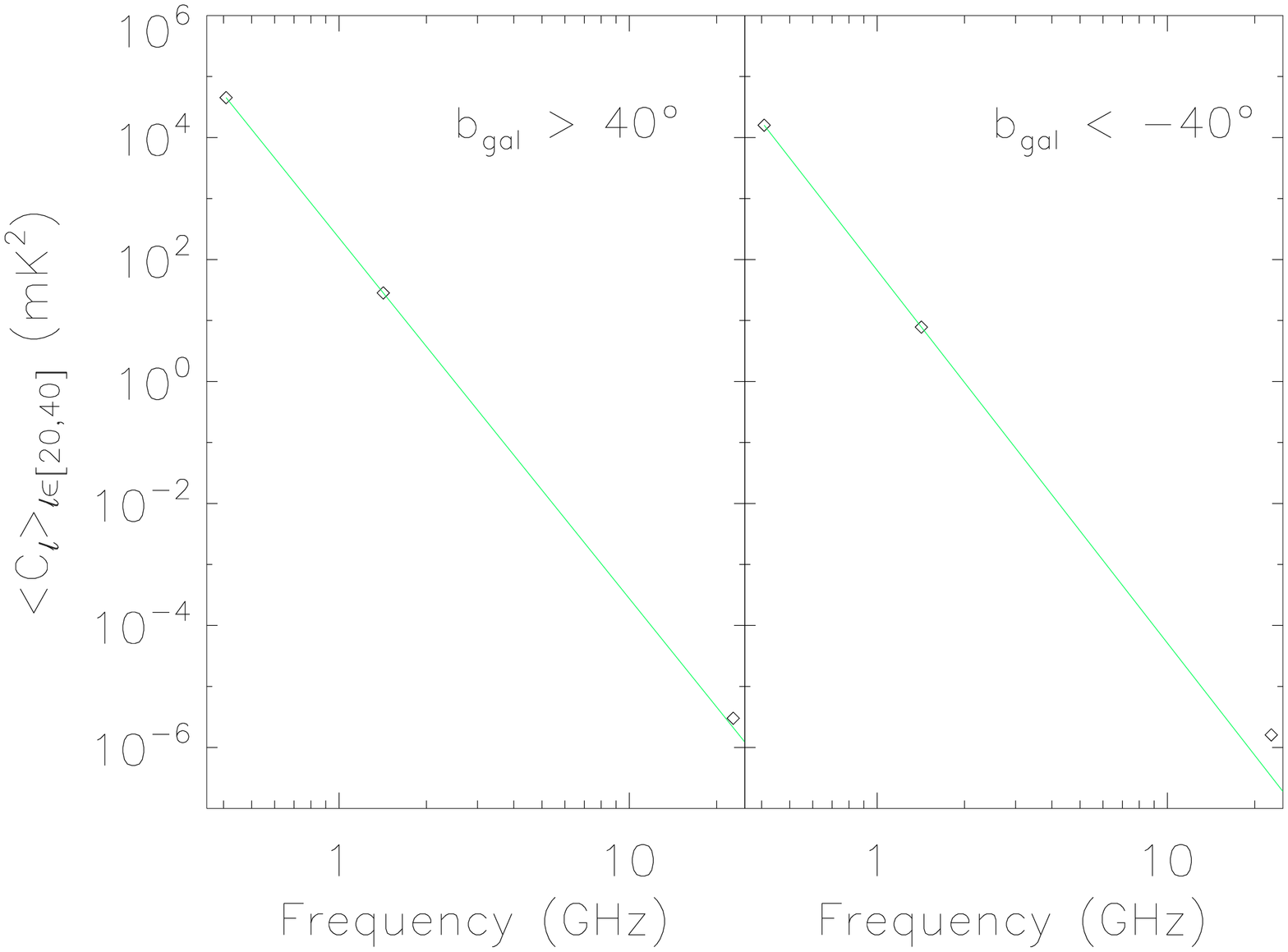}
   \vskip -0.4cm
   \caption{ Mean angular power spectra, $<C_{\ell}(\nu)>_{\ell\in[20,40]}$, of    
             high latitude cuts ($b_{cut} = \pm 40^{\circ}$) against frequency. 
	     The best fit of the mean APS of the two lower frequencies 
	     is also plotted. 
             }
   \label{beta_radio_wmap}
\end{figure}

We first carried out a source subtraction on the 23 GHz 
map at intermediate and high latitudes ($|b_{gal}| \gtrsim 40^{\circ}$), 
similar to that performed on the radio surveys. 
The comparison of the map of subtracted sources with 
the mask of sources produced by the {\sc WMAP} team 
shows that most ($\sim 80\%$) of the objects have been 
 identified 
and subtracted. 
We have derived and compared the APS of the  
original, source-subtracted and source map at 23 GHz 
for some northern and southern cuts. Namely, we considered   
$|b_{cut}|=40^{\circ},50^{\circ},60^{\circ}$ and 
verified that $C_{\ell}^{orig.\;map} \sim c^{src}$ over 
 the significant multipole range (i.e. $\ell \gtrsim 20$). This means that in 
the 23 GHz map the source contribution dominates the high latitude cut  
APS at all angular scales. In Figure~\ref{radioaps_vs_wmap} 
we show the APS of the 23 GHz map after source 
subtraction for the asymmetric cuts with $b_{cut}=\pm40^{\circ}$. 
For comparison, we also display the extrapolated radio APS 
(from Tables ~\ref{BFpar_tab1420} and \ref{BFpar_tab408}), derived as 
$$C_{\ell}(23)=C_{\ell}(\nu_{radio})\cdot (23/\nu_{radio})^{2\beta}$$ 
where $\nu_{radio}=0.408,1.420$~GHz. The frequency spectral  
index $\beta$ is chosen case by case as the value that 
brings the radio APS to overlay the {\sc WMAP} one
 at the lower multipoles (the exact values are reported 
 in the figure caption). 
The extrapolated angular power spectra 
are very similar to each other, but 
 steeper than those of {\sc WMAP}. Furthermore,  
the frequency spectral indices needed in the extrapolation 
suggest the existence of a steeper spectral behaviour   
between 408 MHz and 23 GHz than between 1420 MHz and 23 GHz. 
For the Galactic diffuse synchrotron emission we would 
expect instead 
$\beta_{(0.408-23){\rm GHz}} \ge \beta_{(1.4-23){\rm GHz}}$  
because of the possible steepening of the cosmic ray 
energy spectrum.  
In order to obtain a more quantitative estimate of 
the mean spectral index between the lower frequencies 
and 23 GHz, we used the mean value of the APS 
of the source-subtracted maps for $\ell \in [20,40]$. 
In fact, in this range the APS is dominated 
by the diffuse synchrotron emission, whereas at higher 
multipoles the APS could still be influenced by the 
contribution of unsubtracted sources. 
Figure~\ref{beta_radio_wmap} shows the mean APS as a function 
of frequency for the cuts with $b_{cut}=\pm 40^{\circ}$. 
The spectral index $\beta$ is obtained as \\ 
$$<C_{\ell}(\nu_{1})>_{\ell\in[20,40]}=<C_{\ell}(\nu_{2})>_{\ell\in[20,40]} (\nu_{1}/\nu_{2})^{2\beta}\;{\rm .}$$ 
For the northern cut we found $\beta_{(0.408-23){\rm GHz}} \sim -2.92$  
and $\beta_{(1.42-23){\rm GHz}} \sim -2.85$.  
For the southern cuts, $\beta_{(0.408-23){\rm GHz}} \sim -2.76$ 
and $\beta_{(1.42-23){\rm GHz}} \sim -2.59$. 
These values of $\beta$ have a typical uncertainty 
of a few percent according to the  
choice of the multipole range adopted to compute 
the mean value of the APS. However, it turns out that  
$\beta_{(0.408-23){\rm GHz}} < \beta_{(1.42-23){\rm GHz}}$ 
for all reasonable choices of the multipole range. 
One possible explanation for such a result is that 
the 23 GHz map includes one or more astrophysical components 
beside synchrotron emission. On the one hand, the 23 GHz 
map could still include some Galactic free-free emission,  
residual from the component separation. Another and likely 
more relevant candidate is Galactic spinning dust emission. 
A rough estimate of the excess signal in the 23 GHz 
map can be obtained by extrapolating  
the radio results to 23 GHz, as shown in Fig.~\ref{beta_radio_wmap}. 
The observed and extrapolated mean APS differ by a 
factor of $\sim 2.5$ in the northern cut 
and $\sim 8.4$ in the southern cut. Thus, they differ by 
factors of $\sim 1.6$ and $\sim 2.9$ respectively 
in terms of signal in the map. We repeated this 
calculation by computing the mean APS over other 
reasonable intervals of multipoles and found in this way 
that the uncertainty for the given values of the excess 
signal is about 20\%. 
The difference between the observed and extrapolated 
values of the mean APS is smaller in the northern 
hemisphere, which is likely due to the compensatory 
contribution of the NPS.  

Finally, we find that the angular power spectra 
of the northern and southern sky are almost superimposed 
 at 23 GHz for $|b_{cut}|\gtrsim 40^{\circ}$. 
This can be interpreted as the combination of two effects. 
The synchrotron emission of the NPS has a steeper frequency 
spectrum than the average one \citep{reich88_betasynchr}  
and the contribution of emission processes 
other than synchrotron may be significant 
at 23~GHz. Therefore, the relative importance of the NPS with 
respect to the overall diffuse background diminishes from 
408 MHz and 1420 MHz to microwave frequencies.

\begin{table}[!t] 
\begin{center} 
\begin{tabular}{|c|c|c|c|c|c|c|c|} 
\hline 
 \multicolumn{8}{|c|}{$\beta_{(0.408-1.420)GHz}$}\\ 
\hline 
$|b_{cut}|$ & $5^{\circ}$ & $10^{\circ}$ & $20^{\circ}$ & $30^{\circ}$ & $40^{\circ}$ & $50^{\circ}$ & $60^{\circ}$ \\ 
\hline 
$b_{gal} > |b_{cut}|$ & -2.9 & -2.9 & -2.9 & -3.0 & -3.1 &  -3.2 & -3.2 \\
\hline 
$b_{gal} < -|b_{cut}|$ & -2.9 & -2.9 & -2.9 & -2.9 & -3.0 & -3.1 & -3.1 \\ 
\hline
\end{tabular}
\end{center}
\caption{Frequency spectral index of the Galactic synchrotron  
          fluctuations between 408 MHz and 1420 MHz 
	  derived from the APS.} 
\label{beta_408_1420}
\end{table}

\subsection{Synchrotron contamination of the CMB anisotropies} 

\begin{figure*}
\vskip +0.8cm
\hskip +4cm
\includegraphics[width=8cm,height=12cm,angle=90]{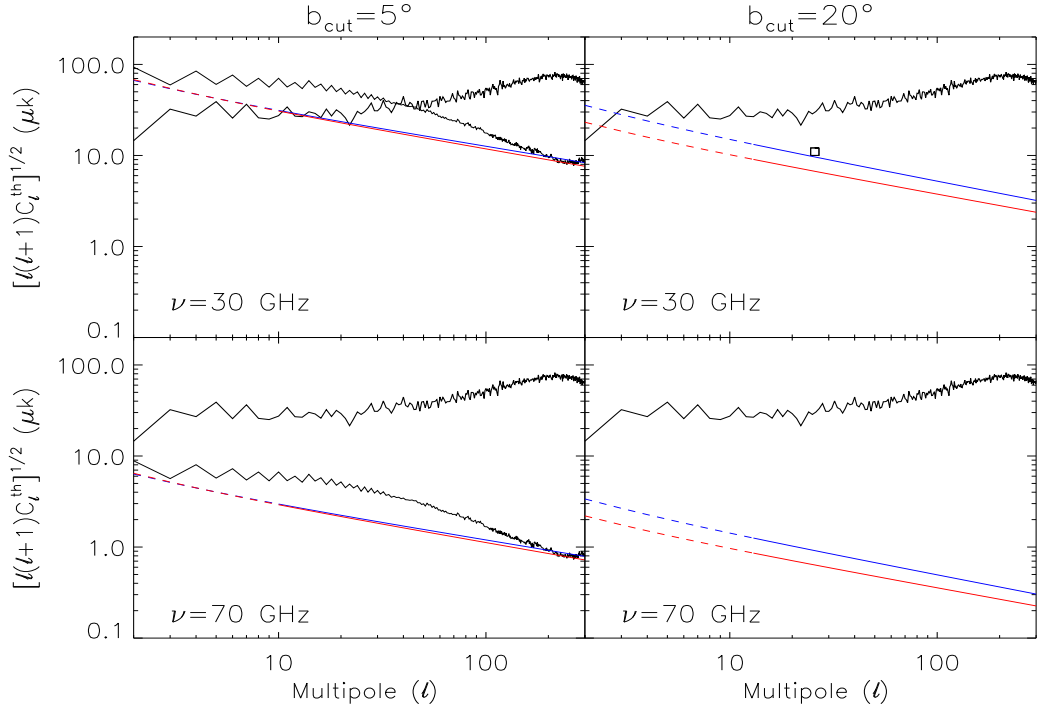}
\vskip +0.7cm
\caption{ Comparison between the CMB APS retrieved by \citet{hins06_wmap_3yr_temp} 
          and the synchrotron angular power spectra derived for some cuts of the 
          1420 MHz map (from Table~\ref{BFpar_tab1420} and Fig.~\ref{bfit1420}), 
          extrapolated to 30 GHz and 70 GHz with a spectral index of -2.9~. 
      Color legend (see online version): 
      blue (upper straight lines) $\to$ $C_{\ell}^{N}$ (northern cut), 
      red (lower) $\to$ $C_{\ell}^{S}$ (southern cut). 
      The left panels also display the APS for the all-sky (black line) 
      extrapolated as above. 
      The empty square in the top right panel marks the upper limit 
      on synchrotron contamination inferred from COBE-DMR observations 
      \citep{kogut96b_spinndust}. 
}
\label{cmbTT_vs_radio} 
\end{figure*}

It is a standard practice to estimate the foreground 
contamination of CMB anisotropies by means of the 
corresponding APS, which is usually extrapolated 
from the frequency range where the foreground component 
is best observed. For the Galactic synchrotron emission 
a constant spectral index in the interval $\sim[-2.5,-3.0]$ 
is commonly adopted, as suggested by the spectral 
behaviour of the Galactic diffuse emission at radio frequencies.  
We instead derive the spectral index directly from the 
results of our APS analysis, thus identifying a proper 
value for each cut considered. 
 As in the previous section, we compute the mean value 
of the APS at 408 MHz and at 1420 MHz over the lower 
multipoles ($\ell \in [20,40]$) and perform a linear 
extrapolation based on the two points. 
We prefer to work with the APS of the source-subtracted 
map rather than with the synchrotron power law   
derived by fitting it, since the former provides us with 
a value that exclusively depends on observed data. 
Our results are reported in Table~\ref{beta_408_1420}.    
The obtained spectral indices vary by a few percent 
for a different choice of the multipole range 
used to calculate the mean APS. 
Figure~\ref{cmbTT_vs_radio} shows the CMB APS recovered by WMAP 
\citep{hins06_wmap_3yr_temp}, together with the synchrotron APS 
 derived from the 1420 MHz survey, extrapolated 
 to 30~GHz and 70~GHz (corresponding to the lowest 
 and the highest {\sc Planck-LFI} channels). We display 
 the results obtained for four coverage cases 
($b_{cut}=\pm 5^{\circ},\pm 20^{\circ}$). 
For comparison, we also extrapolated as above 
the APS directly extracted from the map
for the region at $|b| \ge 5^\circ$
and for the all-sky. 
The foreground dominates over the CMB at 30~GHz for a wide multipole range
if a mask excluding the Galactic plane is not applied.
The frequency spectrum of free-free emission, 
relevant at low latitudes, is flatter than 
that of the synchrotron emission. Thus, the 
extrapolated APS provides a lower limit to the overall 
Galactic foreground, even neglecting dust emission.

The analysis of the low-frequency maps shows that the
 APS amplitude of the northern Galactic hemisphere is
strongly influenced by the presence of the NPS.  
Consequently, the results obtained at 1420 MHz 
for the northern cuts constitute a conservative  
upper limit for the Galactic diffuse synchrotron  
emission and can be used together with those 
of the southern cuts to bracket the synchrotron 
APS at microwave frequencies. 

At 30 GHz, a severe contamination is expected from the 
synchrotron emission up to $\ell \sim 50$ for 
an almost complete sky coverage ($b_{cut}=5^{\circ}$). 
A mask excluding the region with $|b_{gal}| \le 20^{\circ}$
 reduces the expected synchrotron signal to about half of 
the CMB anisotropies for $\ell \gtrsim 10$, whereas  
for lower multipoles the two are comparable. 
\citet{kogut96b_spinndust} examined the COBE-DMR 
results at 31.5~GHz for $|b_{gal}| \gtrsim 20^{\circ}$ 
and derived an upper limit 
of $\sim 11 \mu$K on the temperature fluctuations 
due to synchrotron emission on angular scales 
of $\sim 7^{\circ}$. This value, marked in 
Figure~\ref{cmbTT_vs_radio} by an empty 
square, is in good agreement with the extrapolated APS for 
the northern cut at $20^{\circ}$.

At 70 GHz, which is the most promising channel 
for CMB anisotropy measurements 
since 
the overall foreground emission reaches a 
minimum for $\nu \sim [60,80]$~GHz 
(Bennett et al.~2003), the contribution of the 
Galactic synchrotron emission to the microwave sky 
fluctuation field is small over the multipole 
range explored in our analysis ($\ell \gtrsim 10$). 
The CMB anisotropies are larger than 
the foreseen foreground fluctuations by a factor 
$\gtrsim 10$ for a cut at $5^{\circ}$. 
For $b_{cut} \sim 20^{\circ}$ the foreground signal 
further decreases by a factor $\sim 2$. 
The extrapolation of our results to $\ell \lesssim 10$ 
indicates that the cosmological signal should be a factor 
$\gtrsim 2$ larger than the foreground at the 
largest angular scales. 
The precise recovery of the CMB APS for $\ell \lesssim 10$ 
therefore remains a delicate issue, since the 
foreground emission is a competitive signal.  
However, we note that the APS extracted directly from the map 
shows a certain flattening toward lowest multipoles, slightly 
improving the situation with respect to the above 
power law extrapolation.

\section{The APS dependence on sky position: the local analysis}

\begin{figure}
  \hskip +0.6cm
   \includegraphics[width=4cm,height=9cm,angle=90]{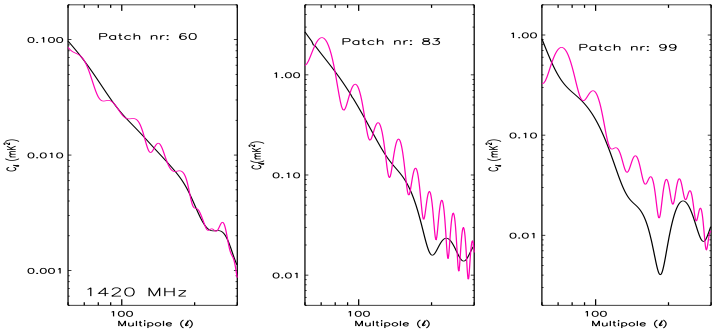}
   \vskip +0.2cm
   \caption{
    Comparison between the angular power spectra derived using
    {\tt Anafast} (fuxia) and via integration of the correlation
    function. Some examples of good (left panel), fair and
    poor (right) agreement are shown for the
    map at 1420 MHz after DSs subtraction.
    }
   \label{anaf_vs_cf_ex}
\end{figure}

We have also carried out the analysis of the APS on patches 
 of roughly $14\fdg7 \times 14\fdg7$, in 
 order to describe the local variations of the Galactic
emission at 408 MHz and 1420 MHz. 
Significant changes in the amplitude of the synchrotron APS
with the considered portion of the sky are expected, 
since the diffuse radio background gradually increases 
toward the Galactic plane, where it reaches maximum intensity. \\
These patches correspond to the pixels of an {\tt HEALPix} map
at $n_{side}=4$ and allow the study of the  angular power spectra  
 on the multipoles range $\sim[60,200-300]$. 
An angular size $\theta_{patch}\sim 14\fdg7$ is a good compromise
between the wish to divide the sky in a large number of
areas and the need to preserve a relatively wide 
interval of statistically relevant multipoles 
($\ell \sim 180^{\circ}/\theta$). 
We have computed the patch angular power spectra 
for all the versions of the radio maps (original, 
DS-subtracted and DSs only), both by using  
the {\tt HEALPix} facility {\tt Anafast} and by integrating 
the two point correlation function (see Appendix~D 
of La Porta~2007 for details).
Despite the differences found in individual cases, 
on the average there is a good agreement between 
the  angular power spectra derived with the two methods 
(in Fig.~\ref{anaf_vs_cf_ex} some examples of bad, 
fair and good cases are shown for the map at 1420 MHz 
after DS subtraction).\\
The angular power spectra obtained by using {\tt Anafast} 
typically present more oscillations\footnote{
{\tt Anafast} computes the APS in the Fourier space 
by expressing the temperature fluctuation 
field in spherical harmonics. 
The APS is obtained working over the whole sky, 
even if the map is zero outside 
the patch taken into account (the derived APS is 
then renormalized to the case of a full sky coverage).
This operation is heuristically equivalent 
to computing the Fourier transform of a discontinuos function, 
thus implying a Gibbs effect \citep{arfken01_book}.} 
and tend to be slightly flatter than the correlation function 
 angular power spectra at $\ell \gtrsim 200$. However, 
the correlation function results are less reliable at higher 
multipoles, where the choice of the window function might 
have a non negligible influence. 
Consequently, we exploited the {\tt Anafast}  
angular power spectra in the following analysis. 

\subsection{Results}

\begin{figure*}
   \hskip -0.5cm
   \begin{tabular}{ccc}
   Fit no SRC & Best Model & Fit no Noise \\
   \includegraphics[width=4.0cm,height=6cm,angle=90]{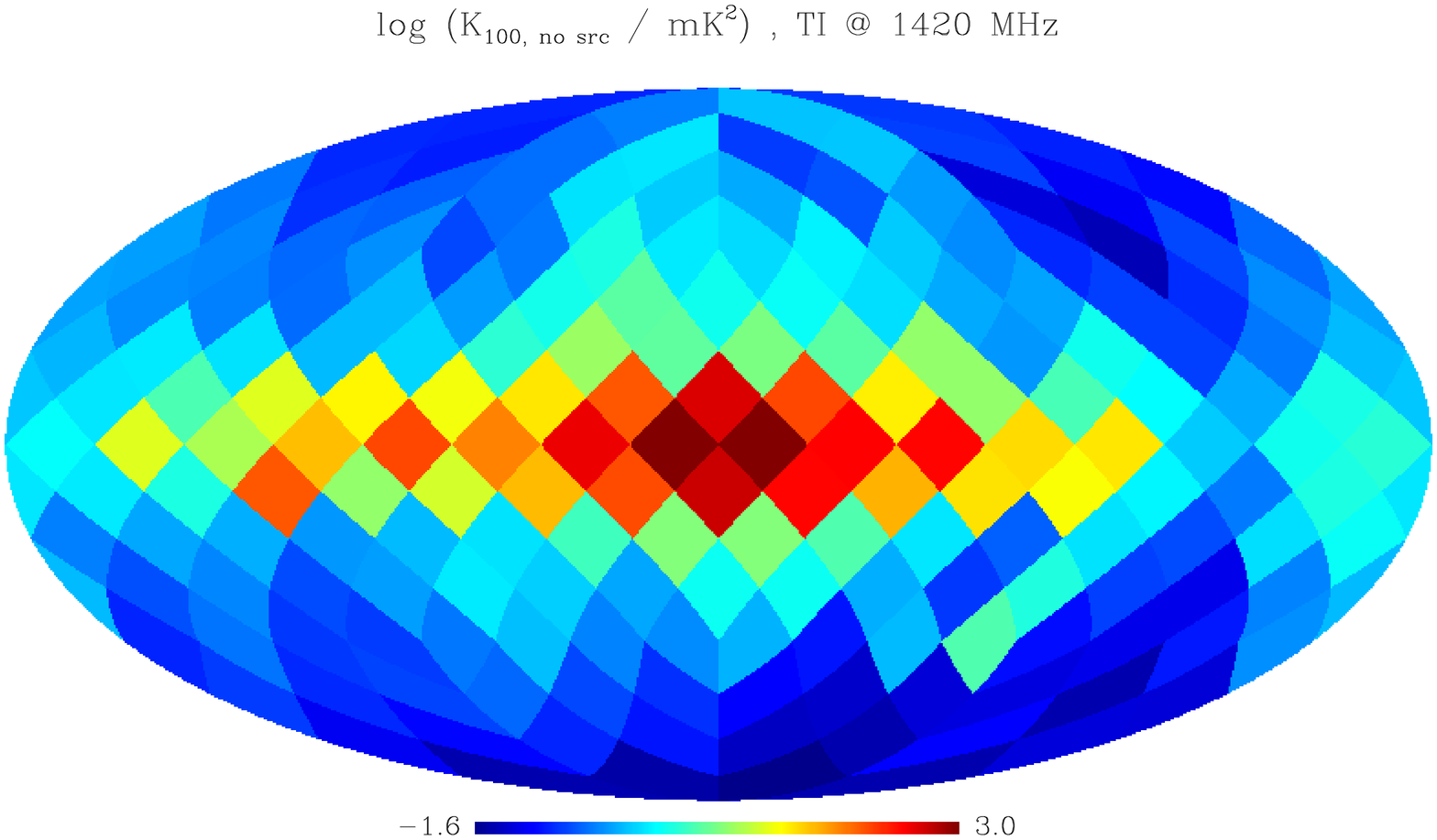}&
   \includegraphics[width=4.0cm,height=6cm,angle=90]{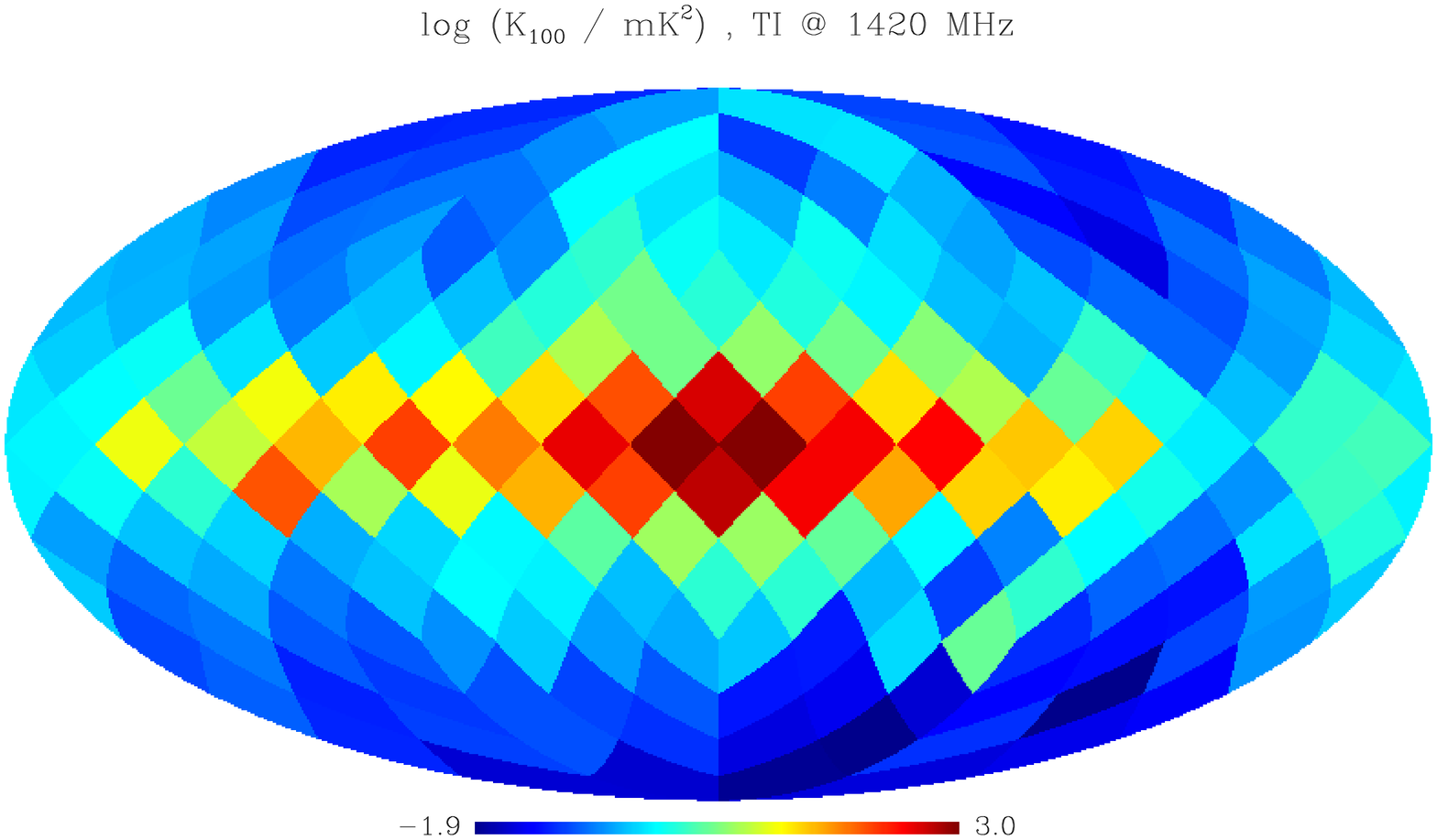}&
   \includegraphics[width=4.0cm,height=6cm,angle=90]{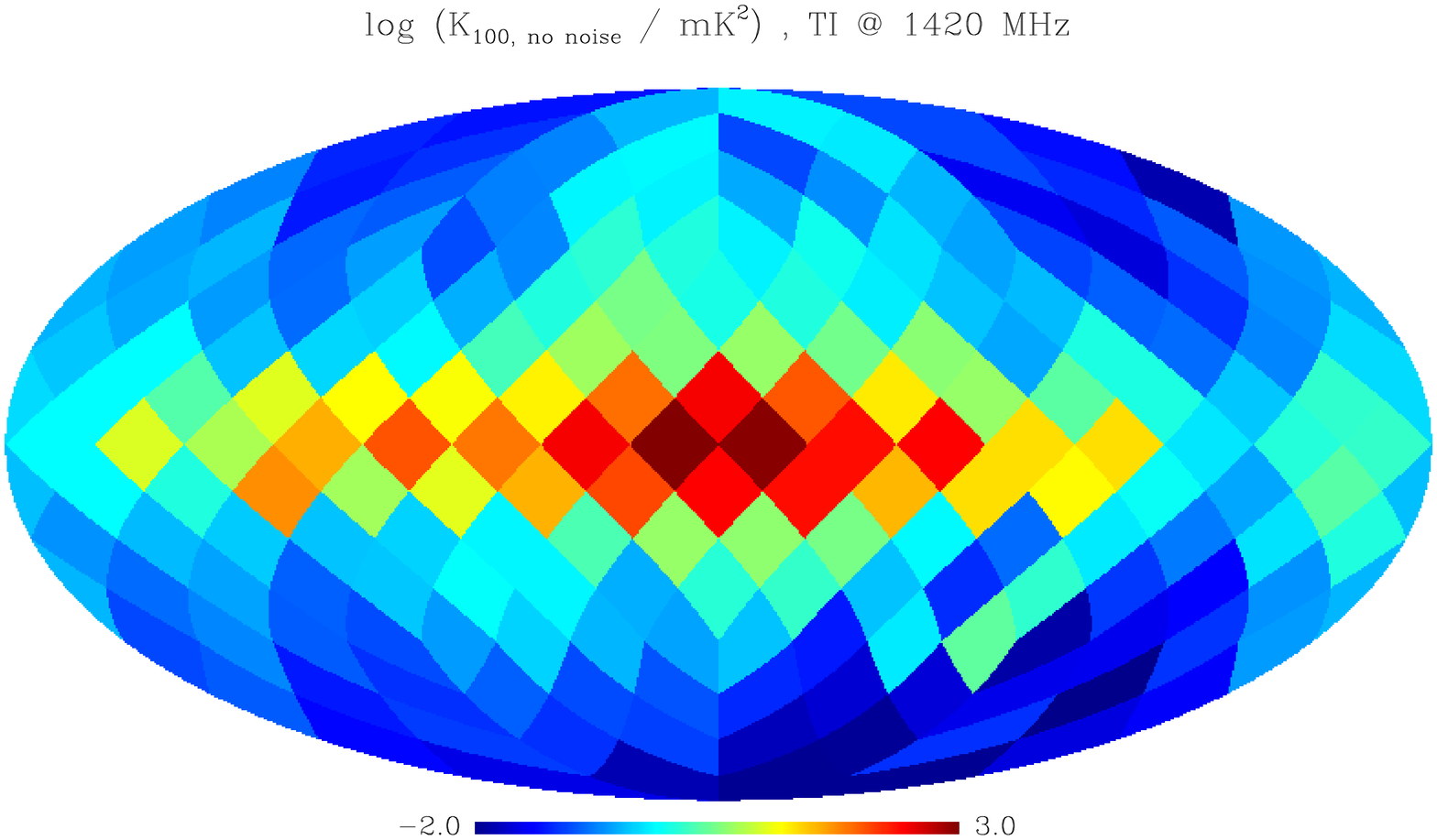}\\
   \includegraphics[width=4.0cm,height=6cm,angle=90]{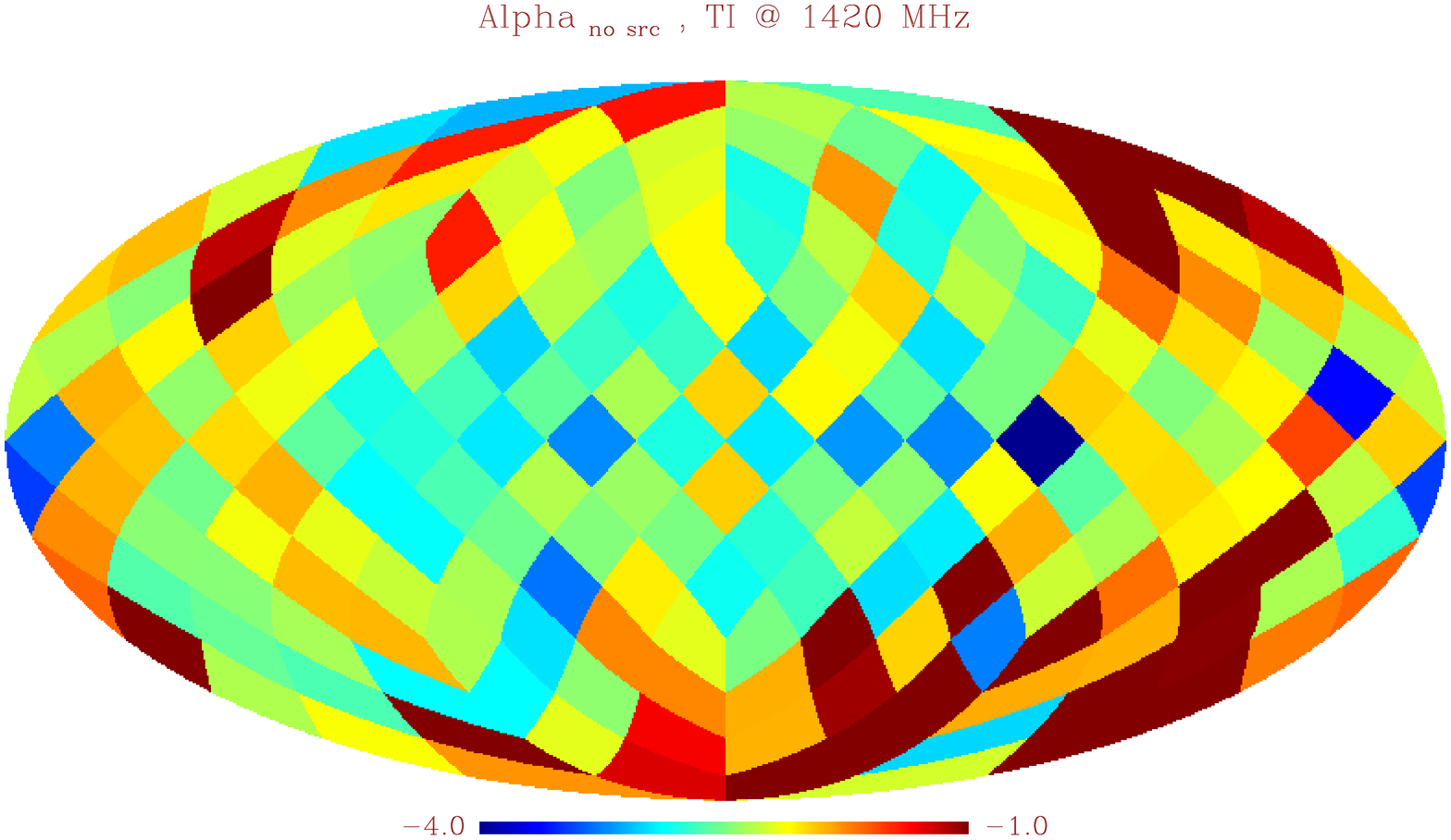}&
   \includegraphics[width=4.0cm,height=6cm,angle=90]{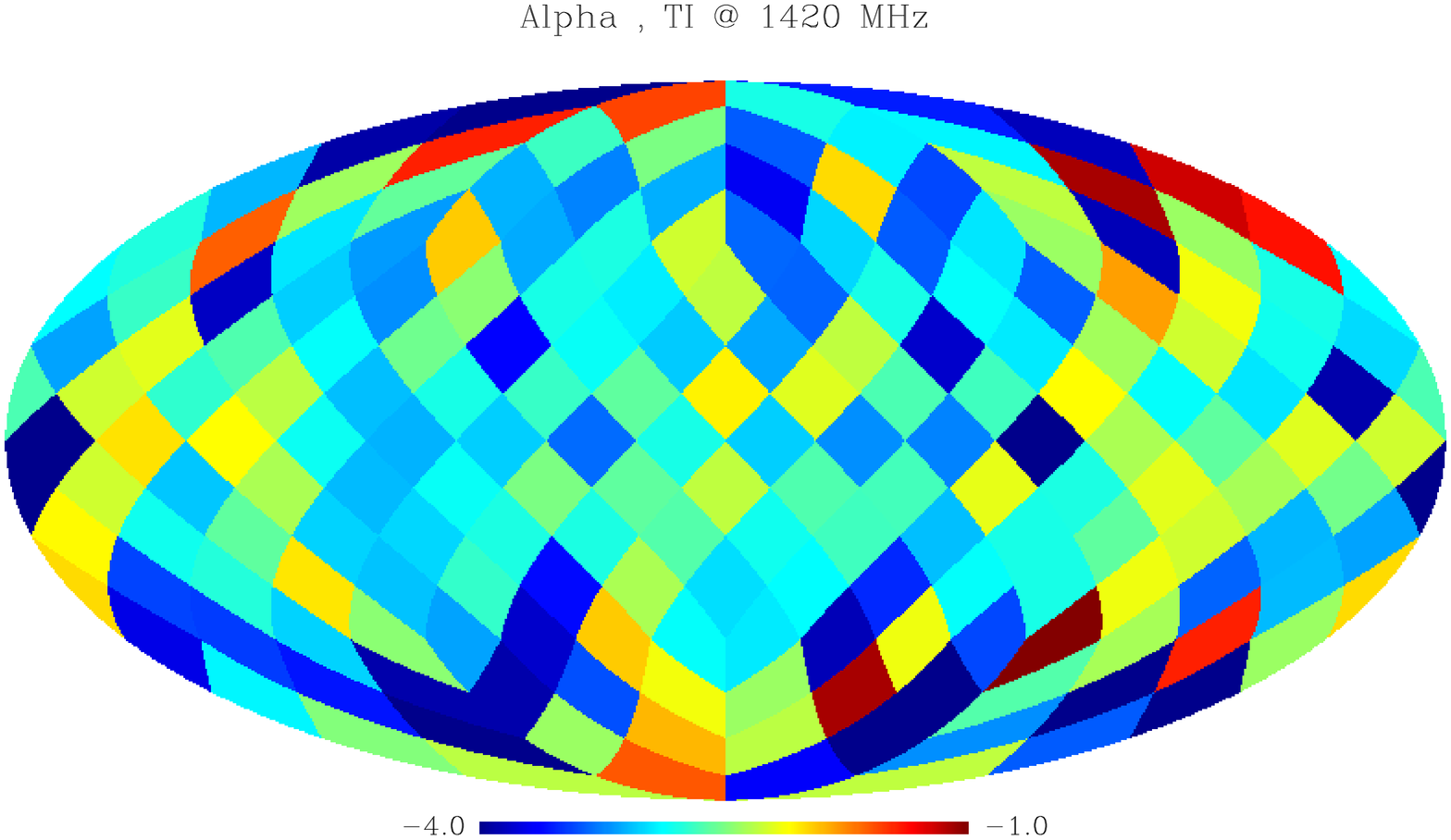}&
   \includegraphics[width=4.0cm,height=6cm,angle=90]{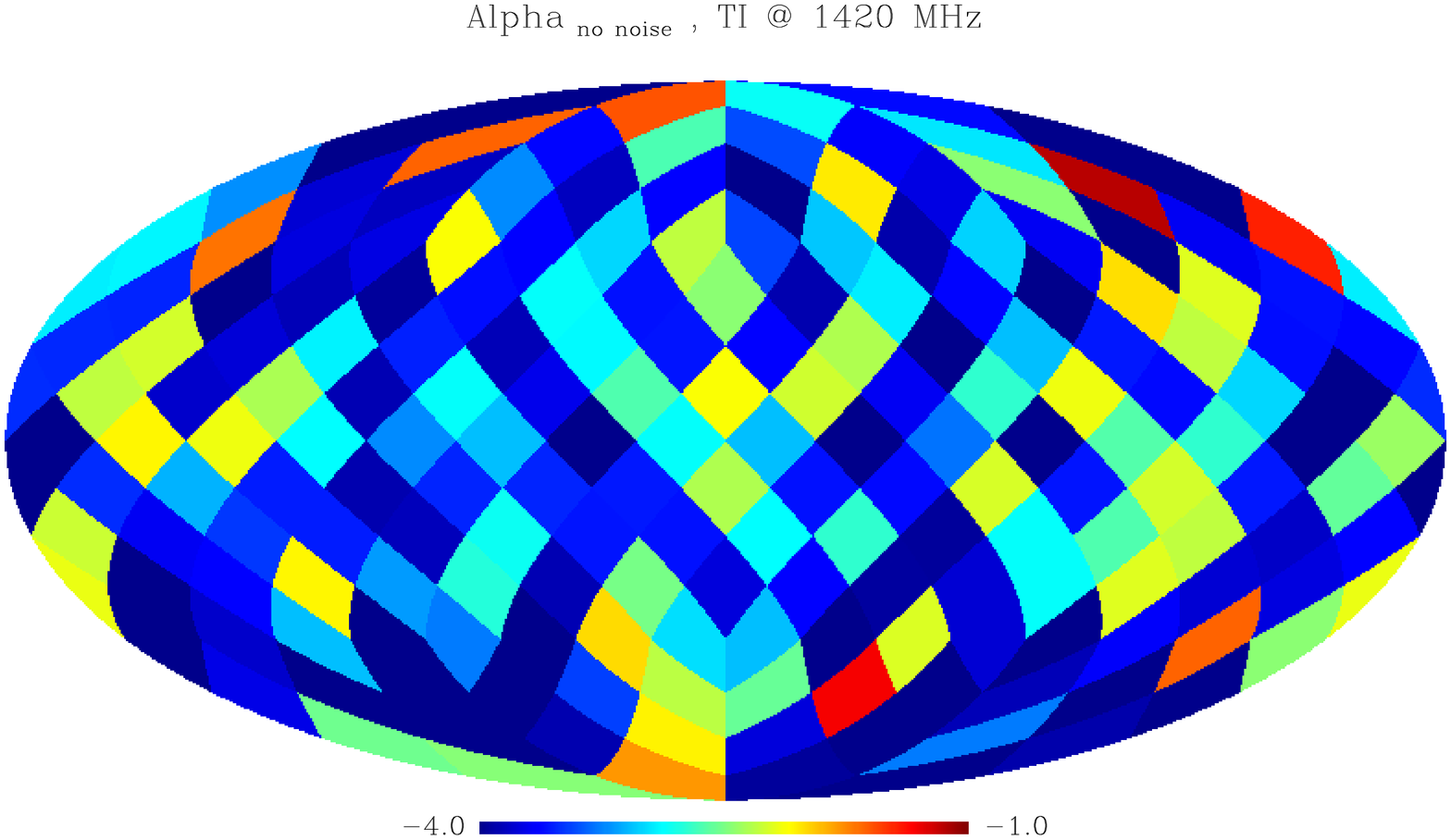}\\&
   \includegraphics[width=4.0cm,height=6cm,angle=90]{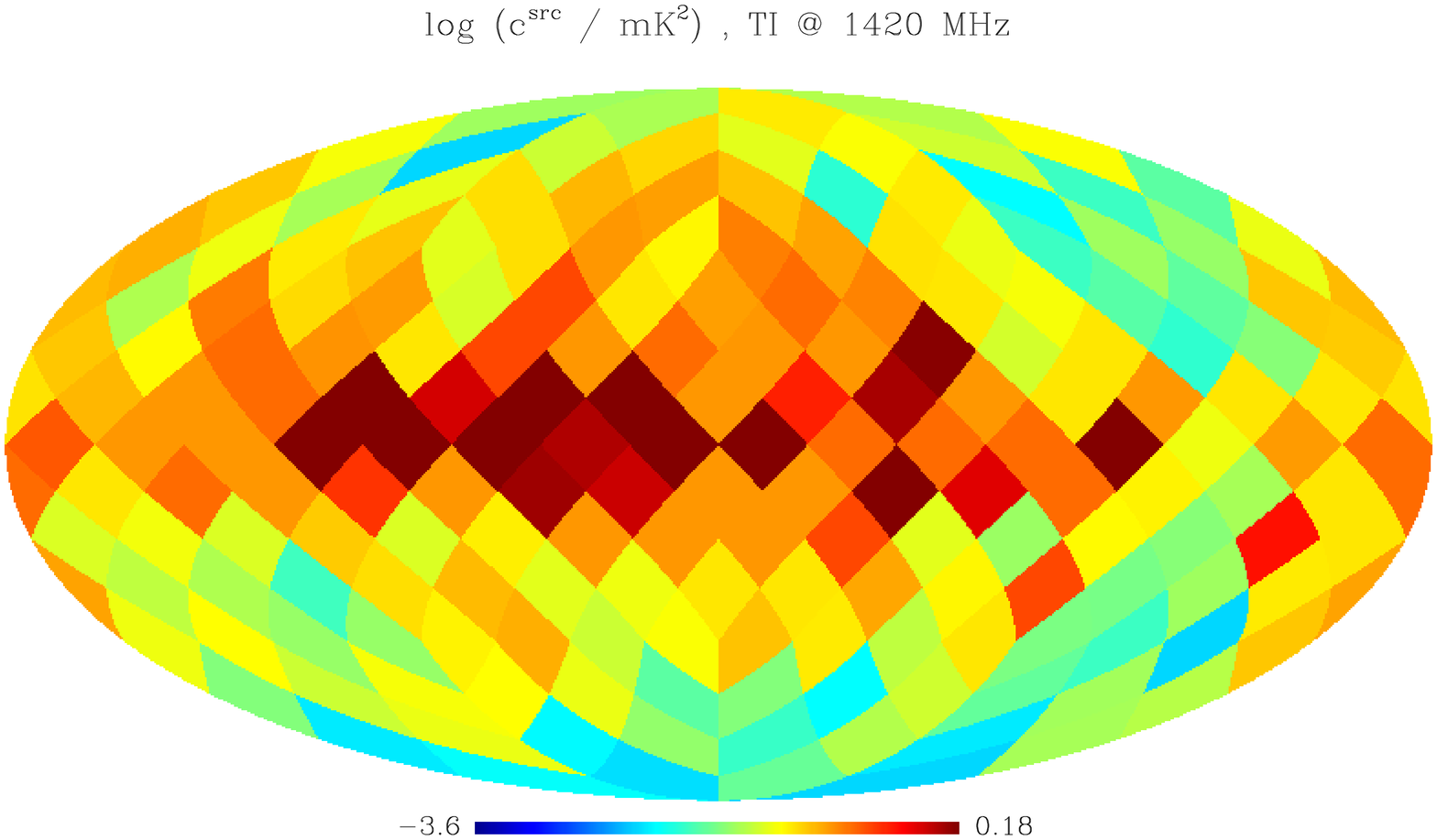}&
   \includegraphics[width=4.0cm,height=6cm,angle=90]{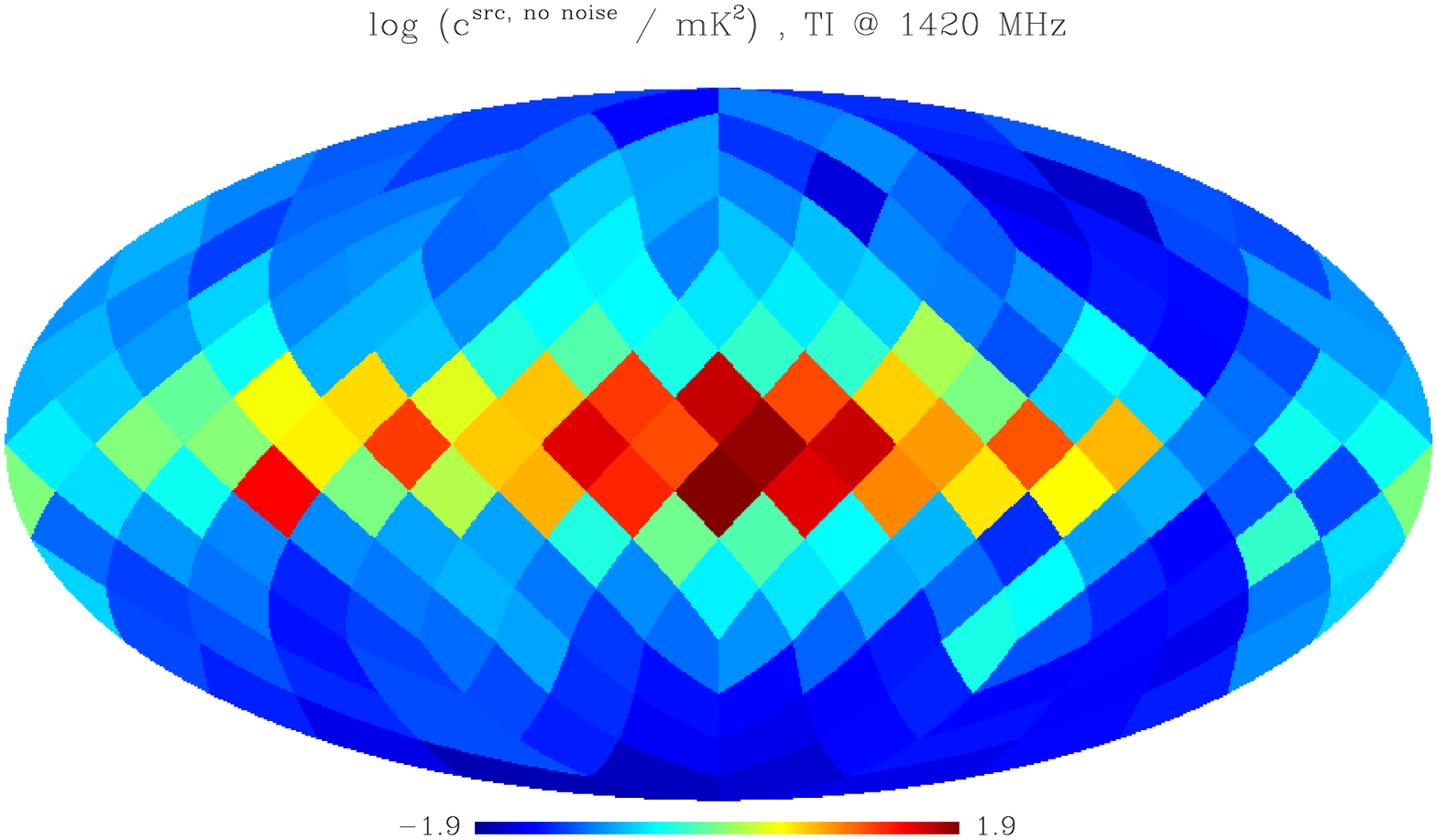}\\
   \includegraphics[width=4.0cm,height=6cm,angle=90]{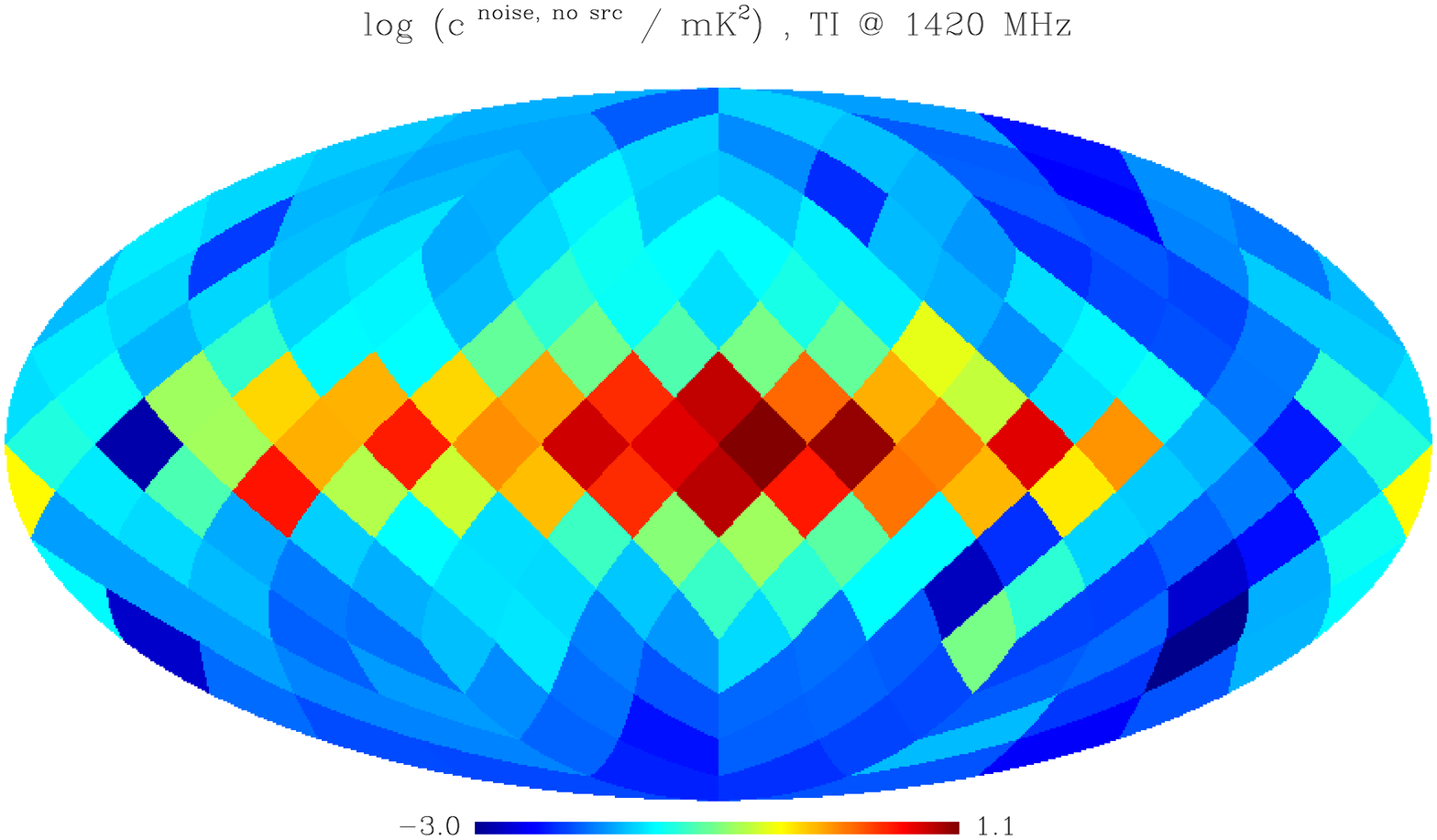} &
   \includegraphics[width=4.0cm,height=6cm,angle=90]{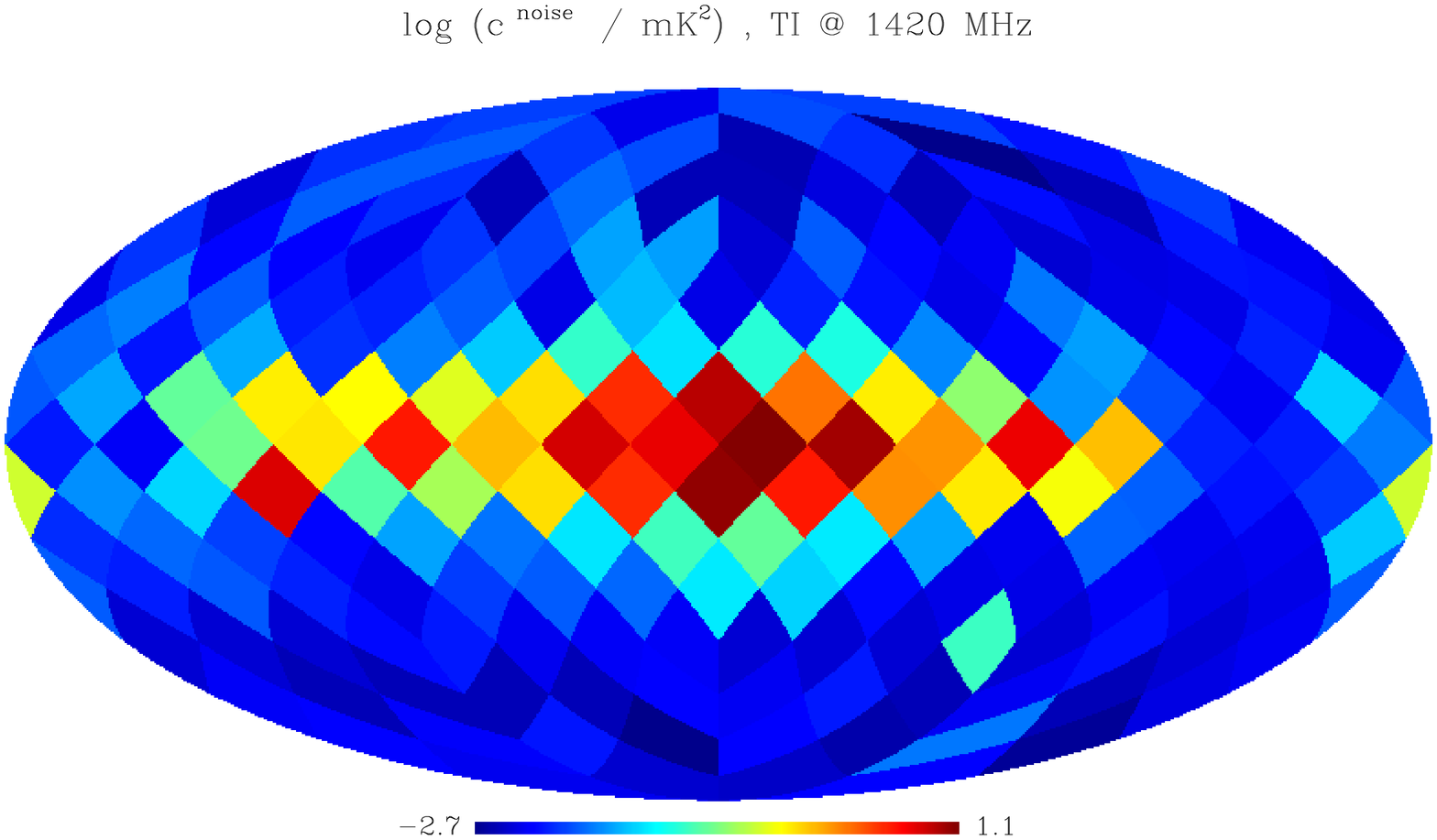}  &  \\
   \end{tabular}
   \caption{
   Maps of the best fit parameters obtained by fitting the 
   angular power spectra of the local analysis patches at 1420 MHz. 
   The maps aligned along each row refer to the same parameter. 
   From the top, ${\rm log}(k_{100}^{synch}/{\rm mK}^2)$, $\alpha^{synch}$ , 
   ${\rm log} (c^{src}/{\rm mK}^2)$ and  ${\rm log}(c^{noise}/{\rm mK}^2)$.
   The first and third columns correspond to the extreme cases,
   assuming respectively that the source contribution 
   or the noise contamination is negligible.
   }
   \label{BFparmap_ti1420}
\end{figure*}

\begin{figure*}
   \hskip -0.5cm
   \begin{tabular}{ccc}
   Fit no SRC & Best Model & Fit no Noise \\
   \includegraphics[width=4cm,height=6cm,angle=90]{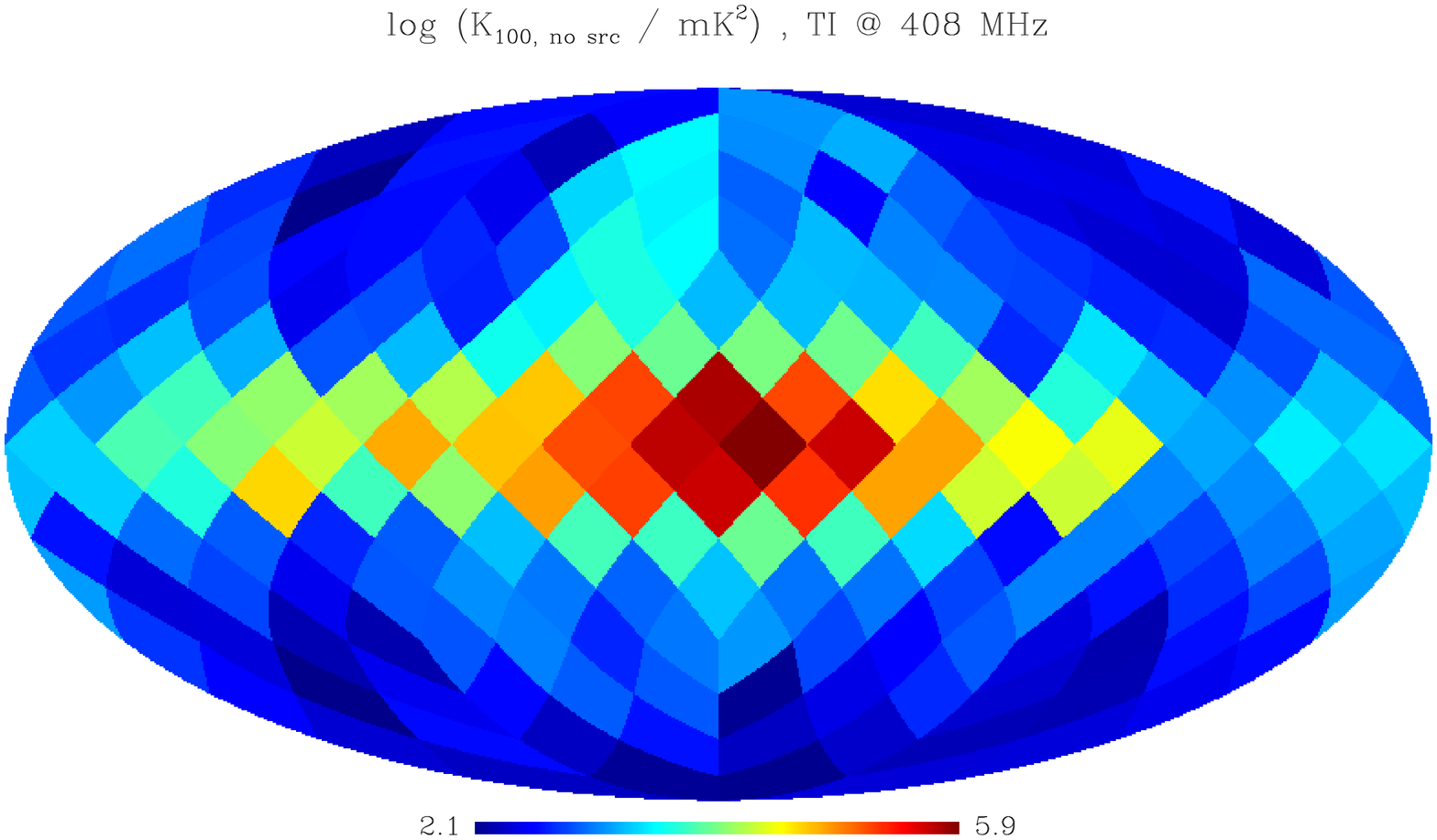}&
   \includegraphics[width=4cm,height=6cm,angle=90]{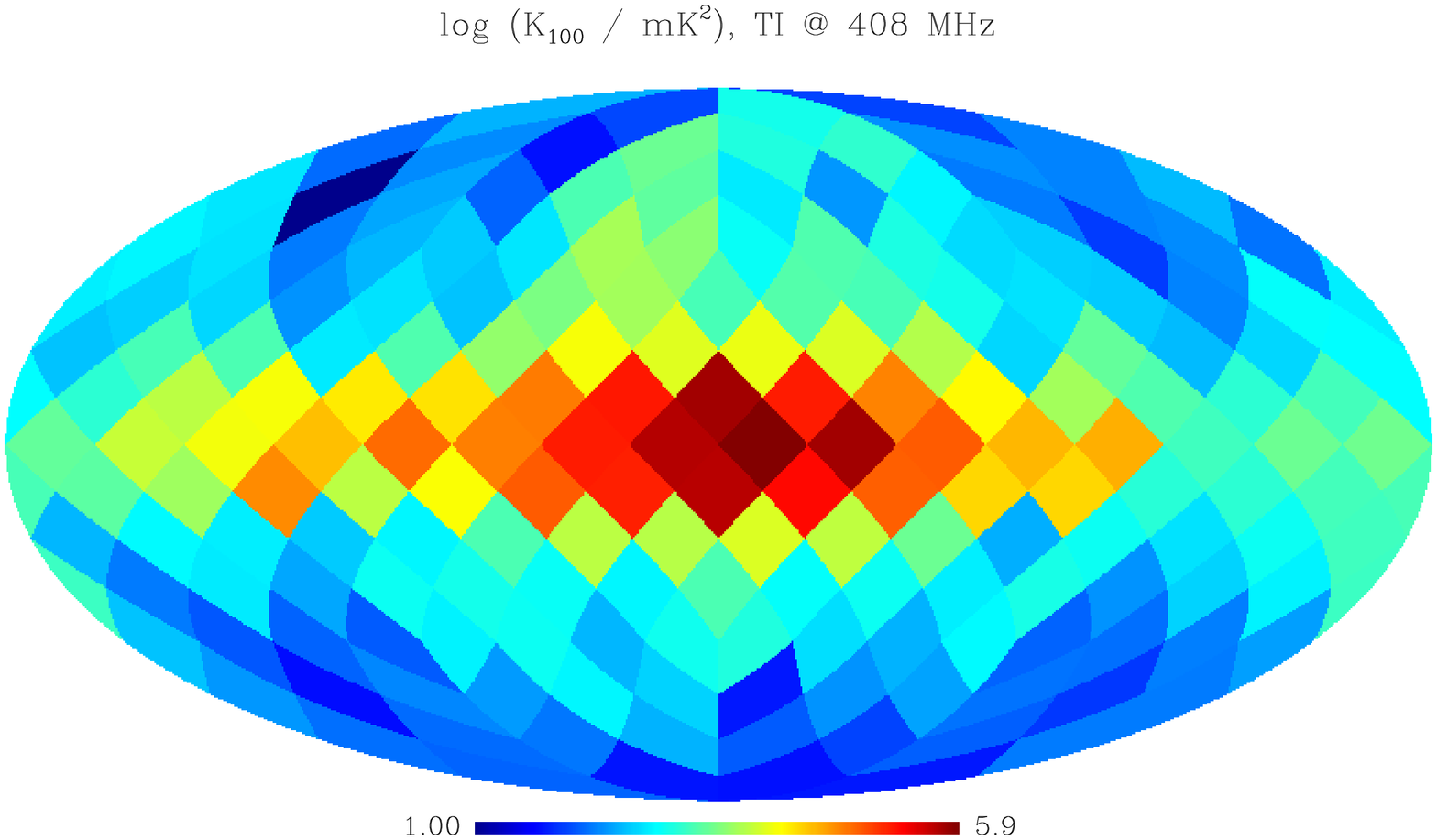}&
   \includegraphics[width=4cm,height=6cm,angle=90]{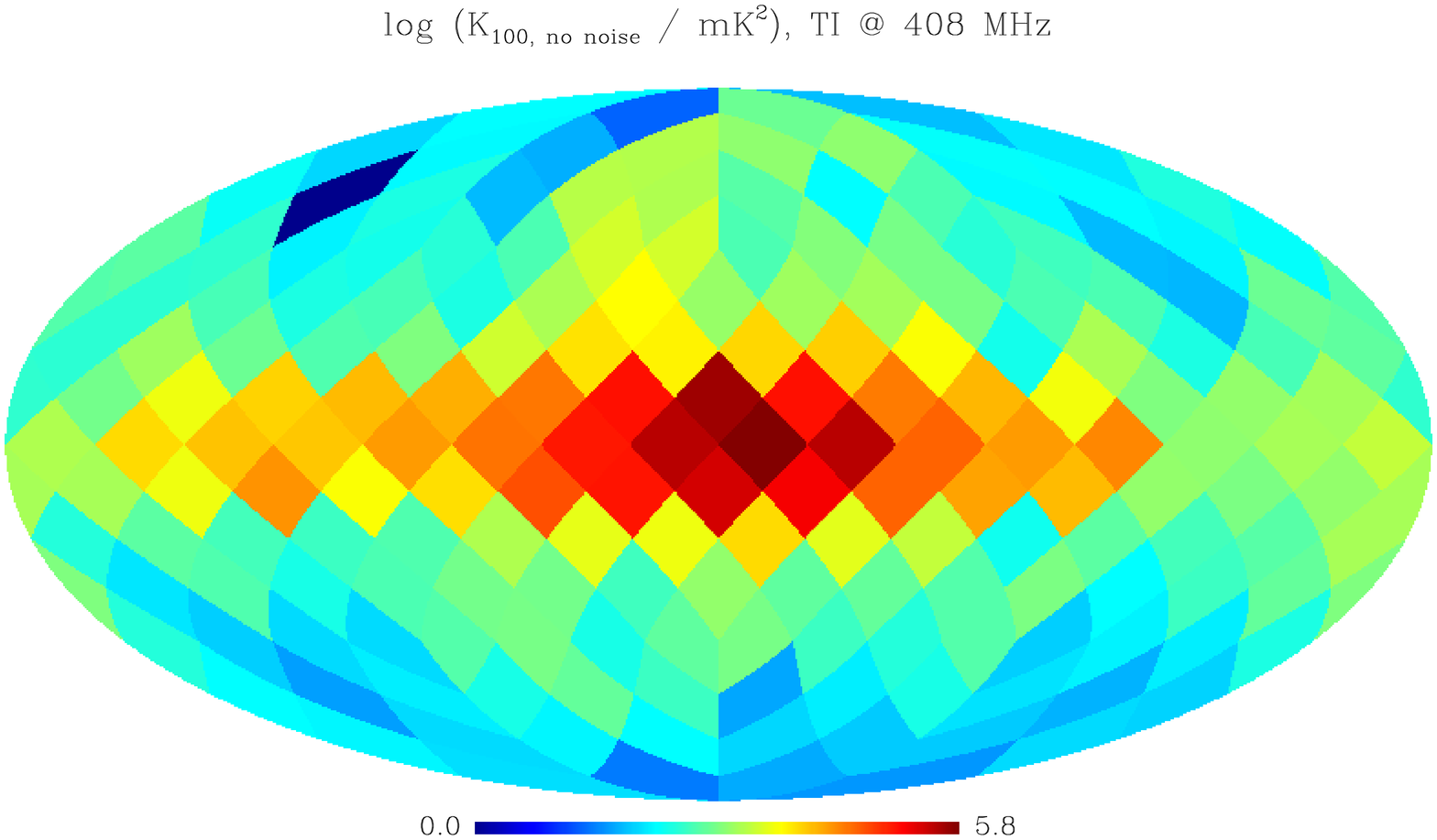}\\
   \includegraphics[width=4cm,height=6cm,angle=90]{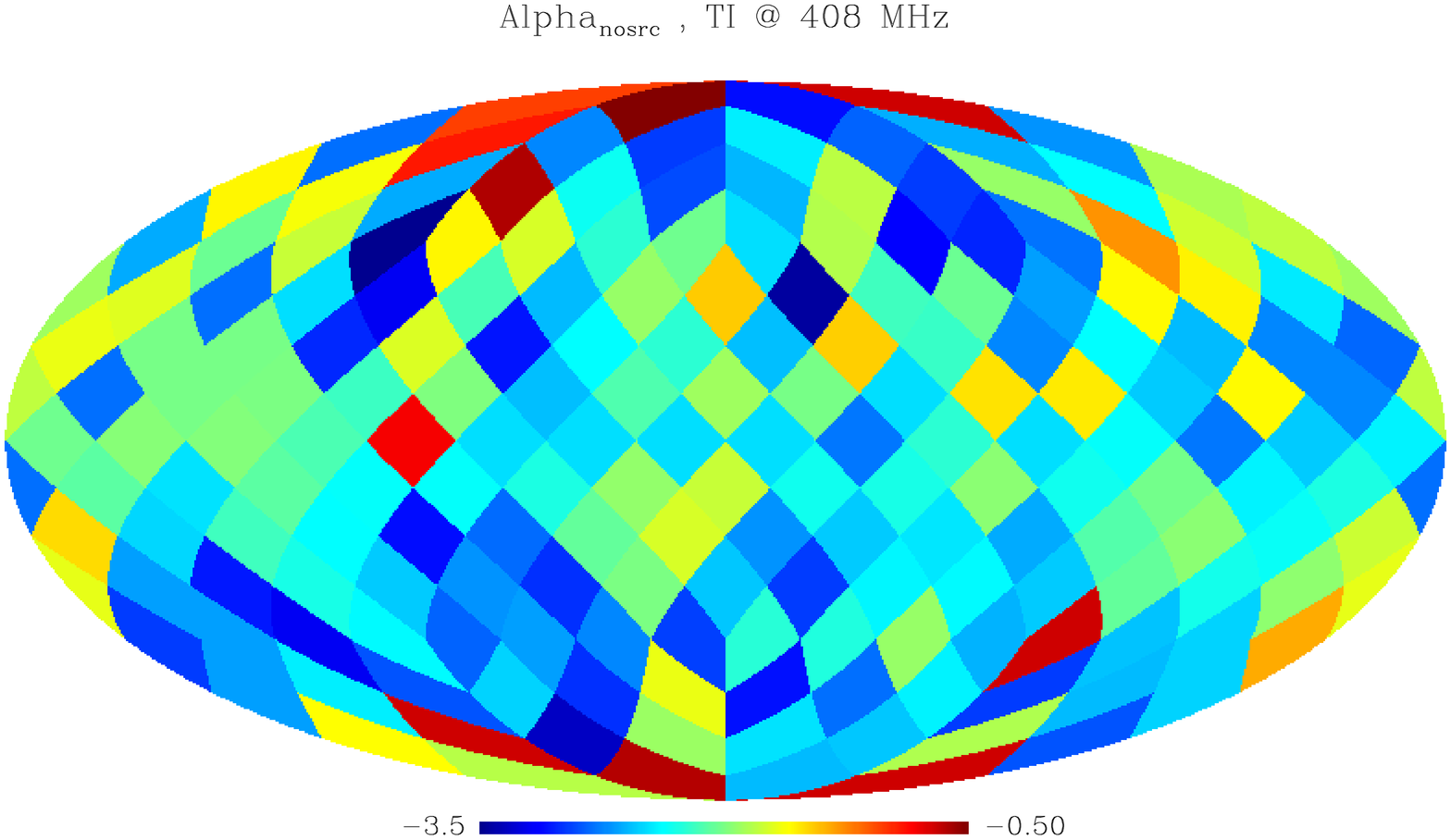}&
   \includegraphics[width=4cm,height=6cm,angle=90]{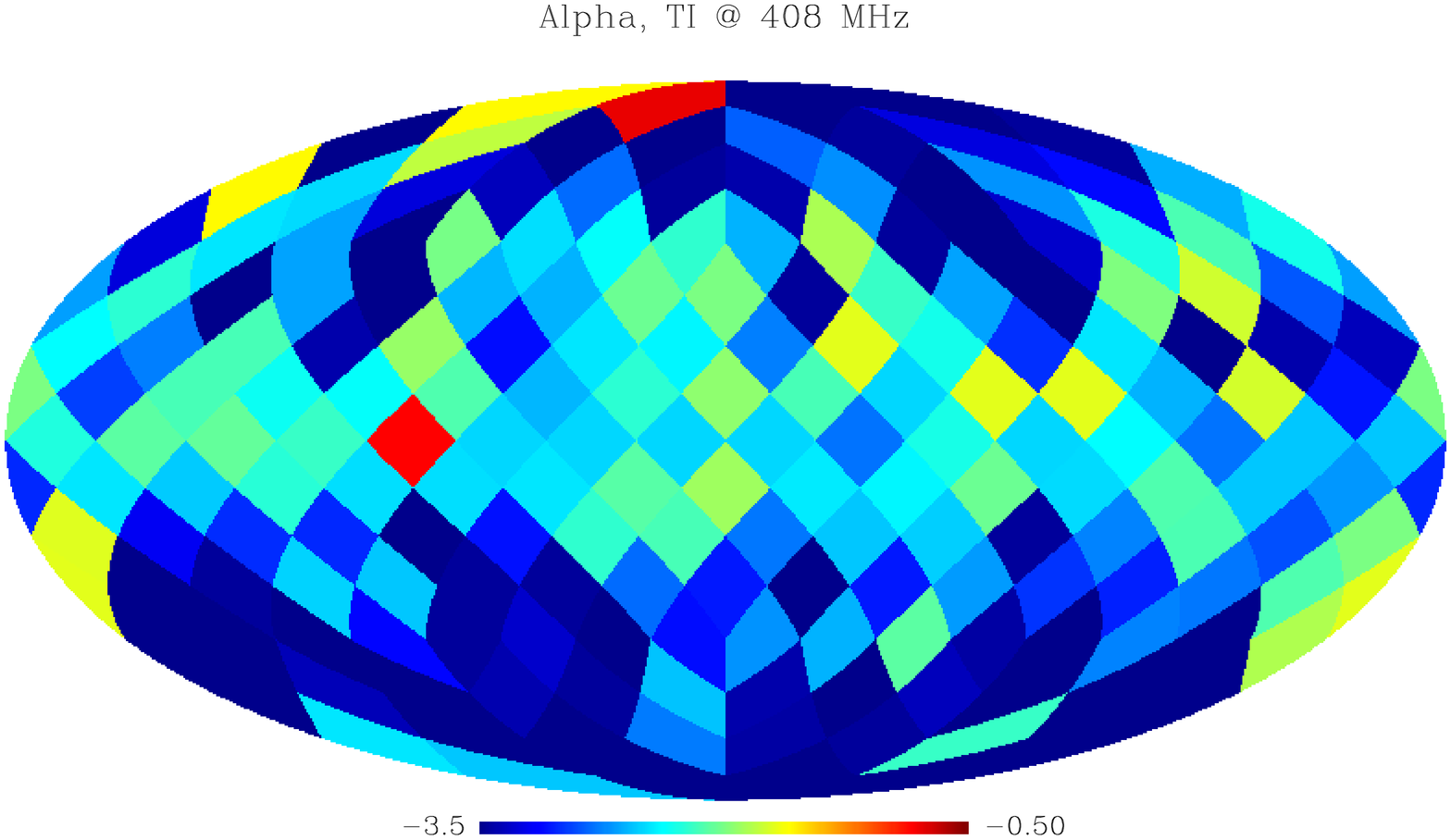}&
   \includegraphics[width=4cm,height=6cm,angle=90]{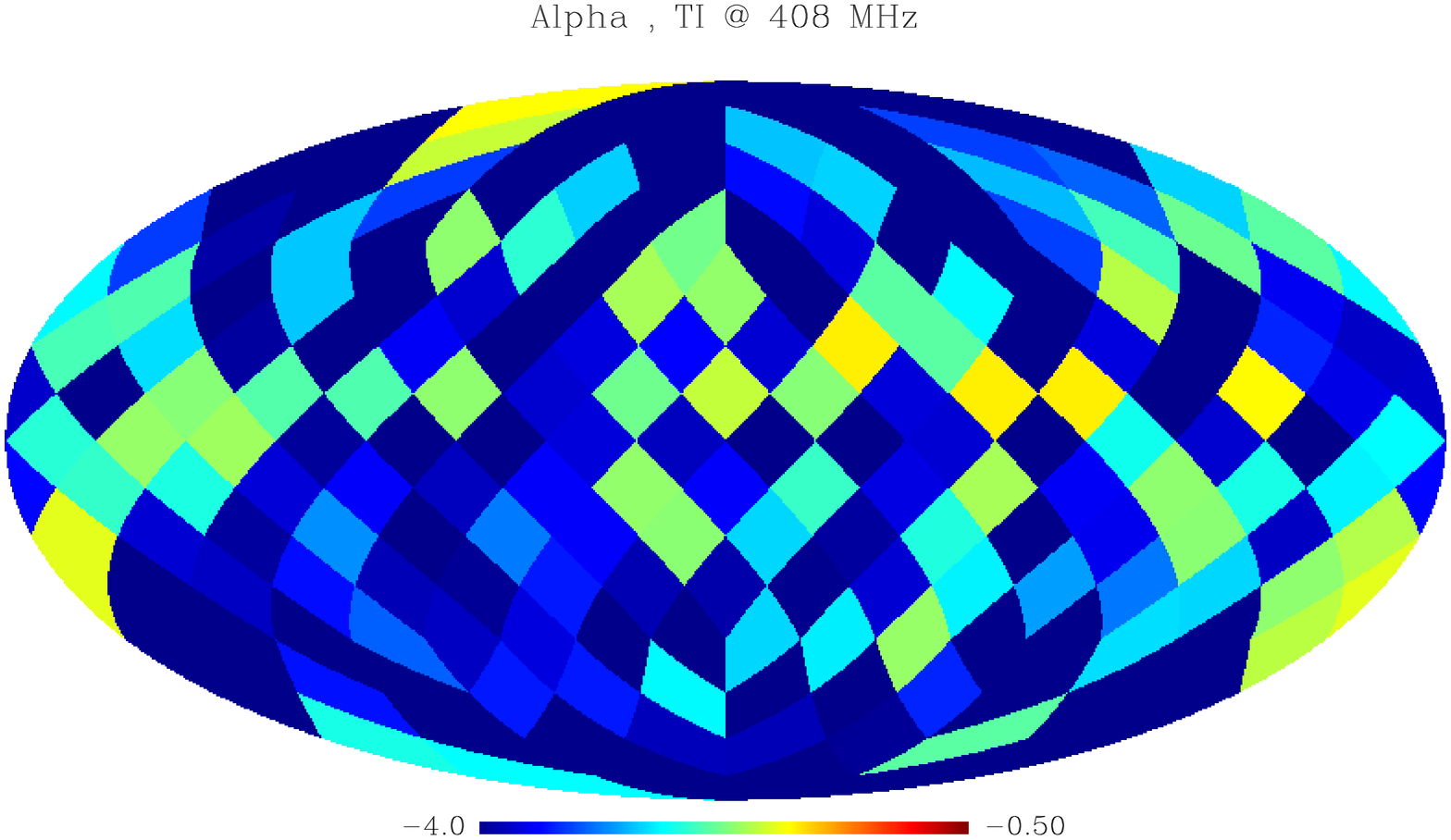}\\
 & \includegraphics[width=4cm,height=6cm,angle=90]{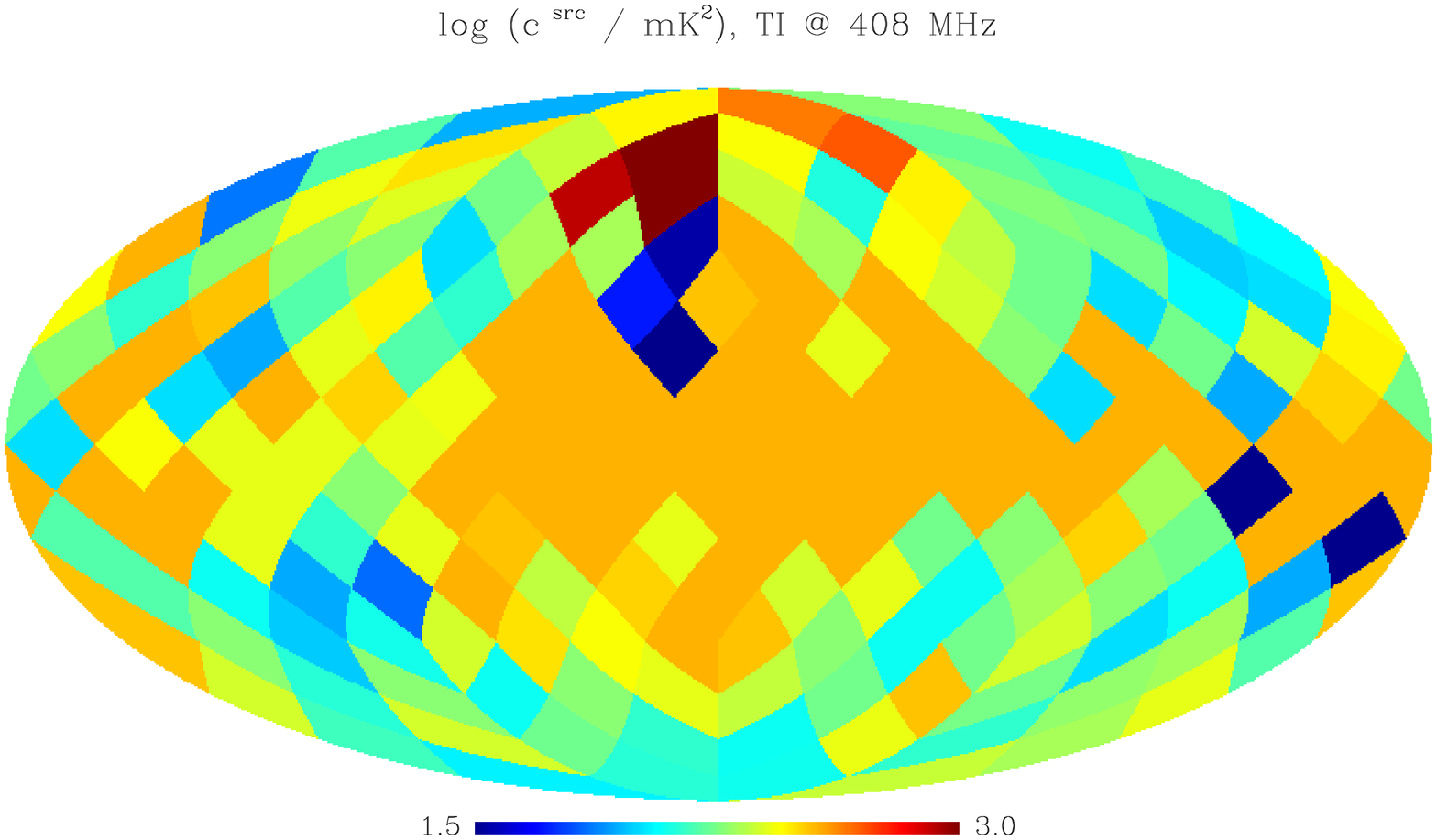}&
   \includegraphics[width=4cm,height=6cm,angle=90]{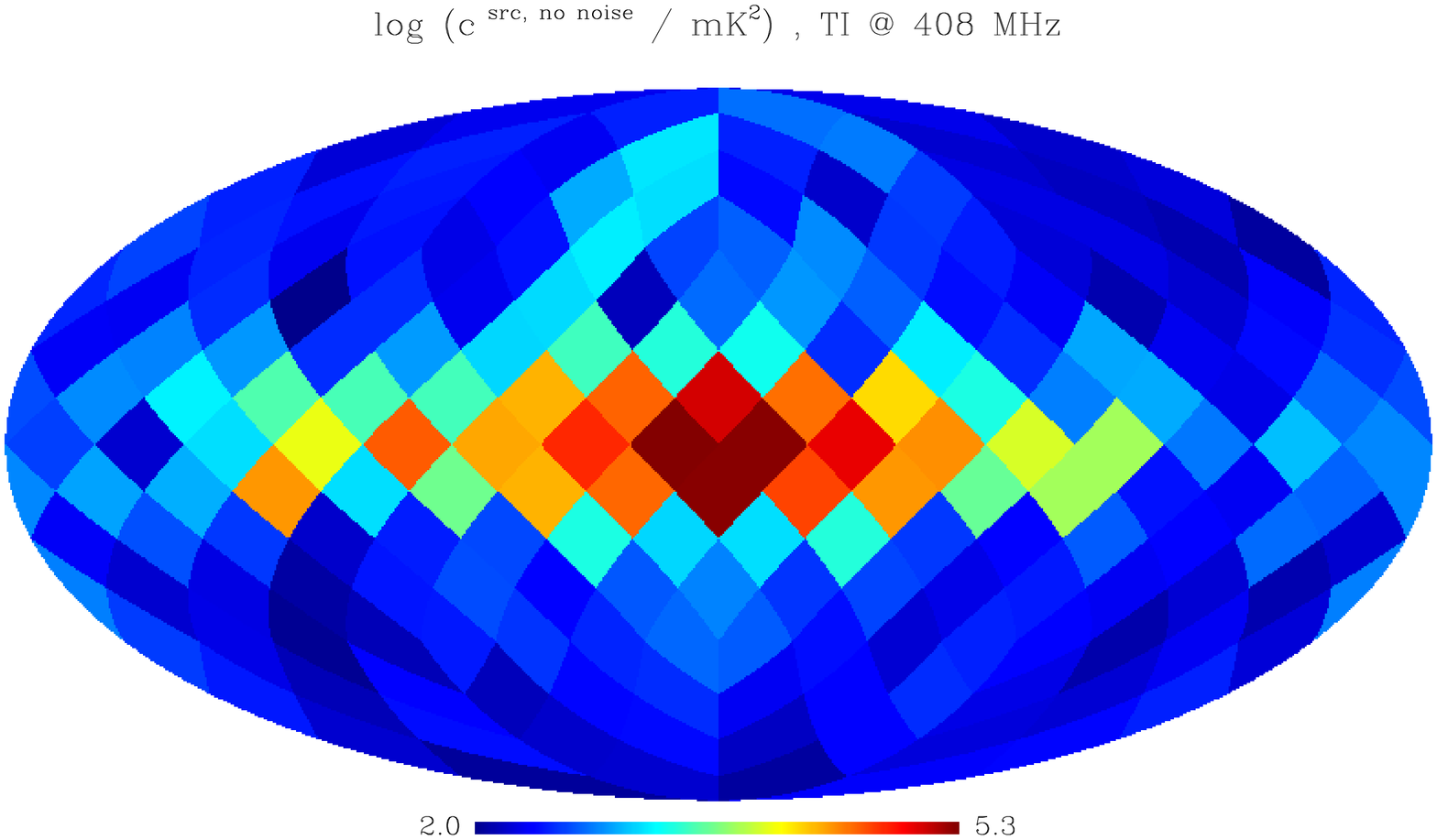}\\
   \includegraphics[width=4cm,height=6cm,angle=90]{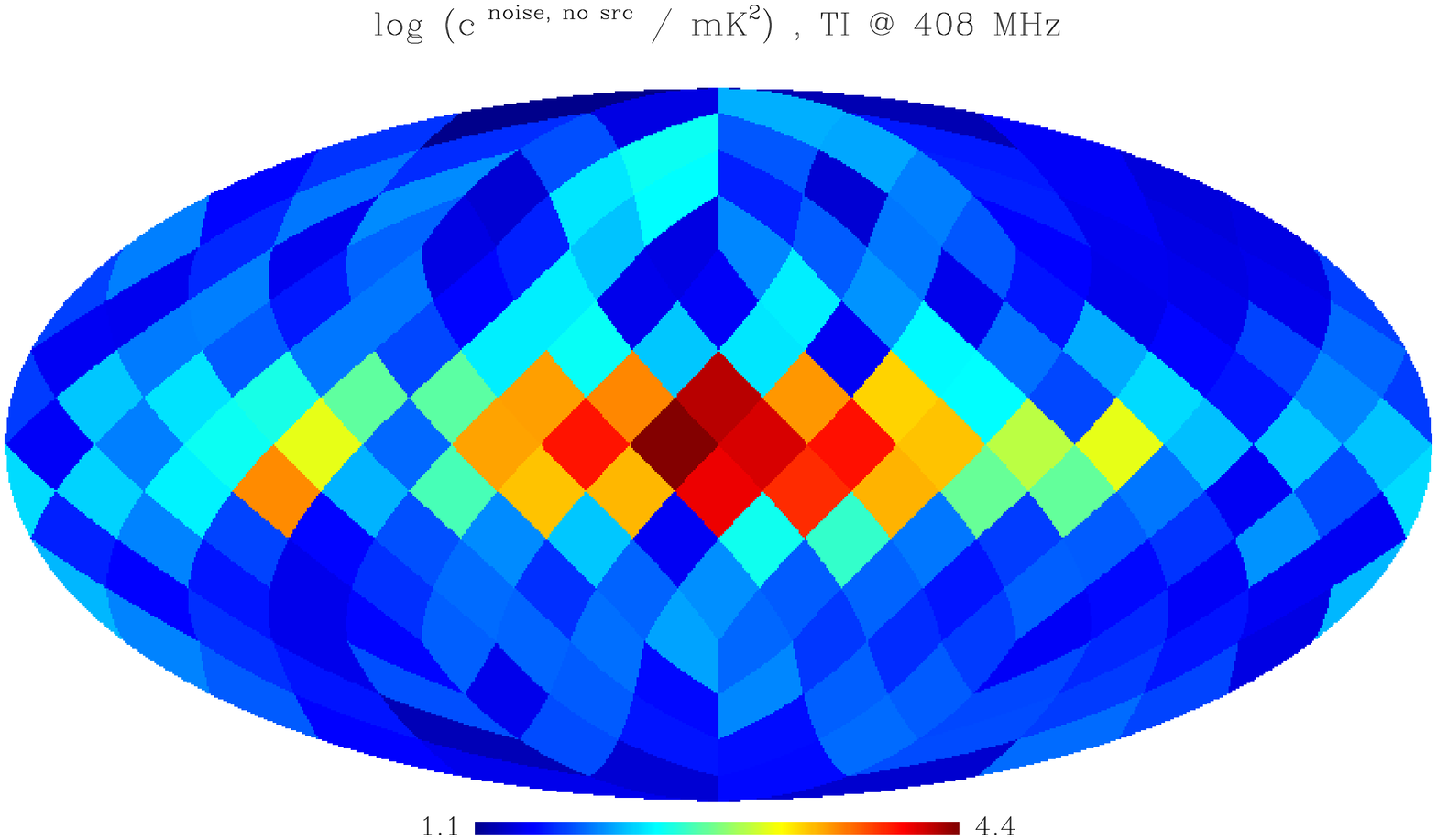} &
   \includegraphics[width=4cm,height=6cm,angle=90]{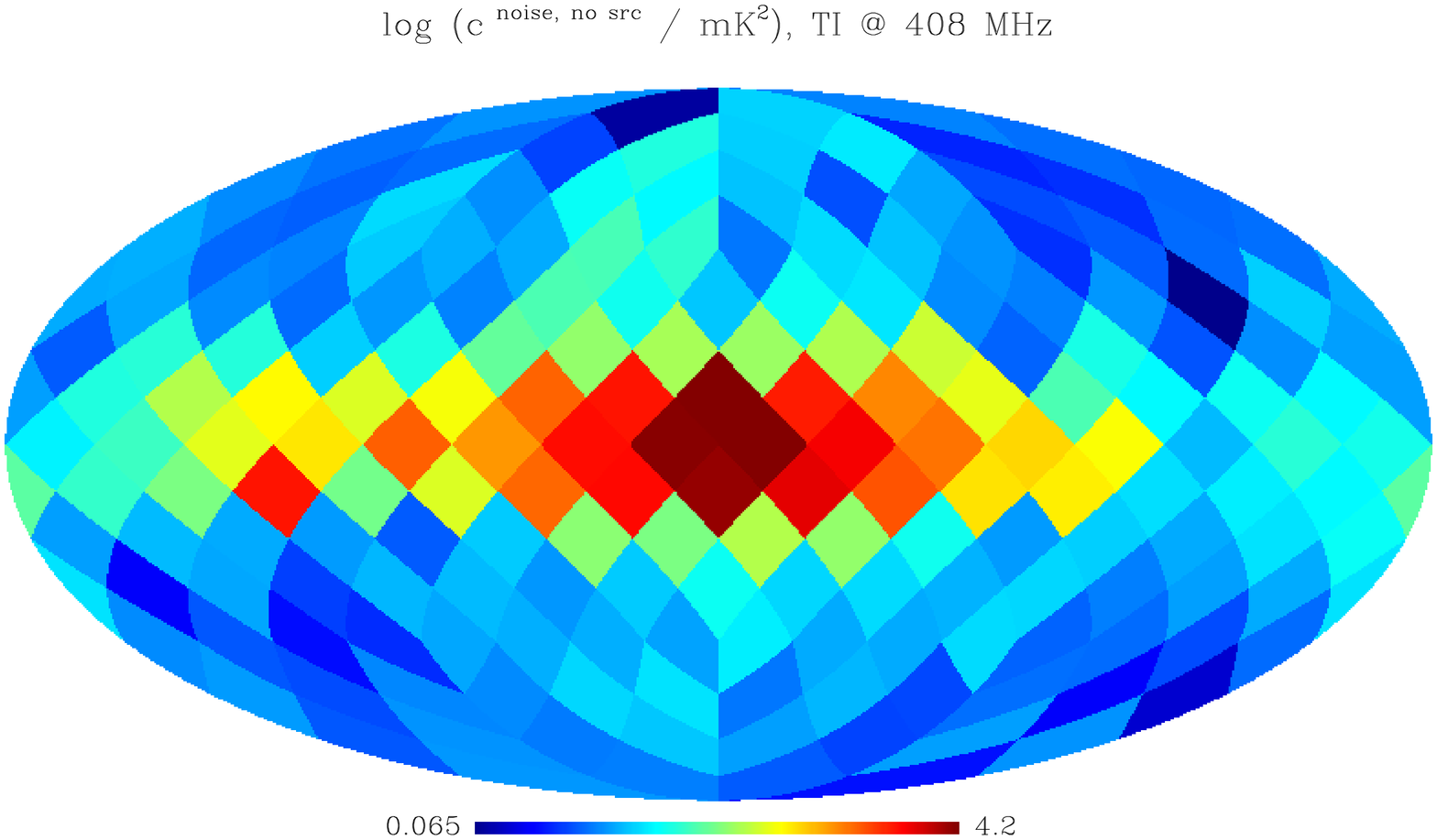} & \\
   \end{tabular} 
   \caption{ As in Fig.~\ref{BFparmap_ti1420}, but at 408 MHz.} 
   \label{BFparmap_ti408}
 \end{figure*}

The patch angular power spectra for the map after DS subtraction
 are fitted exactly as done in the case of the 
 Galactic cuts (see Sect.~\ref{sect_fit_aps_nos}).  
The maps of the obtained parameters are shown in 
Figs.~\ref{BFparmap_ti1420} and \ref{BFparmap_ti408}. 
The results derived by using the best model are summarized 
in Tables~\ref{tab_local_res_04} and \ref{tab_local_res_14}. 
The estimated relative error of the synchrotron APS slope 
averaged over the ensamble of patches is 
$|\Delta\alpha/\alpha| \sim 25\%$ at 408~MHz and 22\% at 1420~MHz.  
The mean relative error of the normalized 
amplitude, $k_{100}$, is $\sim 25\%$ at 408~MHz 
and $\sim 20\%$ at 1420~MHz. 
\begin{table}[!t]
\begin{center}
\begin{tabular}{|c|c|c|c|c|c|} 
\hline 
 parameter, $x$ & $x_{min}$ & $x_{max}$ & $< x >$ & $\sigma_{x}$ & $\%\,(x \in < x > \pm \sigma_{x})$ \\
\hline
 $\alpha$ & -3.50 & -0.70 & -2.70 & 0.60 & 54 \\
 ${\rm log}(k_{100}^{1}/{\rm mK}^2)$ & 2.90 & 5.90 & 3.80 & 0.81 & 70 \\
 ${\rm log}(k_{100}^{2}/{\rm mK}^2)$ & 1.00 & 2.90 & 2.40 & 0.36 & 69 \\
\hline 
\end{tabular}
\end{center}
\caption{Characteristics of the synchrotron APS best fit parameters 
derived in the local analysis of the 408 MHz map. $k_{100}^{1}$ refers to 
the patches covering about 
the brightest half of the sky, which includes the Galactic plane and 
the NPS. $k_{100}^{2}$ correspond to the other half with weak 
high latitude emission. } 
\label{tab_local_res_04}
\end{table}
\begin{table}[!t]
\begin{center}
\begin{tabular}{|c|c|c|c|c|c|} 
\hline 
 parameter, $x$ & $x_{min}$ & $x_{max}$ & $< x >$ & $\sigma_{x}$ & $\%\,(x \in < x > \pm \sigma_{x})$ \\
\hline
 $\alpha$ & -4.00 & -1.00 & -2.80 & 0.60 & 70 \\
 ${\rm log}(k_{100}^{1}/{\rm mK}^2)$ & -0.30 & 3.00 & 0.66 & 0.91 & 78 \\
 ${\rm log}(k_{100}^{2}/{\rm mK}^2)$ & -1.90 & -0.31 & -0.85 & 0.38 & 67 \\
\hline 
\end{tabular}
\end{center}
\caption{As in Table~\ref{tab_local_res_04}, but at 1420 MHz.} 
\label{tab_local_res_14}
\end{table}
%
%

\begin{figure}[!t]
  \includegraphics[width=6cm,height=9cm,angle=90]{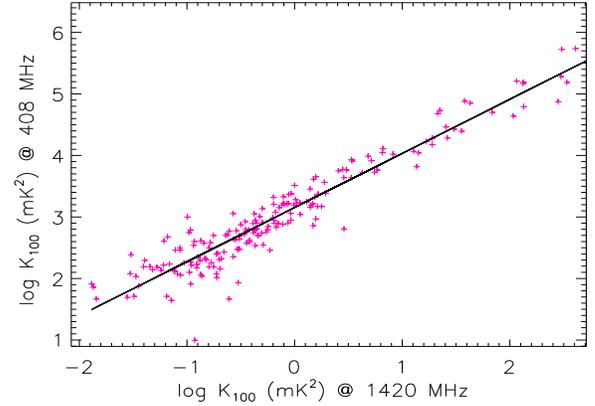}
   \vskip -0.2cm
   \caption{ Correlation between the best fit values obtained 
    for the synchrotron emission APS 
    normalized amplitude ($k_{100}$)
    at 408 MHz and 1420 MHz. 
     }
   \label{par_corr_2freq}
\end{figure}

The most striking result is that at each frequency 
the maps of the corresponding parameters, derived  
by adopting the three different fitting models,  
show a very similar morphology. 
Such a resemblance proves that the parameter patterns 
revealed by the local analysis are reliable, 
despite the uncertainties in the obtained 
parameter values.\\ 
The slope of the synchrotron APS does not show a 
systematic dependence on Galactic latitude, in agreement 
with the findings of Sect.~3.3. 
The normalized amplitude of the synchrotron APS, $k_{100}$, peaks 
close to the Galactic plane, which reflects the 
observed morphology. 
A good correlation is found between the 
normalized amplitude of the synchrotron APS 
at 408 MHz and 1420 MHz, 
which is defined by   
${\rm log}(k_{100}^{408}/{\rm mK}^2) \sim {\rm A} + {\rm B}\,{\rm log}(k_{100}^{1420}/{\rm mK}^2)\,$, 
where ${\rm A}=3.15\pm0.02$ and ${\rm B}=0.88\pm0.02$ 
(see Fig.~\ref{par_corr_2freq}). 
We note that $10^{\rm A} \sim (408/1420)^{2\beta}$ 
with $\beta \sim -2.9$, in agreement with the results 
of Table~\ref{beta_408_1420}. 

The contribution of sources reaches a maximum in 
the vicinity of the Galactic plane,  
mainly because a less complete source subtraction was possible for 
  $|b_{gal}| \lesssim 45^{\circ}$ than at higher 
  latitudes (see Sect.~\ref{DS}). 
The obtained source terms are in fair agreement with 
the values estimated by using source counts.
This comparison is particularly significant  
at 1420 MHz, where such estimates are more reliable. 
We summarize the results obtained in this case in 
Table~\ref{tab_csrc_14}. 
\begin{table}
\begin{center}
\begin{tabular}{|c|c|c|c|c|c|} 
\hline 
 parameter, $x$ & $x_{min}$ & $x_{max}$ & $< x >$ & $\sigma_{x}$ & $\%\,(x \in < x > \pm \sigma_{x})$ \\
\hline 
${\rm log}\,(c^{src}_{1}/{\rm mK}^2)$ & -2.30 & -0.75 & -1.50 & 0.42 & 66 \\ 
\hline
${\rm log}\,(c^{src}_{2}/{\rm mK}^2)$ & -2.00 & 0.18 & -0.86 & 0.53 & 72 \\
\hline
\end{tabular}
\end{center}
\caption{Characteristics of the source term derived in the local 
 analysis at 1420 MHz for $|b_{gal}| \gtrsim 45^{\circ}$ ($c^{src}_{1}$) 
and $|b_{gal}| \lesssim 45^{\circ}$ ($c^{src}_{2}$).}
\label{tab_csrc_14}
\end{table}

\section{Summary and conclusions}

The aim of our analysis is to improve our   
understanding of the Galactic synchrotron 
emission as a foreground for CMB dedicated experiments. 
For this purpose, we carried out an unprecedented detailed 
study of the Galactic radio emission, in terms of its 
angular power spectrum (APS), using total intensity 
all-sky maps at 408 MHz and 1420 MHz. \\
An accurate modeling of the synchrotron APS is missing 
in the literature so far, but is urgently required for a more 
precise and complete exploitation of the information 
awaited from the {\sc Planck} satellite. 
It constitutes a precious input for component separation 
activities, both for the realization of spatial templates 
of the foreground and for the definition of priors on 
its spatial and frequency dependence. 
\begin{itemize} 
\item[{\bf 1.}] Being interested in the diffuse component of the 
 synchrotron emission, the brighter discrete sources (DS) 
 have been eliminated from the radio maps by 
 2-dimensional Gaussian fitting.  
 This approach is very flexible and also permits  
 the removal of extended structures.  
 \item[{\bf 2.}] The APS was computed 
 for both large areas and 
 small patches and several consistency tests were
 used to check the reliability of the recovered
 APS in the case of limited sky coverage (see Appendix~B of 
 La Porta 2007). \\
 The study of the APS for various cuts, i.e. of  
 regions with Galactic latitudes above or below a 
 certain value $b_{cut}$ ($|b_{cut}| \in [5^{\circ},60^{\circ}]$), 
  allowed us to explore a possible dependence of 
 the mean properties of the Galactic synchrotron 
 emission on latitude, preserving at the same time 
 the largest possible coverage, which is important when
 estimating the CMB APS because of the sampling variance. 
 Such cuts provided information for $\ell \in [\ell_{min},\ell_{max}]$,
 where $10 \lesssim \ell_{min} \lesssim 30$ for increasing $b_{cut}$ 
 and $\ell_{max} \sim 200 -300$ at 408 MHz and 1420 MHz, respectively. \\ 
 The patches correspond to the pixels of an {\tt HEALPix} map at 
 $n_{side}=4$, which have an angular dimension of $\sim 15^{\circ}$ 
 and permit us to investigate the local variations of the synchrotron APS 
 for multipoles larger than $\sim 60$. \\
 The derived angular power spectra were modelled in both cases 
 according to Eq.~\ref{aps_best_model} 
and a specific method was set up to find 
the best least square fit on adaptive grids of the parameter 
space and to evaluate the uncertanties on the retrieved 
parameters (see Appendix~C of La Porta~2007 for details).
\item[{\bf 3.}] An indirect cross-check of the fit result reliability 
  was provided in the case of the Galactic cuts 
  by the estimated source terms, which are consistent with the 
  expectations from extragalactic source counts 
  at both frequencies. 
  Nowadays, source counts at 1.4 GHz are well established down to 
  very low flux limits.  
  It is remarkable that the source angular power spectra 
  obtained independently from source counts and from the 
  fit of the survey APS are in good agreement. 
\item[{\bf 4.}] The slope of the synchrotron APS,
 $C_{\ell} \sim k \ell^{\alpha}$, 
 changes with $b_{cut}$ without showing a well-defined regular trend,
 although it is found to be typically 
 steeper for $b_{gal} \gtrsim 20^{\circ}$.
 For the cuts, $\alpha$ varies in the range 
 $\sim [-3.0,-2.6]$ at both frequencies. 
 However, the analysis of the small patches gives evidence
 that locally the synchrotron APS can be much flatter 
 for $\ell \gtrsim 60$, reaching in some cases 
 values of $\alpha \sim -0.8$. 
 The normalized amplitude, 
 $k_{100}=k \times 100^{\alpha}$,  
 gradually increases toward the Galactic plane, 
 following, as expected, the background radio  
 emission gradient. 
 A good correlation exists between the 
 results obtained for $k_{100}$ 
 at 408 MHz and 1420 MHz, for both cuts and patches. 
 This is expected, given that 
 the spectral properties of the electron density distribution 
 responsible for the Galactic diffuse non-thermal emission 
 should be the same in that frequency range 
 and further supports the reliability of 
 the obtained estimates of the synchrotron APS. 
\item[{\bf 5.}] The maps of $k_{100}$ and $\alpha$ resulting 
from the local analysis represent the starting point for the 
simulation of small-scale fluctuation fields to be added to 
the DS-subtracted maps to build phenomenological 
templates of the Galactic synchrotron emission. 
At present, an empirical approach is the most reliable 
 way to proceed in the realization of realistic templates 
 of the foreground, given the poor knowledge 
of the Galactic magnetic field and of the cosmic ray 
electron density distribution needed for a 3-dimensional 
physical modelling. 
\end{itemize}
The issues (discussed in Sect.~\ref{extrap_kmem_comp}) 
related to the K-band synchrotron component retrieved 
by \citet{hins06_wmap_3yr_temp} are a clear example 
of the difficulties encountered in the foreground separation  
due to the lack of adequate priors and guess. 
We performed a source subtraction on that map and produced a 
map of the Galactic diffuse non-thermal emission 
at 23 GHz to be compared with those at lower frequencies.  
\begin{itemize} 
\item[{\bf 6.}] The extrapolation to 23 GHz of the APS 
 obtained at 408 MHz and 1420 MHz for higher latitude regions  
 ($|b_{gal}|\gtrsim 40^{\circ}$) reveals that the 
 mean spectral index 
 ($C_{\ell}(\nu) \propto \nu^{2\beta}$) 
 $\beta_{(0.408-23){\rm GHz}} \lesssim \beta_{(1.4-23){\rm GHz}}$, which 
 is the opposite of what is expected for synchrotron emission.  
 We estimate the excess of the 
 signal in the 23 GHz map to be $\gtrsim 50\%$ by using 
 the mean value of the APS at lower multipoles.  
 This result can be interpreted in terms of additional 
 contributions to the 23 GHz Galactic non-thermal emission, 
 which could be mainly due to anomalous dust. 
\end{itemize}

A direct application of the presented analysis is to  
determine the level of contamination of the CMB anisotropies 
due to the Galactic diffuse synchrotron emission 
at different angular scales.  
The conclusions reported below refer to the cuts, 
which are more relevant for CMB measurements 
because of the large coverage. 
However, similar considerations could be 
repeated using the results of the local analysis, 
thus allowing the identification of the clearest 
sky areas, which is essential for the success of 
ground-based CMB experiments. 
\begin{itemize}
\item[{\bf 7.}] An important - although not unexpected - outcome
 of the cut analysis is that 
 the amplitude of the APS for the northern hemisphere cuts
 above $\sim 20^{\circ}$ is raised 
 by the presence of the NPS. 
 Consequently, the results obtained separately for the 
 two Galactic hemispheres were used to bracket the APS 
 of the synchrotron emission. 
\item[{\bf 8.}] We used the APS results at 408 MHz and 1420 
 MHz to determine the frequency spectral index to be 
 adopted in the extrapolation to the microwave range 
 for each coverage case. In particular, we found that 
 $\beta_{(0.408-1.4){\rm GHz}} \in [-3.2,-2.9]$ with an  
 uncertainty of a few percent.     
\item[{\bf 9.}] The extrapolation to 30~GHz of the 
 synchrotron APS obtained at 1420 MHz 
 led to a signal in good agreement 
 with the upper limit fixed by COBE-DMR, thus 
 supporting the reliability of our results. 
 At this frequency the synchrotron emission 
 constitutes a severe contamination of  
 CMB anisotropies at the largest angular scales 
 ($\ell \lesssim 40$). Nevertheless, a cut at $\sim 20^{\circ}$  
 would reduce the synchrotron fluctuations to about 
 half the cosmological ones. The same holds 
 at $\nu \sim 70$~GHz for a cut at $\sim 5^{\circ}$ and 
 the situation further improves when excluding a larger 
 portion of the sky around the Galactic plane. 
 This implies that even though the current treatment of the
 foregrounds does not permit an accurate removal of the
 synchrotron emission, the latter does not prevent 
 the recovery of the bulk of the cosmological information 
 encoded in the CMB temperature APS.\\ 
\end{itemize}
A deeper understanding of the foreground 
is indispensable to settle other important issues  
 in the perspective of the forthcoming {\sc Planck} 
 mission. {\sc Planck} should achieve a sensitivity comparable 
 to the cosmic variance and its performance should 
be limited mainly by foregrounds, thanks to the 
extremely accurate control of all instrumental systematic 
effects \citep{buri04_PafterW,menn04_lfi}. \\
A first issue is the estimate of the CMB 
temperature-polarization APS, which carries information  
about the reionization history and the tensor-to-scalar 
ratio (see, e.g. Kogut~2003, Kogut et al.~2003, and 
references therein). Possible foreground residual 
contamination in the total intensity CMB anisotropy 
map would affect fine analysis based on 
the estimate of the cross-correlation APS, 
also because the polarized component of the 
cosmological signal is orders of magnitude lower. \\ 
Another issue is the evaluation of the Gaussianity 
of the primordial fluctuations, which in the standard 
inflationary paradigm\footnote{
Simple (standard) inflationary scenarios
(see the reviews by Lyth \& Riotto~1999 and Linde~2005)
predict the existence of Gaussian density fluctuations
and of gravitational waves
with a nearly scale-invariant spectrum. }
generate the
structures observed in the Universe today.
Gaussianity tests are a powerful tool, complementary to
 the tests exploiting the APS, which allow us 
 to probe the ``concordance''
model \citep{sperg06_wmap_3yr_param} 
and also to distinguish among inflationary
models \citep{bart04_nonG}. 
The level of non-Gaussianity predicted
 by the ``concordance'' model
cannot be detected by {\sc WMAP}, 
but in principle it should be observable with {\sc Planck}. 
Galactic foregrounds are non-Gaussian and anisotropic 
and even low-level contamination in the maps 
can produce detectable non-Gaussianities (see, e.g.  
Naselsky et al.~2005), although they have minimal 
effects on the estimated APS \citep{hin03_wmap_1yr_aps}.  
Consequently, the foreground removal has to be 
extremely accurate, so as not to limit {\sc Planck} 
in verifying this crucial\footnote{Standard cosmologies
predict a minimum level of non-Gaussianity for the
the primordial perturbations. In alternative
cosmologies \citep{lyth03_curvaton,ark04_ghostinfl,alish04_altercosmol} 
such a lower limit is even  higher. Consequently,
detection or non-detection of non-Gaussianities
 sheds light on the physics of the early Universe.}
prediction. 

\begin{acknowledgements} 

We are grateful to R. Wielebinski for a careful reading of the 
original manuscript. We wish to thank G. De Zotti, L. Toffolatti 
and R. Rebolo for helpful discussions.   
L.L.P. warmly thanks R. Wielebinski and A. Zensus for 
granting a post-doc fellowship. 
C.B. acknowledge the support by 
the ASI contract "Planck LFI Activity of Phase E2".
We are grateful to M. Genghini for technical support. 
Some of the results in this paper have been derived 
using the HEALPix \citep{gorski05_healpix} package. 
The availability of the WMAP 3-yr maps is acknowledged. 
We warmly thank the anonymous referee for useful comments.

\end{acknowledgements}

\end{document}